\documentclass[prl,aps,epsfig,superscriptaddress,twocolumn]{revtex4-2}
\setcitestyle{super}
\usepackage{amsfonts,amssymb,amscd,amsthm}
\usepackage{graphicx}
\usepackage{mathrsfs}
\usepackage{etoolbox}
\usepackage{upgreek}

\usepackage{amsfonts}
\usepackage{amssymb}
\usepackage{hyperref}
\usepackage{graphicx}
\usepackage{dcolumn}
\usepackage{bm,amsmath,verbatim}
\usepackage{mathrsfs}
\usepackage{color}
\usepackage{dsfont}
\usepackage{amsmath,amssymb,amsthm}
\usepackage{mathrsfs}

\usepackage[T1]{fontenc}
\usepackage[latin9]{inputenc}
\usepackage{booktabs}

\hypersetup{colorlinks,linkcolor=blue,citecolor=blue,filecolor=blue,urlcolor=blue}
\newcommand{\ket}[1]{|#1\rangle}
\newcommand{\bra}[1]{\langle #1 |}

\begin{document}


\setlength\columnsep{25pt}

\title{Observation of average topological phase in disordered Rydberg atom array}
\author{Zongpei Yue}
\thanks{These authors contributed equally to this work.}
\affiliation{State Key Laboratory of Low Dimensional Quantum Physics,
Department of Physics, Tsinghua University, Beijing 100084, China}
\author{Yu-Feng Mao}
\thanks{These authors contributed equally to this work.}
\affiliation{Center for Quantum Information, IIIS, Tsinghua University, Beijing 100084, China}
\author{Xinhui Liang}
\thanks{These authors contributed equally to this work.}
\affiliation{State Key Laboratory of Low Dimensional Quantum Physics,
Department of Physics, Tsinghua University, Beijing 100084, China}
\affiliation{Beijing Academy of Quantum Information Sciences, Beijing 100193, China}
\author{Zhen-Xing Hua}
\affiliation{State Key Laboratory of Low Dimensional Quantum Physics,
Department of Physics, Tsinghua University, Beijing 100084, China}
\affiliation{Frontier Science Center for Quantum Information, Beijing, China}
\author{Peiyun Ge}
\affiliation{State Key Laboratory of Low Dimensional Quantum Physics,
Department of Physics, Tsinghua University, Beijing 100084, China}
\affiliation{Frontier Science Center for Quantum Information, Beijing, China}
\author{Yu-Xin Chao}
\affiliation{State Key Laboratory of Low Dimensional Quantum Physics,
Department of Physics, Tsinghua University, Beijing 100084, China}
\affiliation{Frontier Science Center for Quantum Information, Beijing, China}
\author{Kai Li}
\affiliation{Center for Quantum Information, IIIS, Tsinghua University, Beijing 100084, China}
\affiliation{RIKEN Center for Emergent Matter Science (CEMS), Wako, Saitama 351-0198, Japan}
\author{Chen Jia}
\affiliation{State Key Laboratory of Low Dimensional Quantum Physics,
Department of Physics, Tsinghua University, Beijing 100084, China}
\affiliation{Frontier Science Center for Quantum Information, Beijing, China}
\author{Meng Khoon Tey}
\email{mengkhoon_tey@tsinghua.edu.cn}
\affiliation{State Key Laboratory of Low Dimensional Quantum Physics,
Department of Physics, Tsinghua University, Beijing 100084, China}
\affiliation{Frontier Science Center for Quantum Information, Beijing, China}
\affiliation{Hefei National Laboratory, Hefei, Anhui 230088, China}
\author{Yong Xu}
\email{yongxuphy@tsinghua.edu.cn}
\affiliation{Center for Quantum Information, IIIS, Tsinghua University, Beijing 100084, China}
\affiliation{Hefei National Laboratory, Hefei, Anhui 230088, China}
\author{Li You}
\email{lyou@mail.tsinghua.edu.cn}
\affiliation{State Key Laboratory of Low Dimensional Quantum Physics,
Department of Physics, Tsinghua University, Beijing 100084, China}
\affiliation{Beijing Academy of Quantum Information Sciences, Beijing 100193, China}
\affiliation{Frontier Science Center for Quantum Information, Beijing, China}
\affiliation{Hefei National Laboratory, Hefei, Anhui 230088, China}

\begin{abstract}
Topological phases have been extensively studied 
over the past two decades, primarily in quantum 
pure states, where they are protected by exact symmetries. Recently, 
numerous studies have theoretically demonstrated the existence of 
average symmetry-protected topological (SPT) phases in mixed quantum 
states, which naturally arise in real systems due to decoherence or disorder. 
Despite extensive experimental observations of exact SPT phases 
in various systems, ranging from solid-state materials to synthetic matters, 
average SPT phases are yet to be observed until this work.
Here we report direct observations of disorder-induced many-body interacting average SPT phase in an atom array at half-filling, whereby random offsets to tweezer locations forming a lattice implement structural disorder, resulting in fluctuating long-range dipolar interactions between tweezer confined single atoms. 
The induced topological phase is vindicated by the spatially resolved atom-atom correlation functions for different forms of dimer compositions.
The ground state degeneracy in disordered configurations is detected and compared to the regular lattice without disorder. By probing the quench dynamics of a highly excited state, we observe markedly slower decay of edge spin magnetization in comparison to the bulk spin, consistent with the presence of topologically protected edge modes in disordered lattices.
\end{abstract}   

\maketitle

The classification of topological phases, such as topological 
insulators~\cite{RevModPhys.82.3045,RevModPhys.83.1057} and the Haldane phase~\cite{haldane1983nonlinear}, relies critically on symmetries 
in pure quantum states~\cite{chen2011classification,schuch2011classifying,PhysRevB.81.064439,
doi:10.1126/science.1227224,ChenPhysRevB.87.155114,Senthil_2015}. 
Yet, real systems--from solid-state materials 
with inherent disorder to noisy intermediate-scale quantum (NISQ) simulators~\cite{preskill2018quantum} 
with decoherence--host mixed quantum states that evade the pure-state framework.
This has spurred significant recent interest in defining and understanding 
topological phases in mixed states~\cite{Stern2012PhysRevB,Moore2012PRL,fu2012topology,fulga2014statistical,wang_structural-disorder-induced_2021,ma2023average,lee2025symmetry,BiZhen2022arXiv,Ma2025PRX,Abhinav2025PRL,Alex2025PRXQuantum,Yang2025PRB,Zhen2025PRX,YIzhi2025PRXQuantum}. A key concept developed concerns average (or weak) 
symmetry, which ensures that a mixed state's density matrix is invariant under a 
symmetry's action even if its constituent pure states are not~\cite{fu2012topology,fulga2014statistical,ma2023average,Ma2025PRX},
in stark contrast to exact symmetry. 
Such symmetry can give rise to average  
symmetry-protected topological (SPT) phases in both noninteracting systems~\cite{Stern2012PhysRevB,Moore2012PRL,fu2012topology,fulga2014statistical,wang_structural-disorder-induced_2021} and quantum spin models~\cite{ma2023average,lee2025symmetry,BiZhen2022arXiv,Ma2025PRX,
Alex2025PRXQuantum,Zhen2025PRX,YIzhi2025PRXQuantum}.
While the exact SPT phases in both noninteracting systems and quantum spin 
models have been experimentally observed in various solid-state~\cite{konig2007quantum,hsieh2008topological,hsieh2009observation,tanaka2012experimental,chang2013experimental} and 
synthetic platforms~\cite{atala2013direct,jotzu2014experimental,de2019observation,sompet2022realizing,mishra2021observation,zhao2024tunable,kiczynski2022engineering,wang2024construction}, an average SPT phase remains experimentally 
elusive despite extensive theoretical studies.

Contrary to the expectation that disorder is harmful to topological phases, 
earlier studies have shown that it can actually induce exact SPT phases 
from a trivial one in both noninteracting systems
(e.g., the well-known topological Anderson insulator~\cite{li2009topological})
and quantum spin models~\cite{PhysRevLett.127.263004}, provided that disorder does not
break the protecting symmetry. 
This counterintuitive phenomenon has motivated experimental studies across a range of systems, from the classical ones of waveguide arrays~\cite{stutzer2018photonic,daiProgrammableTopologicalPhotonic2024,chenRealizationTimeReversalInvariant2024}, photonic crystals~\cite{liu2020topological,ren2024realization}, and
acoustics~\cite{zangeneh2020disorder,liu2023acoustic}, to the quantum ones of atomic condensates in momentum lattices~\cite{meier2018observation} and superconducting circuit~\cite{PhysRevResearch.6.L042038}. 
However, all realizations are concentrated so far on exact SPT phases for
noninteracting particles or are fully captured by band topology of single-particle states.

\begin{figure*}
	\centering
	\includegraphics[width=1\textwidth]{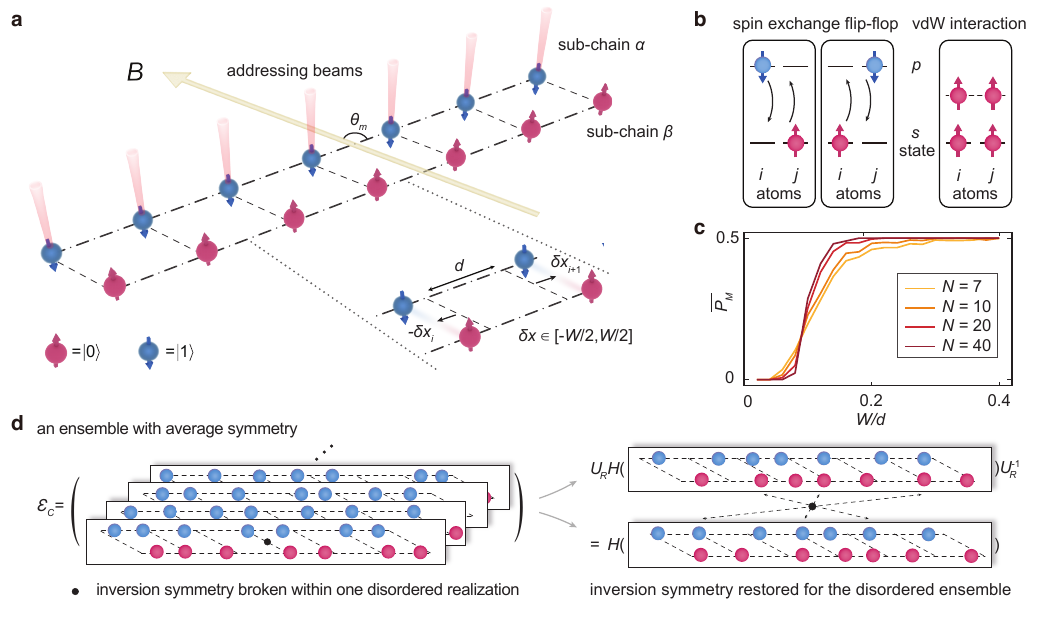} 
	\caption{\textbf{Hard-core boson model with structural disorder.}
		\textbf{a,} Schematics of a dimerized 1D regular lattice with two sub-chains corresponding to a topologically trivial phase.
	Structural disorder is introduced by randomly displacing each unit cell 
        from their regular lattice sites along the sub-chain direction, as illustrated in the inset
        with $d=13\text{ }\mu \text{m}$ in our experiments. 	
        The light red tubes denote the 1013 nm addressing laser beams used to prepare a
        N\'eel state, and $B$ represents the magnetic field.
        \textbf{b,} Interaction between Rydberg atoms: the flip-flop hopings between atoms $i$ and $j$ result from the dipolar exchange interaction that de-excites atom $i$ from $p$ to $s$ states while simultaneously excites atom $j$ from $s$ to $p$ states, or vice versa. Two Rydberg $s$ atoms also experience van der Waals (vdW) interaction, which shifts the energy of the pair states $ss$. 
		\textbf{c,} Numerically calculated topological invariant $P_{\mathrm{M}}$ 
		(averaged over more than $150$ random configurations) for our model at half-filling
		as a function of the structural disorder strength $W$,
		illustrating a topological phase transition from a trivial state at $\overline{P}_{\mathrm{M}}=0$ to 
		a nontrivial phase at $\overline{P}_{\mathrm{M}}=0.5$.
		\textbf{d,} Illustration of the emergence of inversion symmetry on average over an 
        ensemble of all disordered Hamiltonians, in contrast to the absence of exact
        inversion symmetry for a Hamiltonian on a single disordered configuration.
         This occurs because every 
        configuration $C$ is equally probable as its inversion partner $\mathcal{R}C$, 
        rendering the ensemble of all disordered Hamiltonians statistically symmetric under inversion.
        }
	\label{fig1} 
\end{figure*}

Here we theoretically predict and experimentally realize a disorder-induced bosonic average SPT phase
in a Rydberg atom array.
Specifically, we implement 
a 1D long-range Su-Schrieffer-Heeger (SSH) model for hard-core bosons (of infinite onsite interaction) 
with strong dipolar besides van der Waals interactions between atoms. 
The initial atomic array forms a regular chain in a topologically trivial phase for either single or many particles. Structural disorder is introduced through random displacements of optical tweezers~\cite{doi:10.1126/science.aah3778, doi:10.1126/science.aah3752,kimSituSingleatomArray2016} otherwise centered on regular lattice sites. 
At the single-particle level, edge mode emerges once disorder strength exceeds a critical threshold, confirming the structural disorder-induced topological state. 
At near half-filling with many particles, ground state degeneracy and correlation functions are observed, both substantiating the earlier prediction of an induced interacting topological phase~\cite{PhysRevLett.127.263004}. 
The observed quench dynamics from a highly excited state further implicate the presence of edge states by their significantly slower decay compared to atoms in the bulk. 
We therefore provide comprehensive evidence for the disorder-induced many-body interacting topological phase, supported by measurements of the density distribution, correlation functions, microwave response, and quench dynamics.

Our system is described by the following 1D staggered model involving two sub-chains $\alpha$ and $\beta$ (see Fig.~\ref{fig1}a), 
\begin{equation}
	\hat{H}=\sum_{i < j}^{2N} J_{ij}(\hat{b}_i^{\dagger}\hat{b}_j+\hat{b}_j^{\dagger}\hat{b}_i) +\hat{H}_{\mathrm{vdW}},
	\label{eq:boson}
\end{equation} 
with the vacuum and one-particle excitation states
encoded into the Rydberg $|s\rangle=|0\rangle$ and $|p\rangle=\hat{b}^\dagger|0\rangle$ states, respectively~\cite{chenContinuousSymmetryBreaking2023}.
$\hat{b}_j$ ($\hat{b}_j^\dagger$) is the annihilation (creation) operator of a boson at site $j$, constrained by the hard-core condition $\hat{b}_j^2=(\hat{b}_j^\dagger)^2=0$.
$\hat{n}_i=\hat{b}_i^\dagger \hat{b}_i$ is the particle number operator.
The terms in the parentheses describe hopping from dipolar exchanges between two Rydberg atoms at sites $i$ and $j$ (see Fig.~\ref{fig1}b) with strength given by 
$J_{ij}=C_3 (1-3\cos^2\theta_{ij})/(2 r_{ij}^3)$~\cite{Browaeys_2016_dipole_dipole,Weber_2017} which fluctuates with the random vector ${\bm r}_{ij}$ as structural disorder is introduced.
$C_3$ is the dipolar interaction coefficient, 
and the external magnetic field ${\bm B}$ makes an angle $\theta_{ij}$ with respect to ${\bm r}_{ij}$.
$ \hat{H}_{\mathrm{vdW}}=
\sum_{i<j} V_{ij}^{\mathrm{vdW} } (1-\hat{n}_i)(1-\hat{n}_j)$
denotes van der Waals interactions between Rydberg $s$ states, with $V_{ij}^{\mathrm{vdW}}=-C_{6}/r_{ij}^6$ and $C_{6}$ the  corresponding coefficient. The much weaker interactions between $p$ states are ignored. 
$V_{ij}^{\mathrm{vdW}}$ is comparable to $J_{ij}$ in our experiment, 
and contributes an Ising term of the form $\sum_{i<j}\sigma_i^z \sigma_j^z$~\cite{urbanObservationRydbergBlockade2009, doi:10.1126/science.abi8794,doi:10.1126/science.abo6587}.

The tweezers can create a regular lattice with each unit cell containing two sites separated by $\Delta x$ along the direction of the chain, as shown in Fig.~\ref{fig1}a. The regular lattice atomic locations in the two sub-chains labeled $\alpha$ and $\beta$ are denoted by $x_{2j-1}=j {d}$ (lattice constant ${d}$) and $x_{2j}=j {d}+\Delta x$ ($j=1,2,\dots, N$), respectively.
Since the intracell hopping significantly exceeds the intercell one (e.g., $0.8$ vs $0.1$ MHz 
in our experiment), a topologically trivial phase results.
Structural disorder introduces random displacements to the centers
of tweezers from their regular lattice sites according to  
$x_{2j-1} \rightarrow x_{2j-1}^\prime=x_{2j-1} +\delta x_j$ 
and $x_{2j} \rightarrow x_{2j}^\prime=x_{2j}+\delta x_j$, i.e., each unit cell incurs a random offset $\delta x_j$ along the chain, which is uniformly 
sampled from an interval $[-W/2,W/2]$, with $W$ characterizing the disorder strength (Fig.~\ref{fig1}a).

\begin{figure} 
	\centering
	\includegraphics[width=1\linewidth]{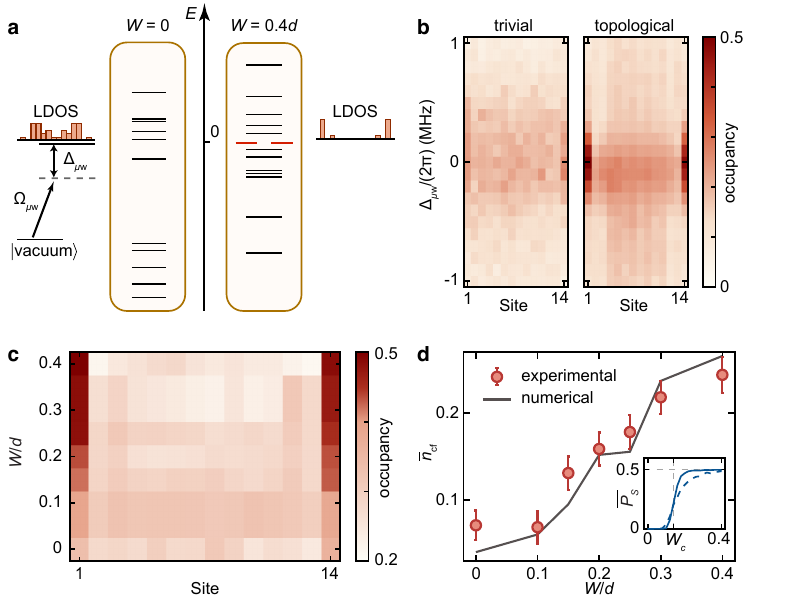} 
	\caption{\textbf{Observation of a structural disorder-induced topological phase transition at the single-particle level.}
		\textbf{a,} Single-particle spectra for regular and disordered 
		lattices ($W=0.4 {d}$).    
		In the regular lattice, no edge state near zero energy is found, while in the disordered lattice, two-fold statistically degenerate edge states emerge  (dark red lines), as shown by the local density of states (LDOS). Microwaves with a Rabi frequency $\Omega_{\mathrm{\mu w}}$ and detuning $\Delta_{\mathrm{\mu w}}$ probe the edge states.		  
		\textbf{b,} The dependence of measured site occupancy on the detuning $\Delta_{\mathrm{\mu w}}$ for a regular lattice without edge states (left) and for disordered lattices with edge states at $W=0.4 {d}$ (right).
		\textbf{c,} The measured dependence of site occupancy at zero detuning on structural disorder strength, showing a clear increase at two edges.
		\textbf{d,} Contrast between the excitation of edge sites and bulk sites 
		as a function of disorder strength from the same data in \textbf{c}.
		The gray solid line represents the numerical result incorporating various experimental errors.
		Error bars are SEMs averaged over $15$ random configurations. 
		Inset: numerically computed polarization $\overline{P}_{\mathrm{S}}$
		with $N=7$ (dashed line) and $N=40$ (solid line) versus disorder strength, averaged over $1000$ realizations. 
	}
	\label{fig2} 
\end{figure}

For a single atom excited into the $p$ state, the physics can be discussed within the single-particle subspace 
with the corresponding Hamiltonian reducing to 
$[H^{\mathrm{S}}]_{ij}=J_{ij}(1-\delta_{ij})-V_i^{\mathrm{vdW}} \delta_{ij}$ ($1 \le i,j \le 2N$), where $V_i^{\mathrm{vdW}}=-C_6 \sum_{j\neq i} 1/r_{ij}^6$, apart from a constant. 
At the "magic angle" $\theta_m\approx 54.7^{\circ}$ or $125.3^{\circ}$, hoppings vanish along the two sub-chains. 
When $V_i^{\mathrm{vdW}}=0$, the Hamiltonian
respects the sublattice symmetry, leading to the $\mathbb{Z}$ classification~\cite{ChiuRevModPhys}. 
A nonzero $V_i^{\mathrm{vdW}}$ due to Van der Waals interactions breaks the sublattice symmetry.
Nevertheless, for our structural disorder model, all the Hamiltonians on random lattice configurations ($\mathcal{E}_C$) constitute an ensemble 
$\mathcal{E}_H \equiv \left\{ H^{\mathrm{S}}(C) : C\in \mathcal{E}_C \right\}$ respecting 
an average inversion symmetry
with respect to $x_c$, which is marked by the black filled circle in Fig.~\ref{fig1}d.
The average symmetry arises because the configuration
$C=\{x_j^\prime:j=1,2,\dots,2N\}$ occurs with the same probability as its inversion partner $\mathcal{R}C=\{\mathcal{R}x_j^\prime: j=1,2,\dots,2N \}$, where 
$\mathcal{R} x_j^\prime=2x_c-x_j^\prime$ is the inverted coordinate of $x_j^\prime$.
This allows us to generalize the polarization $P_\mathrm{S}$ (a topological invariant)~\cite{resta1998quantum} to $P_\mathrm{S}(C)$ for 
the ground states of the pair of inversion Hamiltonians $H^{\mathrm{S}}(C)$ and $H^{\mathrm{S}}(\mathcal{R}C)$, 
and prove that $P_\mathrm{S}(C)$ is quantized to $0$ or $0.5$ based on 
the average symmetry as detailed in the Supplementary Information (SI). 
Since $P_\mathrm{S}(C)$ cannot vary continuously as the Hamiltonian pair are continuously deformed, their ground states have a topological feature with $P_\mathrm{S}(C)=0$ for the trivial phase and $P_\mathrm{S}(C)=0.5$ for the nontrivial one. 
Accordingly, in the single-particle case, we find increasing structural disorder strength indeed induces an average SPT phase, as illustrated in Fig.~\ref{fig2}d.

At half-filling with $N$ atoms in the $p$ state, we rewrite the hard-core boson model as the 
Heisenberg model with a $\sigma^z$ term~\cite{RevModPhys.63.1}, which
represents a genuine interacting many-body system as it cannot be mapped to a free fermion model (see SI).
The total spin along $z$, $\sigma^z=\sum_i \sigma_i^z$, or the $U(1)$ symmetry, is preserved as $[\hat{H},\sigma^z]=0$, which confines the system to the fixed $\sigma^z$ subspaces. 
When $V_i^{\mathrm{vdW} }=0$, the Hamiltonian also respects additional symmetries: rotations about $x$ or $y$ and time-reversal~\cite{de2019observation,PhysRevLett.127.263004}.
In a regular lattice without disorder, the resulting SPT phase is topologically equivalent to the Haldane phase and features four-fold degenerate ground states~\cite{hida1992crossover,de2019observation}.
This phase is also characterized by a $\mathbb{Z}_2$ topological invariant $P_\mathrm{M}$~\cite{nakamura2002order,tasaki2018topological,PhysRevLett.127.263004}. 
A non-vanishing $\sigma^z$ term breaks all the above mentioned symmetries except for the $U(1)$ symmetry. 
As a result of the average inversion symmetry mentioned before and detailed in the SI, the many-body topological invariant $P_\mathrm{M}$, analogous to the single-particle case as detailed in the SI, is generalized to $P_\mathrm{M}(C)$ defined as
\begin{equation}
	P_\mathrm{M}(C) =\left[ \frac{1}{2\pi} \mathrm{Im} \ln 
	\sum_{S \in \{C, \mathcal{R}C\}}  \bra{\Psi_S} \hat{\mathcal{P}}_\mathrm{M} (S) \ket{\Psi_S} \right] 
	 \text{ mod } 1.
\end{equation}
Here $\ket{\Psi_S}$ is the many-body ground state of the Hamiltonian 
$\hat{H}$ at half-filling
in the lattice configuration $S$ under periodic boundary conditions, 
and $\hat{\mathcal{P}}_\mathrm{M}(S) = \prod_{j=1}^{2N} e^{- \frac{\pi i}{N{d}} x_j \sigma_{j}^z }$ 
is the twist operator~\cite{nakamura2002order}.
Analogously, we prove in the SI that $P_\mathrm{M}(C)$ is quantized to $0$ or $0.5$, 
and thus serves as a well-defined
topological invariant. 
Figure~\ref{fig1}c illustrates that averaged over many independent realizations, a sharp transition with increasing structural disorder strength indeed appears, indicating that a many-body average SPT phase is induced.

Our experiments are carried out on a short array of fourteen $^{87}$Rb atoms ($N=7$)~\cite{liang2024observationanomalousinformationscrambling}, comparable to earlier studies in an atom array~\cite{de2019observation} or quantum dots~\cite{kiczynski2022engineering} or
surface atoms~\cite{wang2024construction}.
Each atom is initially prepared in the $s$ state (denoted by $\left| \uparrow \right\rangle \equiv \left| s \right \rangle \equiv |55S_{1/2},m_J=-1/2\rangle $) via two-photon stimulated Raman adiabatic passage (STIRAP) from the ground state 
$|g\rangle \equiv |5S_{1/2},F=2,m_F=-2\rangle$
through the intermediate state $|e\rangle \equiv \ket{6P_{3/2},F=3}$. 
Further excitation to the $p$ state ($\left| \downarrow \right\rangle \equiv |p\rangle \equiv |55P_{1/2},m_J=1/2\rangle$) is accomplished by a microwave field at a resonant frequency of about $E_0/h \sim 22.1$ GHz.
A bias magnetic field of $B_z \approx 30$ G prevents cross-couplings with other magnetic sub-levels.
Each experiment typically lasts for a few microseconds, during which the atomic spatial motion can be considered frozen.
At the end of an experiment, atoms in the $s$ state are de-excited to the 
ground state, imaged by fluorescence spectroscopy, while
atoms in the $p$ state are lost. 
All experimental results for disordered lattices are averaged over $15$ 
random configurations.

Using microwave spectroscopy, we determine the presence or absence of an excitation at the edge sites, thus affirmatively demonstrating the structural disorder-induced topology at the single-particle level. 
After initializing all atoms into the $s$ state, a weak microwave field is applied, coupling it to the $p$ state in a process described by the Hamiltonian
$ \hat{H}_{\mathrm{\mu w}}=\sum_{i=1}^{2N} [(\hbar \Omega_{\mathrm{\mu w}}/2)(\hat{b}_i^\dagger + \hat{b}_i) 
-\hbar \Delta_{\mathrm{\mu w}} \hat{b}_i^\dagger \hat{b}_i]$ 
with the Rabi frequency $\Omega_\mathrm{\mu w}/(2\pi) \approx 0.2$ MHz and the microwave detuning $\Delta_{\mathrm{\mu w}}$.
An edge excitation is created when the detuning matches the resonance as illustrated in Fig.~\ref{fig2}a after $2.5\text{ } \mathrm{\mu s}$. 
Figure~\ref{fig2}b presents the $p$ state occupancy
at the end of excitation, revealing pronounced peaks near zero detuning at the edges of disordered lattices
with $W=0.4{d}$.
In contrast, the same edge occupancy in the regular lattice is found to resemble closely that of the bulk.
The van der Waals interactions are found to slightly
modify the energy at the edges but significantly shift the bulk states, although they do not break the topological nature of the system (see SI). 
Our observations thus confirm the presence of edge states near zero energy in
the disordered lattice as well as their absence in the regular lattice.

For further investigations, 
we measure the site occupancy at zero detuning as a function of disorder strength,
and Fig.~\ref{fig2}c reveals increased excitations at both edge sites with increasing disorder strength. 
The phase transition is more clearly illustrated by plotting the contrast for the $p$ state excitation $n_i$ between edge and bulk sites,
defined by 
$
	n_{\mathrm{cf}}=(n_{1}+n_{2N})/2-\sum_{i =2}^{2N-1} n_{i}/(2N-2)
$.
Figure~\ref{fig2}d shows its noticeable increase with increasing disorder strength, consistent
with numerical simulations that account for various experimental errors (see SI). The sharp rise around $W_c=0.15$ provides strong evidence for a topological phase 
transition at the single-particle level.

\begin{figure*} 
	\centering
	\includegraphics[width=1\linewidth]{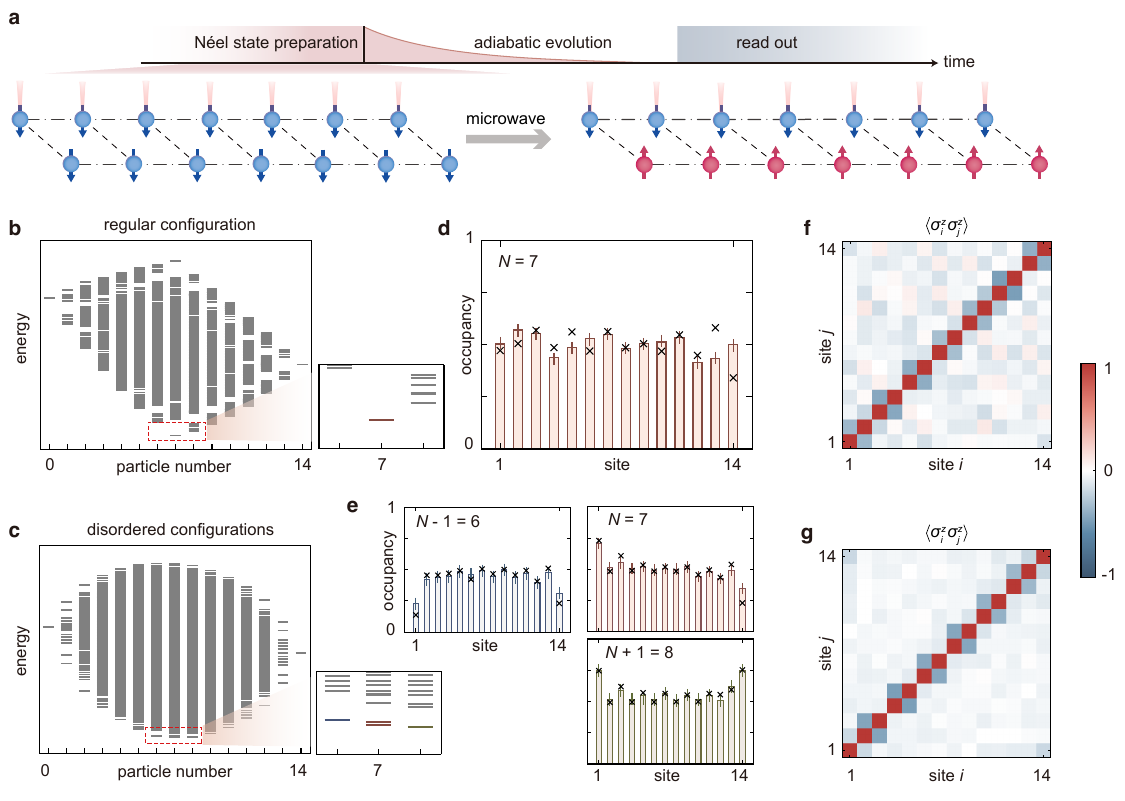} 
	\caption{\textbf{Observation of structural disorder-induced many-body average SPT phase.}
		\textbf{a,} Schematics of the experimental sequence for many-body ground state preparation and detection.
		Initially, all atoms are excited into the Rydberg $p$ state (top left panel). 
		Focused pinning laser beams as in Fig.\ref{fig1} are applied to atoms in the upper sub-chain, and a 
		microwave field is swept across resonance to produce a N{\'{e}}el state (right panel). Adiabatic evolution subsequently prepares the ground state of the Hamiltonian $\hat{H}$ 
		in a fixed particle number subspace. 
		\textbf{b,} Energy spectra at different particle numbers for a regular lattice and 
		\textbf{c,} disordered lattice ($W=0.4 {d}$), with a zoomed-in view near the ground states. The four-fold degeneracy of the ground state is slightly lifted due to the van der Waals interactions. The average inversion symmetry only
		protects the two-fold ground state degeneracy in the half-filling subspace
		(see the SI).
		\textbf{d,} Measured site occupancy of the ground state for 
		the regular lattice compared to (\textbf{e}) the disordered lattice
		near half-fillings of $N-1$ (left), $N$ (right), and $N+1$ (below) particles. 
        The measured results are postselected for specific particle fillings.
		Diagonal crosses represent numerical results incorporating various experimental errors.
		Measured two-point correlation functions of the ground state
		at half-filling for (\textbf{f}) the regular and (\textbf{g}) disordered lattices.}
	\label{fig3} 
\end{figure*}

\begin{figure*} 
	\centering
	\includegraphics[width=1\linewidth]{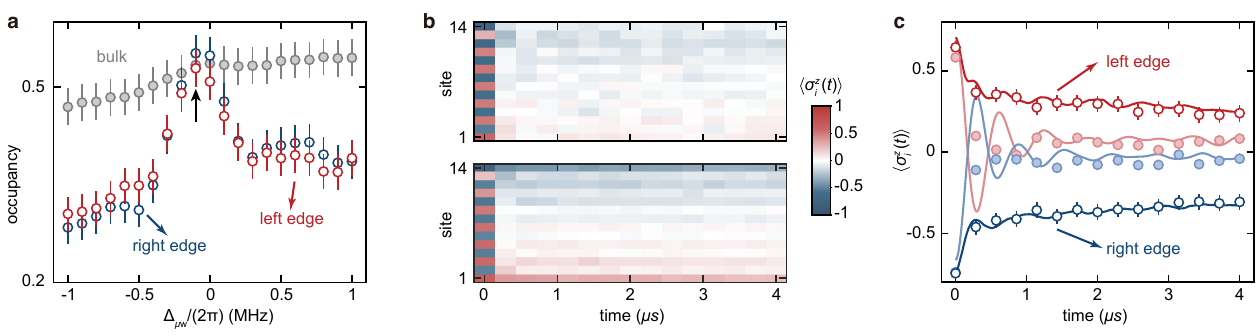} 
	\caption{\textbf{Many-body microwave spectroscopy and temporal decay of quantum spins.}
		\textbf{a,} Measured dependence of $p$ state occupancies for edge sites (open red and blue circles) and bulk sites (filled gray circles)
		on microwave detuning $\Delta_{\mathrm{\mu w}}$ at $W=0.4d$.
		A pronounced peak is observed near zero energy for both edge sites, indicating 
		degeneracy between the two ground states at half-filling. The slight shift of the 
		peak away from $s$ to $p$ resonance is due to van der Waals interactions, 
		which lift the degeneracy between the ground states with $N-1$ and $N$ excitations.
		\textbf{b,} Measured magnetization $\sigma_i^z(t)$ 
		for all 14 sites of the initial N{\'{e}}el state evolved by the realized Hamiltonian in disordered lattices (lower panel) compared to the regular case (upper panel). 
        Rapid loss of magnetization is observed for all sites except the topologically protected edge states of the disordered lattice, whose decays are significantly slower.
		\textbf{c,} Empty circles denote results for two edge sites in \textbf{b} compared to numerical predictions incorporating various errors in solid lines at $W=0.4 {d}$.
		Filled circles correspond to the two edge states in the regular lattice.
		Error bars represent SEMs averaged over 15 random configurations.
        }
	\label{fig4} 
\end{figure*}

We next present results for the many-body interacting average SPT phase induced by structural disorder at half-filling.
Many-body ground states with particle numbers 
near half-filling are prepared experimentally to demonstrate the induced SPT phase. 
Initially, atoms are prepared in a product state $\left| \psi_0 \right\rangle$, 
e.g., $|1010... 10\rangle$, $|0010... 10\rangle$, or $|1010... 11\rangle$, 
respectively in the subspaces with $N$, $N-1$, and $N+1$ particles, 
as illustrated Fig.~\ref{fig3}a and detailed in the SI.
The product state $\left| \psi_0 \right\rangle$ then evolves into the many-body ground state following the  
adiabatic ramping down of the pinning laser beams (see Fig.~\ref{fig3}a and SI).

Figures~\ref{fig3}d and \ref{fig3}e present the measured occupancies for all sites of the prepared ground state in both regular and disordered lattices.  
While their bulk properties are comparable, their behaviors 
at the edge sites are significantly different. At half-filling, the edge sites in the regular lattice behave similarly to the bulk sites consistent with the fact that the spectrum features a single ground state (see Fig.~\ref{fig3}b), whereas the disordered configuration reveals pronounced differences for either the left or right edges, as well as the imbalance between the left and right edges. 
In particular, two statistically degenerate ground states exist in the half-filling subspace (see Fig.~\ref{fig3}c and SI), one of which is presented, and is associated with a higher particle occupation at 
the left edge over the right one. 
We also observe that in the subspace with $N-1$ particles, the occupancies of edge sites
are significantly lower than in the bulk. Conversely, in the subspace with $N+1$ 
particles, the edge site occupancies are notably higher.
These behaviors are consistent with the expected properties of the induced topological ground states.

We further measure the
correlation function $C_{i,j}^z=\langle \sigma_i^z \sigma_j^z \rangle$ for the prepared ground
state at half-filling. Figures~\ref{fig3}f and \ref{fig3}g reveal intracell correlations 
($C^z_{\mathrm{intra} }=\sum_{i=1}^{N}C_{2i-1,2i}^z /N \approx -0.38$) are significantly stronger than
intercell ones ($C^z_{\mathrm{inter} }=\sum_{i=1}^{N-1}C_{2i,2i+1}^z /(N-1) \approx -0.15$)
in the regular lattice, whereas the opposite trends are observed in the disordered lattice 
($C^z_{\mathrm{intra} } \approx -0.13$ and $C^z_{\mathrm{inter} } \approx -0.42$).
They provide smoking gun evidence that the ground state in the regular lattice is topologically trivial, 
while it is nontrivial in the disordered lattice. 

We now show the observed statistical degeneracy of the ground states in the half-filling subspace in the disordered case.
We first prepare a ground state with $N-1$ particles and then 
employ microwave spectroscopy (as in the single-particle case) with a Rabi frequency of
$\Omega_\mathrm{\mu w}/(2\pi) \approx 0.25$ MHz and a pulse duration of $2\text{ }\mathrm{\mu s}$ 
to probe possible excitations by scanning the detuning $\Delta_{\mathrm{\mu w}}$.
Figure~\ref{fig4}a shows a sharp rise near zero detuning, with a 
distinct peak appearing
at $\Delta_{\mathrm{\mu w}}/(2\pi)\approx-0.1$ MHz for both edge sites,
heralding the transition into the half-filling ground state,
by adding a particle from absorption of a resonant microwave photon to create a $p$ state atom from an $s$ state at the detuning of $\sim -0.1$ MHz for both edge sites.
Here, the small detuning arises from van der Waals interactions (more details in SI).

Finally, we provide compelling evidence that the disorder-induced phase exhibits robust edge states or topological protection, 
even for highly excited states~\cite{bahri2015localization,zhang_digital_2022,doi:10.1126/science.abq5769}.
Specifically, we initialize the system into a product state such as the N{\'{e}}el state $\left|0101\dots 01 \right\rangle$, 
a highly excited state, and subsequently evolve it under the 
many-body Hamiltonian $\hat{H}$ as in a quench. In Fig.~\ref{fig4}b, we observe that in the regular lattice, the magnetization 
$ \sigma_{i}^z$ for both bulk and edge sites rapidly decays to zero, implicating a quick loss 
of initial information.
The small residual value of $ \langle \sigma_{i}^z \rangle$ is attributed 
to van der Waals interactions (see SI). 
In contrast, in the disordered lattice, while the bulk sites exhibit a
similar decay to the regular lattice, at the edges $ \langle \sigma_{i}^z \rangle$ exhibits a slower decay and eventually stabilizes at finite values (Figs.~\ref{fig4}b and \ref{fig4}c), in near complete agreement with numerical simulations.

Our work opens a broad avenue to study the interplay between structural disorder and topology in a programable quantum platform. 
At the single-particle level, future explorations could address how structural disorder affects higher-dimensional topological phases, such as the Chern insulators~\cite{PhysRevLett.118.236402,mitchell2018amorphous} and higher-order topological phases~\cite{AgarwalaPRRAmorphous,wang_structural-disorder-induced_2021}. 
At the many-body level with interactions, it is now possible to examine many-body topological phenomena in higher dimensions, 
including spin liquids in two-dimensional amorphous lattices~\cite{grushin2023amorphous,cassella2023exact}. 
Furthermore, the Rydberg atom array platform can be employed to explore amorphous Floquet topological phases~\cite{WangFloquentWang} by incorporating periodic driving. In addition, it also offers new opportunities to study the many-body localization induced by structural disorder~\cite{aramthottil2024phenomenology} and the interplay between localization and topology, such as topologically protected quantum dynamics in disordered systems~\cite{bahri2015localization}.

\textbf{Note added.} During the completion of this work, we became aware of related experiment on average symmetry protected topology in an optical lattice system~\cite{su2025topologicalphasescriticalitymixed}.

\bibliography{main_ref}

\begin{thebibliography}{80}%
\makeatletter
\providecommand \@ifxundefined [1]{%
 \@ifx{#1\undefined}
}%
\providecommand \@ifnum [1]{%
 \ifnum #1\expandafter \@firstoftwo
 \else \expandafter \@secondoftwo
 \fi
}%
\providecommand \@ifx [1]{%
 \ifx #1\expandafter \@firstoftwo
 \else \expandafter \@secondoftwo
 \fi
}%
\providecommand \natexlab [1]{#1}%
\providecommand \enquote  [1]{``#1''}%
\providecommand \bibnamefont  [1]{#1}%
\providecommand \bibfnamefont [1]{#1}%
\providecommand \citenamefont [1]{#1}%
\providecommand \href@noop [0]{\@secondoftwo}%
\providecommand \href [0]{\begingroup \@sanitize@url \@href}%
\providecommand \@href[1]{\@@startlink{#1}\@@href}%
\providecommand \@@href[1]{\endgroup#1\@@endlink}%
\providecommand \@sanitize@url [0]{\catcode `\\12\catcode `\$12\catcode `\&12\catcode `\#12\catcode `\^12\catcode `\_12\catcode `\%12\relax}%
\providecommand \@@startlink[1]{}%
\providecommand \@@endlink[0]{}%
\providecommand \url  [0]{\begingroup\@sanitize@url \@url }%
\providecommand \@url [1]{\endgroup\@href {#1}{\urlprefix }}%
\providecommand \urlprefix  [0]{URL }%
\providecommand \Eprint [0]{\href }%
\providecommand \doibase [0]{https://doi.org/}%
\providecommand \selectlanguage [0]{\@gobble}%
\providecommand \bibinfo  [0]{\@secondoftwo}%
\providecommand \bibfield  [0]{\@secondoftwo}%
\providecommand \translation [1]{[#1]}%
\providecommand \BibitemOpen [0]{}%
\providecommand \bibitemStop [0]{}%
\providecommand \bibitemNoStop [0]{.\EOS\space}%
\providecommand \EOS [0]{\spacefactor3000\relax}%
\providecommand \BibitemShut  [1]{\csname bibitem#1\endcsname}%
\let\auto@bib@innerbib\@empty
\bibitem [{\citenamefont {Hasan}\ and\ \citenamefont {Kane}(2010)}]{RevModPhys.82.3045}%
  \BibitemOpen
  \bibfield  {author} {\bibinfo {author} {\bibfnamefont {M.~Z.}\ \bibnamefont {Hasan}}\ and\ \bibinfo {author} {\bibfnamefont {C.~L.}\ \bibnamefont {Kane}},\ }\bibfield  {title} {\bibinfo {title} {Colloquium: {{Topological}} insulators},\ }\href {https://doi.org/10.1103/RevModPhys.82.3045} {\bibfield  {journal} {\bibinfo  {journal} {Reviews of Modern Physics}\ }\textbf {\bibinfo {volume} {82}},\ \bibinfo {pages} {3045} (\bibinfo {year} {2010})}\BibitemShut {NoStop}%
\bibitem [{\citenamefont {Qi}\ and\ \citenamefont {Zhang}(2011)}]{RevModPhys.83.1057}%
  \BibitemOpen
  \bibfield  {author} {\bibinfo {author} {\bibfnamefont {X.-L.}\ \bibnamefont {Qi}}\ and\ \bibinfo {author} {\bibfnamefont {S.-C.}\ \bibnamefont {Zhang}},\ }\bibfield  {title} {\bibinfo {title} {Topological insulators and superconductors},\ }\href {https://doi.org/10.1103/RevModPhys.83.1057} {\bibfield  {journal} {\bibinfo  {journal} {Reviews of Modern Physics}\ }\textbf {\bibinfo {volume} {83}},\ \bibinfo {pages} {1057} (\bibinfo {year} {2011})}\BibitemShut {NoStop}%
\bibitem [{\citenamefont {Haldane}(1983)}]{haldane1983nonlinear}%
  \BibitemOpen
  \bibfield  {author} {\bibinfo {author} {\bibfnamefont {F.~D.~M.}\ \bibnamefont {Haldane}},\ }\bibfield  {title} {\bibinfo {title} {Nonlinear field theory of large-spin heisenberg antiferromagnets: Semiclassically quantized solitons of the one-dimensional easy-axis n{\'e}el state},\ }\href {https://doi.org/10.1103/PhysRevLett.50.1153} {\bibfield  {journal} {\bibinfo  {journal} {Physical Review Letters}\ }\textbf {\bibinfo {volume} {50}},\ \bibinfo {pages} {1153} (\bibinfo {year} {1983})}\BibitemShut {NoStop}%
\bibitem [{\citenamefont {Chen}\ \emph {et~al.}(2011{\natexlab{a}})\citenamefont {Chen}, \citenamefont {Gu},\ and\ \citenamefont {Wen}}]{chen2011classification}%
  \BibitemOpen
  \bibfield  {author} {\bibinfo {author} {\bibfnamefont {X.}~\bibnamefont {Chen}}, \bibinfo {author} {\bibfnamefont {Z.-C.}\ \bibnamefont {Gu}},\ and\ \bibinfo {author} {\bibfnamefont {X.-G.}\ \bibnamefont {Wen}},\ }\bibfield  {title} {\bibinfo {title} {Classification of gapped symmetric phases in one-dimensional spin systems},\ }\href {https://doi.org/10.1103/PhysRevB.83.035107} {\bibfield  {journal} {\bibinfo  {journal} {Physical Review B}\ }\textbf {\bibinfo {volume} {83}},\ \bibinfo {pages} {035107} (\bibinfo {year} {2011}{\natexlab{a}})}\BibitemShut {NoStop}%
\bibitem [{\citenamefont {Schuch}\ \emph {et~al.}(2011)\citenamefont {Schuch}, \citenamefont {P{\'e}rez-Garc{\'\i}a},\ and\ \citenamefont {Cirac}}]{schuch2011classifying}%
  \BibitemOpen
  \bibfield  {author} {\bibinfo {author} {\bibfnamefont {N.}~\bibnamefont {Schuch}}, \bibinfo {author} {\bibfnamefont {D.}~\bibnamefont {P{\'e}rez-Garc{\'\i}a}},\ and\ \bibinfo {author} {\bibfnamefont {I.}~\bibnamefont {Cirac}},\ }\bibfield  {title} {\bibinfo {title} {Classifying quantum phases using matrix product states and projected entangled pair states},\ }\href {https://doi.org/10.1103/PhysRevB.84.165139} {\bibfield  {journal} {\bibinfo  {journal} {Physical Review B}\ }\textbf {\bibinfo {volume} {84}},\ \bibinfo {pages} {165139} (\bibinfo {year} {2011})}\BibitemShut {NoStop}%
\bibitem [{\citenamefont {Pollmann}\ \emph {et~al.}(2010)\citenamefont {Pollmann}, \citenamefont {Turner}, \citenamefont {Berg},\ and\ \citenamefont {Oshikawa}}]{PhysRevB.81.064439}%
  \BibitemOpen
  \bibfield  {author} {\bibinfo {author} {\bibfnamefont {F.}~\bibnamefont {Pollmann}}, \bibinfo {author} {\bibfnamefont {A.~M.}\ \bibnamefont {Turner}}, \bibinfo {author} {\bibfnamefont {E.}~\bibnamefont {Berg}},\ and\ \bibinfo {author} {\bibfnamefont {M.}~\bibnamefont {Oshikawa}},\ }\bibfield  {title} {\bibinfo {title} {Entanglement spectrum of a topological phase in one dimension},\ }\href {https://doi.org/10.1103/PhysRevB.81.064439} {\bibfield  {journal} {\bibinfo  {journal} {Physical Review B}\ }\textbf {\bibinfo {volume} {81}},\ \bibinfo {pages} {064439} (\bibinfo {year} {2010})}\BibitemShut {NoStop}%
\bibitem [{\citenamefont {Chen}\ \emph {et~al.}(2012)\citenamefont {Chen}, \citenamefont {Gu}, \citenamefont {Liu},\ and\ \citenamefont {Wen}}]{doi:10.1126/science.1227224}%
  \BibitemOpen
  \bibfield  {author} {\bibinfo {author} {\bibfnamefont {X.}~\bibnamefont {Chen}}, \bibinfo {author} {\bibfnamefont {Z.-C.}\ \bibnamefont {Gu}}, \bibinfo {author} {\bibfnamefont {Z.-X.}\ \bibnamefont {Liu}},\ and\ \bibinfo {author} {\bibfnamefont {X.-G.}\ \bibnamefont {Wen}},\ }\bibfield  {title} {\bibinfo {title} {Symmetry-protected topological orders in interacting bosonic systems},\ }\href {https://doi.org/10.1126/science.1227224} {\bibfield  {journal} {\bibinfo  {journal} {Science}\ }\textbf {\bibinfo {volume} {338}},\ \bibinfo {pages} {1604} (\bibinfo {year} {2012})}\BibitemShut {NoStop}%
\bibitem [{\citenamefont {Chen}\ \emph {et~al.}(2013)\citenamefont {Chen}, \citenamefont {Gu}, \citenamefont {Liu},\ and\ \citenamefont {Wen}}]{ChenPhysRevB.87.155114}%
  \BibitemOpen
  \bibfield  {author} {\bibinfo {author} {\bibfnamefont {X.}~\bibnamefont {Chen}}, \bibinfo {author} {\bibfnamefont {Z.-C.}\ \bibnamefont {Gu}}, \bibinfo {author} {\bibfnamefont {Z.-X.}\ \bibnamefont {Liu}},\ and\ \bibinfo {author} {\bibfnamefont {X.-G.}\ \bibnamefont {Wen}},\ }\bibfield  {title} {\bibinfo {title} {Symmetry protected topological orders and the group cohomology of their symmetry group},\ }\href {https://doi.org/10.1103/PhysRevB.87.155114} {\bibfield  {journal} {\bibinfo  {journal} {Physical Review B}\ }\textbf {\bibinfo {volume} {87}},\ \bibinfo {pages} {155114} (\bibinfo {year} {2013})}\BibitemShut {NoStop}%
\bibitem [{\citenamefont {Senthil}(2015)}]{Senthil_2015}%
  \BibitemOpen
  \bibfield  {author} {\bibinfo {author} {\bibfnamefont {T.}~\bibnamefont {Senthil}},\ }\bibfield  {title} {\bibinfo {title} {Symmetry-protected topological phases of quantum matter},\ }\href {https://doi.org/10.1146/annurev-conmatphys-031214-014740} {\bibfield  {journal} {\bibinfo  {journal} {Annual Review of Condensed Matter Physics}\ }\textbf {\bibinfo {volume} {6}},\ \bibinfo {pages} {299} (\bibinfo {year} {2015})}\BibitemShut {NoStop}%
\bibitem [{\citenamefont {Preskill}(2018)}]{preskill2018quantum}%
  \BibitemOpen
  \bibfield  {author} {\bibinfo {author} {\bibfnamefont {J.}~\bibnamefont {Preskill}},\ }\bibfield  {title} {\bibinfo {title} {Quantum computing in the nisq era and beyond},\ }\href {https://doi.org/10.22331/q-2018-08-06-79} {\bibfield  {journal} {\bibinfo  {journal} {Quantum}\ }\textbf {\bibinfo {volume} {2}},\ \bibinfo {pages} {79} (\bibinfo {year} {2018})}\BibitemShut {NoStop}%
\bibitem [{\citenamefont {Ringel}\ \emph {et~al.}(2012)\citenamefont {Ringel}, \citenamefont {Kraus},\ and\ \citenamefont {Stern}}]{Stern2012PhysRevB}%
  \BibitemOpen
  \bibfield  {author} {\bibinfo {author} {\bibfnamefont {Z.}~\bibnamefont {Ringel}}, \bibinfo {author} {\bibfnamefont {Y.~E.}\ \bibnamefont {Kraus}},\ and\ \bibinfo {author} {\bibfnamefont {A.}~\bibnamefont {Stern}},\ }\bibfield  {title} {\bibinfo {title} {Strong side of weak topological insulators},\ }\href {https://doi.org/10.1103/PhysRevB.86.045102} {\bibfield  {journal} {\bibinfo  {journal} {Phys. Rev. B}\ }\textbf {\bibinfo {volume} {86}},\ \bibinfo {pages} {045102} (\bibinfo {year} {2012})}\BibitemShut {NoStop}%
\bibitem [{\citenamefont {Mong}\ \emph {et~al.}(2012)\citenamefont {Mong}, \citenamefont {Bardarson},\ and\ \citenamefont {Moore}}]{Moore2012PRL}%
  \BibitemOpen
  \bibfield  {author} {\bibinfo {author} {\bibfnamefont {R.~S.~K.}\ \bibnamefont {Mong}}, \bibinfo {author} {\bibfnamefont {J.~H.}\ \bibnamefont {Bardarson}},\ and\ \bibinfo {author} {\bibfnamefont {J.~E.}\ \bibnamefont {Moore}},\ }\bibfield  {title} {\bibinfo {title} {Quantum transport and two-parameter scaling at the surface of a weak topological insulator},\ }\href {https://doi.org/10.1103/PhysRevLett.108.076804} {\bibfield  {journal} {\bibinfo  {journal} {Phys. Rev. Lett.}\ }\textbf {\bibinfo {volume} {108}},\ \bibinfo {pages} {076804} (\bibinfo {year} {2012})}\BibitemShut {NoStop}%
\bibitem [{\citenamefont {Fu}\ and\ \citenamefont {Kane}(2012)}]{fu2012topology}%
  \BibitemOpen
  \bibfield  {author} {\bibinfo {author} {\bibfnamefont {L.}~\bibnamefont {Fu}}\ and\ \bibinfo {author} {\bibfnamefont {C.~L.}\ \bibnamefont {Kane}},\ }\bibfield  {title} {\bibinfo {title} {Topology, delocalization via average symmetry and the symplectic anderson transition},\ }\href {https://doi.org/10.1103/PhysRevLett.109.246605} {\bibfield  {journal} {\bibinfo  {journal} {Physical Review Letters}\ }\textbf {\bibinfo {volume} {109}},\ \bibinfo {pages} {246605} (\bibinfo {year} {2012})}\BibitemShut {NoStop}%
\bibitem [{\citenamefont {Fulga}\ \emph {et~al.}(2014)\citenamefont {Fulga}, \citenamefont {Van~Heck}, \citenamefont {Edge},\ and\ \citenamefont {Akhmerov}}]{fulga2014statistical}%
  \BibitemOpen
  \bibfield  {author} {\bibinfo {author} {\bibfnamefont {I.}~\bibnamefont {Fulga}}, \bibinfo {author} {\bibfnamefont {B.}~\bibnamefont {Van~Heck}}, \bibinfo {author} {\bibfnamefont {J.}~\bibnamefont {Edge}},\ and\ \bibinfo {author} {\bibfnamefont {A.}~\bibnamefont {Akhmerov}},\ }\bibfield  {title} {\bibinfo {title} {Statistical topological insulators},\ }\href {https://doi.org/10.1103/PhysRevB.89.155424} {\bibfield  {journal} {\bibinfo  {journal} {Physical Review B}\ }\textbf {\bibinfo {volume} {89}},\ \bibinfo {pages} {155424} (\bibinfo {year} {2014})}\BibitemShut {NoStop}%
\bibitem [{\citenamefont {Wang}\ \emph {et~al.}(2021)\citenamefont {Wang}, \citenamefont {Yang}, \citenamefont {Dai},\ and\ \citenamefont {Xu}}]{wang_structural-disorder-induced_2021}%
  \BibitemOpen
  \bibfield  {author} {\bibinfo {author} {\bibfnamefont {J.-H.}\ \bibnamefont {Wang}}, \bibinfo {author} {\bibfnamefont {Y.-B.}\ \bibnamefont {Yang}}, \bibinfo {author} {\bibfnamefont {N.}~\bibnamefont {Dai}},\ and\ \bibinfo {author} {\bibfnamefont {Y.}~\bibnamefont {Xu}},\ }\bibfield  {title} {\bibinfo {title} {Structural-{Disorder}-{Induced} {Second}-{Order} {Topological} {Insulators} in {Three} {Dimensions}},\ }\href {https://doi.org/10.1103/PhysRevLett.126.206404} {\bibfield  {journal} {\bibinfo  {journal} {Physical Review Letters}\ }\textbf {\bibinfo {volume} {126}},\ \bibinfo {pages} {206404} (\bibinfo {year} {2021})}\BibitemShut {NoStop}%
\bibitem [{\citenamefont {Ma}\ and\ \citenamefont {Wang}(2023)}]{ma2023average}%
  \BibitemOpen
  \bibfield  {author} {\bibinfo {author} {\bibfnamefont {R.}~\bibnamefont {Ma}}\ and\ \bibinfo {author} {\bibfnamefont {C.}~\bibnamefont {Wang}},\ }\bibfield  {title} {\bibinfo {title} {Average symmetry-protected topological phases},\ }\href {https://doi.org/10.1103/PhysRevX.13.031016} {\bibfield  {journal} {\bibinfo  {journal} {Physical Review X}\ }\textbf {\bibinfo {volume} {13}},\ \bibinfo {pages} {031016} (\bibinfo {year} {2023})}\BibitemShut {NoStop}%
\bibitem [{\citenamefont {Lee}\ \emph {et~al.}(2025)\citenamefont {Lee}, \citenamefont {You},\ and\ \citenamefont {Xu}}]{lee2025symmetry}%
  \BibitemOpen
  \bibfield  {author} {\bibinfo {author} {\bibfnamefont {J.~Y.}\ \bibnamefont {Lee}}, \bibinfo {author} {\bibfnamefont {Y.-Z.}\ \bibnamefont {You}},\ and\ \bibinfo {author} {\bibfnamefont {C.}~\bibnamefont {Xu}},\ }\bibfield  {title} {\bibinfo {title} {Symmetry protected topological phases under decoherence},\ }\href {https://doi.org/10.22331/q-2025-01-23-1607} {\bibfield  {journal} {\bibinfo  {journal} {Quantum}\ }\textbf {\bibinfo {volume} {9}},\ \bibinfo {pages} {1607} (\bibinfo {year} {2025})}\BibitemShut {NoStop}%
\bibitem [{\citenamefont {Zhang}\ \emph {et~al.}(2022{\natexlab{a}})\citenamefont {Zhang}, \citenamefont {Qi},\ and\ \citenamefont {Bi}}]{BiZhen2022arXiv}%
  \BibitemOpen
  \bibfield  {author} {\bibinfo {author} {\bibfnamefont {J.-H.}\ \bibnamefont {Zhang}}, \bibinfo {author} {\bibfnamefont {Y.}~\bibnamefont {Qi}},\ and\ \bibinfo {author} {\bibfnamefont {Z.}~\bibnamefont {Bi}},\ }\href {https://arxiv.org/abs/2210.17485} {\bibinfo {title} {Fidelity strange correlators for average symmetry-protected topological phases}} (\bibinfo {year} {2022}{\natexlab{a}}),\ \Eprint {https://arxiv.org/abs/2210.17485} {arXiv:2210.17485} \BibitemShut {NoStop}%
\bibitem [{\citenamefont {Ma}\ \emph {et~al.}(2025)\citenamefont {Ma}, \citenamefont {Zhang}, \citenamefont {Bi}, \citenamefont {Cheng},\ and\ \citenamefont {Wang}}]{Ma2025PRX}%
  \BibitemOpen
  \bibfield  {author} {\bibinfo {author} {\bibfnamefont {R.}~\bibnamefont {Ma}}, \bibinfo {author} {\bibfnamefont {J.-H.}\ \bibnamefont {Zhang}}, \bibinfo {author} {\bibfnamefont {Z.}~\bibnamefont {Bi}}, \bibinfo {author} {\bibfnamefont {M.}~\bibnamefont {Cheng}},\ and\ \bibinfo {author} {\bibfnamefont {C.}~\bibnamefont {Wang}},\ }\bibfield  {title} {\bibinfo {title} {Topological phases with average symmetries: The decohered, the disordered, and the intrinsic},\ }\href {https://doi.org/10.1103/PhysRevX.15.021062} {\bibfield  {journal} {\bibinfo  {journal} {Phys. Rev. X}\ }\textbf {\bibinfo {volume} {15}},\ \bibinfo {pages} {021062} (\bibinfo {year} {2025})}\BibitemShut {NoStop}%
\bibitem [{\citenamefont {Chirame}\ \emph {et~al.}(2025)\citenamefont {Chirame}, \citenamefont {Burnell}, \citenamefont {Gopalakrishnan},\ and\ \citenamefont {Prem}}]{Abhinav2025PRL}%
  \BibitemOpen
  \bibfield  {author} {\bibinfo {author} {\bibfnamefont {S.}~\bibnamefont {Chirame}}, \bibinfo {author} {\bibfnamefont {F.~J.}\ \bibnamefont {Burnell}}, \bibinfo {author} {\bibfnamefont {S.}~\bibnamefont {Gopalakrishnan}},\ and\ \bibinfo {author} {\bibfnamefont {A.}~\bibnamefont {Prem}},\ }\bibfield  {title} {\bibinfo {title} {Stable symmetry-protected topological phases in systems with heralded noise},\ }\href {https://doi.org/10.1103/PhysRevLett.134.010403} {\bibfield  {journal} {\bibinfo  {journal} {Phys. Rev. Lett.}\ }\textbf {\bibinfo {volume} {134}},\ \bibinfo {pages} {010403} (\bibinfo {year} {2025})}\BibitemShut {NoStop}%
\bibitem [{\citenamefont {Ma}\ and\ \citenamefont {Turzillo}(2025)}]{Alex2025PRXQuantum}%
  \BibitemOpen
  \bibfield  {author} {\bibinfo {author} {\bibfnamefont {R.}~\bibnamefont {Ma}}\ and\ \bibinfo {author} {\bibfnamefont {A.}~\bibnamefont {Turzillo}},\ }\bibfield  {title} {\bibinfo {title} {Symmetry-protected topological phases of mixed states in the doubled space},\ }\href {https://doi.org/10.1103/PRXQuantum.6.010348} {\bibfield  {journal} {\bibinfo  {journal} {PRX Quantum}\ }\textbf {\bibinfo {volume} {6}},\ \bibinfo {pages} {010348} (\bibinfo {year} {2025})}\BibitemShut {NoStop}%
\bibitem [{\citenamefont {Guo}\ and\ \citenamefont {Yang}(2025)}]{Yang2025PRB}%
  \BibitemOpen
  \bibfield  {author} {\bibinfo {author} {\bibfnamefont {Y.}~\bibnamefont {Guo}}\ and\ \bibinfo {author} {\bibfnamefont {S.}~\bibnamefont {Yang}},\ }\bibfield  {title} {\bibinfo {title} {Strong-to-weak spontaneous symmetry breaking meets average symmetry-protected topological order},\ }\href {https://doi.org/10.1103/PhysRevB.111.L201108} {\bibfield  {journal} {\bibinfo  {journal} {Phys. Rev. B}\ }\textbf {\bibinfo {volume} {111}},\ \bibinfo {pages} {L201108} (\bibinfo {year} {2025})}\BibitemShut {NoStop}%
\bibitem [{\citenamefont {Guo}\ \emph {et~al.}(2025)\citenamefont {Guo}, \citenamefont {Zhang}, \citenamefont {Zhang}, \citenamefont {Yang},\ and\ \citenamefont {Bi}}]{Zhen2025PRX}%
  \BibitemOpen
  \bibfield  {author} {\bibinfo {author} {\bibfnamefont {Y.}~\bibnamefont {Guo}}, \bibinfo {author} {\bibfnamefont {J.-H.}\ \bibnamefont {Zhang}}, \bibinfo {author} {\bibfnamefont {H.-R.}\ \bibnamefont {Zhang}}, \bibinfo {author} {\bibfnamefont {S.}~\bibnamefont {Yang}},\ and\ \bibinfo {author} {\bibfnamefont {Z.}~\bibnamefont {Bi}},\ }\bibfield  {title} {\bibinfo {title} {Locally purified density operators for symmetry-protected topological phases in mixed states},\ }\href {https://doi.org/10.1103/PhysRevX.15.021060} {\bibfield  {journal} {\bibinfo  {journal} {Phys. Rev. X}\ }\textbf {\bibinfo {volume} {15}},\ \bibinfo {pages} {021060} (\bibinfo {year} {2025})}\BibitemShut {NoStop}%
\bibitem [{\citenamefont {Sun}\ \emph {et~al.}(2025)\citenamefont {Sun}, \citenamefont {Zhang}, \citenamefont {Bi},\ and\ \citenamefont {You}}]{YIzhi2025PRXQuantum}%
  \BibitemOpen
  \bibfield  {author} {\bibinfo {author} {\bibfnamefont {S.}~\bibnamefont {Sun}}, \bibinfo {author} {\bibfnamefont {J.-H.}\ \bibnamefont {Zhang}}, \bibinfo {author} {\bibfnamefont {Z.}~\bibnamefont {Bi}},\ and\ \bibinfo {author} {\bibfnamefont {Y.}~\bibnamefont {You}},\ }\bibfield  {title} {\bibinfo {title} {Holographic view of mixed-state symmetry-protected topological phases in open quantum systems},\ }\href {https://doi.org/10.1103/PRXQuantum.6.020333} {\bibfield  {journal} {\bibinfo  {journal} {PRX Quantum}\ }\textbf {\bibinfo {volume} {6}},\ \bibinfo {pages} {020333} (\bibinfo {year} {2025})}\BibitemShut {NoStop}%
\bibitem [{\citenamefont {Konig}\ \emph {et~al.}(2007)\citenamefont {Konig}, \citenamefont {Wiedmann}, \citenamefont {Brune}, \citenamefont {Roth}, \citenamefont {Buhmann}, \citenamefont {Molenkamp}, \citenamefont {Qi},\ and\ \citenamefont {Zhang}}]{konig2007quantum}%
  \BibitemOpen
  \bibfield  {author} {\bibinfo {author} {\bibfnamefont {M.}~\bibnamefont {Konig}}, \bibinfo {author} {\bibfnamefont {S.}~\bibnamefont {Wiedmann}}, \bibinfo {author} {\bibfnamefont {C.}~\bibnamefont {Brune}}, \bibinfo {author} {\bibfnamefont {A.}~\bibnamefont {Roth}}, \bibinfo {author} {\bibfnamefont {H.}~\bibnamefont {Buhmann}}, \bibinfo {author} {\bibfnamefont {L.~W.}\ \bibnamefont {Molenkamp}}, \bibinfo {author} {\bibfnamefont {X.-L.}\ \bibnamefont {Qi}},\ and\ \bibinfo {author} {\bibfnamefont {S.-C.}\ \bibnamefont {Zhang}},\ }\bibfield  {title} {\bibinfo {title} {Quantum spin hall insulator state in hgte quantum wells},\ }\href {https://doi.org/10.1126/science.1148047} {\bibfield  {journal} {\bibinfo  {journal} {Science}\ }\textbf {\bibinfo {volume} {318}},\ \bibinfo {pages} {766} (\bibinfo {year} {2007})}\BibitemShut {NoStop}%
\bibitem [{\citenamefont {Hsieh}\ \emph {et~al.}(2008)\citenamefont {Hsieh}, \citenamefont {Qian}, \citenamefont {Wray}, \citenamefont {Xia}, \citenamefont {Hor}, \citenamefont {Cava},\ and\ \citenamefont {Hasan}}]{hsieh2008topological}%
  \BibitemOpen
  \bibfield  {author} {\bibinfo {author} {\bibfnamefont {D.}~\bibnamefont {Hsieh}}, \bibinfo {author} {\bibfnamefont {D.}~\bibnamefont {Qian}}, \bibinfo {author} {\bibfnamefont {L.}~\bibnamefont {Wray}}, \bibinfo {author} {\bibfnamefont {Y.}~\bibnamefont {Xia}}, \bibinfo {author} {\bibfnamefont {Y.~S.}\ \bibnamefont {Hor}}, \bibinfo {author} {\bibfnamefont {R.~J.}\ \bibnamefont {Cava}},\ and\ \bibinfo {author} {\bibfnamefont {M.~Z.}\ \bibnamefont {Hasan}},\ }\bibfield  {title} {\bibinfo {title} {A topological dirac insulator in a quantum spin hall phase},\ }\href {https://doi.org/10.1038/nature06843} {\bibfield  {journal} {\bibinfo  {journal} {Nature}\ }\textbf {\bibinfo {volume} {452}},\ \bibinfo {pages} {970} (\bibinfo {year} {2008})}\BibitemShut {NoStop}%
\bibitem [{\citenamefont {Hsieh}\ \emph {et~al.}(2009)\citenamefont {Hsieh}, \citenamefont {Xia}, \citenamefont {Wray}, \citenamefont {Qian}, \citenamefont {Pal}, \citenamefont {Dil}, \citenamefont {Osterwalder}, \citenamefont {Meier}, \citenamefont {Bihlmayer}, \citenamefont {Kane} \emph {et~al.}}]{hsieh2009observation}%
  \BibitemOpen
  \bibfield  {author} {\bibinfo {author} {\bibfnamefont {D.}~\bibnamefont {Hsieh}}, \bibinfo {author} {\bibfnamefont {Y.}~\bibnamefont {Xia}}, \bibinfo {author} {\bibfnamefont {L.}~\bibnamefont {Wray}}, \bibinfo {author} {\bibfnamefont {D.}~\bibnamefont {Qian}}, \bibinfo {author} {\bibfnamefont {A.}~\bibnamefont {Pal}}, \bibinfo {author} {\bibfnamefont {J.~H.}\ \bibnamefont {Dil}}, \bibinfo {author} {\bibfnamefont {J.}~\bibnamefont {Osterwalder}}, \bibinfo {author} {\bibfnamefont {F.}~\bibnamefont {Meier}}, \bibinfo {author} {\bibfnamefont {G.}~\bibnamefont {Bihlmayer}}, \bibinfo {author} {\bibfnamefont {C.~L.}\ \bibnamefont {Kane}}, \emph {et~al.},\ }\bibfield  {title} {\bibinfo {title} {Observation of unconventional quantum spin textures in topological insulators},\ }\href {https://doi.org/10.1126/science.1167733} {\bibfield  {journal} {\bibinfo  {journal} {Science}\ }\textbf {\bibinfo {volume} {323}},\ \bibinfo {pages} {919} (\bibinfo {year} {2009})}\BibitemShut {NoStop}%
\bibitem [{\citenamefont {Tanaka}\ \emph {et~al.}(2012)\citenamefont {Tanaka}, \citenamefont {Ren}, \citenamefont {Sato}, \citenamefont {Nakayama}, \citenamefont {Souma}, \citenamefont {Takahashi}, \citenamefont {Segawa},\ and\ \citenamefont {Ando}}]{tanaka2012experimental}%
  \BibitemOpen
  \bibfield  {author} {\bibinfo {author} {\bibfnamefont {Y.}~\bibnamefont {Tanaka}}, \bibinfo {author} {\bibfnamefont {Z.}~\bibnamefont {Ren}}, \bibinfo {author} {\bibfnamefont {T.}~\bibnamefont {Sato}}, \bibinfo {author} {\bibfnamefont {K.}~\bibnamefont {Nakayama}}, \bibinfo {author} {\bibfnamefont {S.}~\bibnamefont {Souma}}, \bibinfo {author} {\bibfnamefont {T.}~\bibnamefont {Takahashi}}, \bibinfo {author} {\bibfnamefont {K.}~\bibnamefont {Segawa}},\ and\ \bibinfo {author} {\bibfnamefont {Y.}~\bibnamefont {Ando}},\ }\bibfield  {title} {\bibinfo {title} {Experimental realization of a topological crystalline insulator in snte},\ }\href {https://doi.org/10.1038/nphys2442} {\bibfield  {journal} {\bibinfo  {journal} {Nature Physics}\ }\textbf {\bibinfo {volume} {8}},\ \bibinfo {pages} {800} (\bibinfo {year} {2012})}\BibitemShut {NoStop}%
\bibitem [{\citenamefont {Chang}\ \emph {et~al.}(2013)\citenamefont {Chang}, \citenamefont {Zhang}, \citenamefont {Feng}, \citenamefont {Shen}, \citenamefont {Zhang}, \citenamefont {Guo}, \citenamefont {Li}, \citenamefont {Ou}, \citenamefont {Wei}, \citenamefont {Wang} \emph {et~al.}}]{chang2013experimental}%
  \BibitemOpen
  \bibfield  {author} {\bibinfo {author} {\bibfnamefont {C.-Z.}\ \bibnamefont {Chang}}, \bibinfo {author} {\bibfnamefont {J.}~\bibnamefont {Zhang}}, \bibinfo {author} {\bibfnamefont {X.}~\bibnamefont {Feng}}, \bibinfo {author} {\bibfnamefont {J.}~\bibnamefont {Shen}}, \bibinfo {author} {\bibfnamefont {Z.}~\bibnamefont {Zhang}}, \bibinfo {author} {\bibfnamefont {M.}~\bibnamefont {Guo}}, \bibinfo {author} {\bibfnamefont {K.}~\bibnamefont {Li}}, \bibinfo {author} {\bibfnamefont {Y.}~\bibnamefont {Ou}}, \bibinfo {author} {\bibfnamefont {P.}~\bibnamefont {Wei}}, \bibinfo {author} {\bibfnamefont {L.-L.}\ \bibnamefont {Wang}}, \emph {et~al.},\ }\bibfield  {title} {\bibinfo {title} {Experimental observation of the quantum anomalous hall effect in a magnetic topological insulator},\ }\href {https://doi.org/10.1126/science.1234414} {\bibfield  {journal} {\bibinfo  {journal} {Science}\ }\textbf {\bibinfo {volume} {340}},\ \bibinfo {pages} {167} (\bibinfo {year} {2013})}\BibitemShut {NoStop}%
\bibitem [{\citenamefont {Atala}\ \emph {et~al.}(2013)\citenamefont {Atala}, \citenamefont {Aidelsburger}, \citenamefont {Barreiro}, \citenamefont {Abanin}, \citenamefont {Kitagawa}, \citenamefont {Demler},\ and\ \citenamefont {Bloch}}]{atala2013direct}%
  \BibitemOpen
  \bibfield  {author} {\bibinfo {author} {\bibfnamefont {M.}~\bibnamefont {Atala}}, \bibinfo {author} {\bibfnamefont {M.}~\bibnamefont {Aidelsburger}}, \bibinfo {author} {\bibfnamefont {J.~T.}\ \bibnamefont {Barreiro}}, \bibinfo {author} {\bibfnamefont {D.}~\bibnamefont {Abanin}}, \bibinfo {author} {\bibfnamefont {T.}~\bibnamefont {Kitagawa}}, \bibinfo {author} {\bibfnamefont {E.}~\bibnamefont {Demler}},\ and\ \bibinfo {author} {\bibfnamefont {I.}~\bibnamefont {Bloch}},\ }\bibfield  {title} {\bibinfo {title} {Direct measurement of the zak phase in topological bloch bands},\ }\href {https://doi.org/10.1038/nphys2790} {\bibfield  {journal} {\bibinfo  {journal} {Nature Physics}\ }\textbf {\bibinfo {volume} {9}},\ \bibinfo {pages} {795} (\bibinfo {year} {2013})}\BibitemShut {NoStop}%
\bibitem [{\citenamefont {Jotzu}\ \emph {et~al.}(2014)\citenamefont {Jotzu}, \citenamefont {Messer}, \citenamefont {Desbuquois}, \citenamefont {Lebrat}, \citenamefont {Uehlinger}, \citenamefont {Greif},\ and\ \citenamefont {Esslinger}}]{jotzu2014experimental}%
  \BibitemOpen
  \bibfield  {author} {\bibinfo {author} {\bibfnamefont {G.}~\bibnamefont {Jotzu}}, \bibinfo {author} {\bibfnamefont {M.}~\bibnamefont {Messer}}, \bibinfo {author} {\bibfnamefont {R.}~\bibnamefont {Desbuquois}}, \bibinfo {author} {\bibfnamefont {M.}~\bibnamefont {Lebrat}}, \bibinfo {author} {\bibfnamefont {T.}~\bibnamefont {Uehlinger}}, \bibinfo {author} {\bibfnamefont {D.}~\bibnamefont {Greif}},\ and\ \bibinfo {author} {\bibfnamefont {T.}~\bibnamefont {Esslinger}},\ }\bibfield  {title} {\bibinfo {title} {Experimental realization of the topological haldane model with ultracold fermions},\ }\href {https://doi.org/10.1038/nature13915} {\bibfield  {journal} {\bibinfo  {journal} {Nature}\ }\textbf {\bibinfo {volume} {515}},\ \bibinfo {pages} {237} (\bibinfo {year} {2014})}\BibitemShut {NoStop}%
\bibitem [{\citenamefont {De~L{\'e}s{\'e}leuc}\ \emph {et~al.}(2019)\citenamefont {De~L{\'e}s{\'e}leuc}, \citenamefont {Lienhard}, \citenamefont {Scholl}, \citenamefont {Barredo}, \citenamefont {Weber}, \citenamefont {Lang}, \citenamefont {B{\"u}chler}, \citenamefont {Lahaye},\ and\ \citenamefont {Browaeys}}]{de2019observation}%
  \BibitemOpen
  \bibfield  {author} {\bibinfo {author} {\bibfnamefont {S.}~\bibnamefont {De~L{\'e}s{\'e}leuc}}, \bibinfo {author} {\bibfnamefont {V.}~\bibnamefont {Lienhard}}, \bibinfo {author} {\bibfnamefont {P.}~\bibnamefont {Scholl}}, \bibinfo {author} {\bibfnamefont {D.}~\bibnamefont {Barredo}}, \bibinfo {author} {\bibfnamefont {S.}~\bibnamefont {Weber}}, \bibinfo {author} {\bibfnamefont {N.}~\bibnamefont {Lang}}, \bibinfo {author} {\bibfnamefont {H.~P.}\ \bibnamefont {B{\"u}chler}}, \bibinfo {author} {\bibfnamefont {T.}~\bibnamefont {Lahaye}},\ and\ \bibinfo {author} {\bibfnamefont {A.}~\bibnamefont {Browaeys}},\ }\bibfield  {title} {\bibinfo {title} {Observation of a symmetry-protected topological phase of interacting bosons with rydberg atoms},\ }\href {https://doi.org/10.1126/science.aav9105} {\bibfield  {journal} {\bibinfo  {journal} {Science}\ }\textbf {\bibinfo {volume} {365}},\ \bibinfo {pages} {775} (\bibinfo {year} {2019})}\BibitemShut {NoStop}%
\bibitem [{\citenamefont {Sompet}\ \emph {et~al.}(2022)\citenamefont {Sompet}, \citenamefont {Hirthe}, \citenamefont {Bourgund}, \citenamefont {Chalopin}, \citenamefont {Bibo}, \citenamefont {Koepsell}, \citenamefont {Bojovi{\'c}}, \citenamefont {Verresen}, \citenamefont {Pollmann}, \citenamefont {Salomon}, \citenamefont {Gross}, \citenamefont {Hilker},\ and\ \citenamefont {Bloch}}]{sompet2022realizing}%
  \BibitemOpen
  \bibfield  {author} {\bibinfo {author} {\bibfnamefont {P.}~\bibnamefont {Sompet}}, \bibinfo {author} {\bibfnamefont {S.}~\bibnamefont {Hirthe}}, \bibinfo {author} {\bibfnamefont {D.}~\bibnamefont {Bourgund}}, \bibinfo {author} {\bibfnamefont {T.}~\bibnamefont {Chalopin}}, \bibinfo {author} {\bibfnamefont {J.}~\bibnamefont {Bibo}}, \bibinfo {author} {\bibfnamefont {J.}~\bibnamefont {Koepsell}}, \bibinfo {author} {\bibfnamefont {P.}~\bibnamefont {Bojovi{\'c}}}, \bibinfo {author} {\bibfnamefont {R.}~\bibnamefont {Verresen}}, \bibinfo {author} {\bibfnamefont {F.}~\bibnamefont {Pollmann}}, \bibinfo {author} {\bibfnamefont {G.}~\bibnamefont {Salomon}}, \bibinfo {author} {\bibfnamefont {C.}~\bibnamefont {Gross}}, \bibinfo {author} {\bibfnamefont {T.~A.}\ \bibnamefont {Hilker}},\ and\ \bibinfo {author} {\bibfnamefont {I.}~\bibnamefont {Bloch}},\ }\bibfield  {title} {\bibinfo {title} {Realizing the symmetry-protected haldane phase in fermi--hubbard ladders},\ }\href {https://doi.org/10.1038/s41586-022-04688-z}
  {\bibfield  {journal} {\bibinfo  {journal} {Nature}\ }\textbf {\bibinfo {volume} {606}},\ \bibinfo {pages} {484} (\bibinfo {year} {2022})}\BibitemShut {NoStop}%
\bibitem [{\citenamefont {Mishra}\ \emph {et~al.}(2021)\citenamefont {Mishra}, \citenamefont {Catarina}, \citenamefont {Wu}, \citenamefont {Ortiz}, \citenamefont {Jacob}, \citenamefont {Eimre}, \citenamefont {Ma}, \citenamefont {Pignedoli}, \citenamefont {Feng}, \citenamefont {Ruffieux} \emph {et~al.}}]{mishra2021observation}%
  \BibitemOpen
  \bibfield  {author} {\bibinfo {author} {\bibfnamefont {S.}~\bibnamefont {Mishra}}, \bibinfo {author} {\bibfnamefont {G.}~\bibnamefont {Catarina}}, \bibinfo {author} {\bibfnamefont {F.}~\bibnamefont {Wu}}, \bibinfo {author} {\bibfnamefont {R.}~\bibnamefont {Ortiz}}, \bibinfo {author} {\bibfnamefont {D.}~\bibnamefont {Jacob}}, \bibinfo {author} {\bibfnamefont {K.}~\bibnamefont {Eimre}}, \bibinfo {author} {\bibfnamefont {J.}~\bibnamefont {Ma}}, \bibinfo {author} {\bibfnamefont {C.~A.}\ \bibnamefont {Pignedoli}}, \bibinfo {author} {\bibfnamefont {X.}~\bibnamefont {Feng}}, \bibinfo {author} {\bibfnamefont {P.}~\bibnamefont {Ruffieux}}, \emph {et~al.},\ }\bibfield  {title} {\bibinfo {title} {Observation of fractional edge excitations in nanographene spin chains},\ }\href {https://doi.org/10.1038/s41586-021-03842-3} {\bibfield  {journal} {\bibinfo  {journal} {Nature}\ }\textbf {\bibinfo {volume} {598}},\ \bibinfo {pages} {287} (\bibinfo {year} {2021})}\BibitemShut {NoStop}%
\bibitem [{\citenamefont {Zhao}\ \emph {et~al.}(2024)\citenamefont {Zhao}, \citenamefont {Catarina}, \citenamefont {Zhang}, \citenamefont {Henriques}, \citenamefont {Yang}, \citenamefont {Ma}, \citenamefont {Feng}, \citenamefont {Gr{\"o}ning}, \citenamefont {Ruffieux}, \citenamefont {Fern{\'a}ndez-Rossier} \emph {et~al.}}]{zhao2024tunable}%
  \BibitemOpen
  \bibfield  {author} {\bibinfo {author} {\bibfnamefont {C.}~\bibnamefont {Zhao}}, \bibinfo {author} {\bibfnamefont {G.}~\bibnamefont {Catarina}}, \bibinfo {author} {\bibfnamefont {J.-J.}\ \bibnamefont {Zhang}}, \bibinfo {author} {\bibfnamefont {J.~C.}\ \bibnamefont {Henriques}}, \bibinfo {author} {\bibfnamefont {L.}~\bibnamefont {Yang}}, \bibinfo {author} {\bibfnamefont {J.}~\bibnamefont {Ma}}, \bibinfo {author} {\bibfnamefont {X.}~\bibnamefont {Feng}}, \bibinfo {author} {\bibfnamefont {O.}~\bibnamefont {Gr{\"o}ning}}, \bibinfo {author} {\bibfnamefont {P.}~\bibnamefont {Ruffieux}}, \bibinfo {author} {\bibfnamefont {J.}~\bibnamefont {Fern{\'a}ndez-Rossier}}, \emph {et~al.},\ }\bibfield  {title} {\bibinfo {title} {Tunable topological phases in nanographene-based spin-1/2 alternating-exchange heisenberg chains},\ }\href {https://doi.org/10.1038/s41565-024-01805-z} {\bibfield  {journal} {\bibinfo  {journal} {Nature Nanotechnology}\ }\textbf {\bibinfo {volume} {19}},\ \bibinfo {pages} {1789} (\bibinfo {year}
  {2024})}\BibitemShut {NoStop}%
\bibitem [{\citenamefont {Kiczynski}\ \emph {et~al.}(2022)\citenamefont {Kiczynski}, \citenamefont {Gorman}, \citenamefont {Geng}, \citenamefont {Donnelly}, \citenamefont {Chung}, \citenamefont {He}, \citenamefont {Keizer},\ and\ \citenamefont {Simmons}}]{kiczynski2022engineering}%
  \BibitemOpen
  \bibfield  {author} {\bibinfo {author} {\bibfnamefont {M.}~\bibnamefont {Kiczynski}}, \bibinfo {author} {\bibfnamefont {S.}~\bibnamefont {Gorman}}, \bibinfo {author} {\bibfnamefont {H.}~\bibnamefont {Geng}}, \bibinfo {author} {\bibfnamefont {M.}~\bibnamefont {Donnelly}}, \bibinfo {author} {\bibfnamefont {Y.}~\bibnamefont {Chung}}, \bibinfo {author} {\bibfnamefont {Y.}~\bibnamefont {He}}, \bibinfo {author} {\bibfnamefont {J.}~\bibnamefont {Keizer}},\ and\ \bibinfo {author} {\bibfnamefont {M.}~\bibnamefont {Simmons}},\ }\bibfield  {title} {\bibinfo {title} {Engineering topological states in atom-based semiconductor quantum dots},\ }\href {https://doi.org/10.1038/s41586-022-04706-0} {\bibfield  {journal} {\bibinfo  {journal} {Nature}\ }\textbf {\bibinfo {volume} {606}},\ \bibinfo {pages} {694} (\bibinfo {year} {2022})}\BibitemShut {NoStop}%
\bibitem [{\citenamefont {Wang}\ \emph {et~al.}(2024)\citenamefont {Wang}, \citenamefont {Fan}, \citenamefont {Chen}, \citenamefont {Jiang}, \citenamefont {Gao}, \citenamefont {Lado},\ and\ \citenamefont {Yang}}]{wang2024construction}%
  \BibitemOpen
  \bibfield  {author} {\bibinfo {author} {\bibfnamefont {H.}~\bibnamefont {Wang}}, \bibinfo {author} {\bibfnamefont {P.}~\bibnamefont {Fan}}, \bibinfo {author} {\bibfnamefont {J.}~\bibnamefont {Chen}}, \bibinfo {author} {\bibfnamefont {L.}~\bibnamefont {Jiang}}, \bibinfo {author} {\bibfnamefont {H.-J.}\ \bibnamefont {Gao}}, \bibinfo {author} {\bibfnamefont {J.~L.}\ \bibnamefont {Lado}},\ and\ \bibinfo {author} {\bibfnamefont {K.}~\bibnamefont {Yang}},\ }\bibfield  {title} {\bibinfo {title} {Construction of topological quantum magnets from atomic spins on surfaces},\ }\href {https://doi.org/10.1038/s41565-024-01775-2} {\bibfield  {journal} {\bibinfo  {journal} {Nature Nanotechnology}\ }\textbf {\bibinfo {volume} {19}},\ \bibinfo {pages} {1782} (\bibinfo {year} {2024})}\BibitemShut {NoStop}%
\bibitem [{\citenamefont {Li}\ \emph {et~al.}(2009)\citenamefont {Li}, \citenamefont {Chu}, \citenamefont {Jain},\ and\ \citenamefont {Shen}}]{li2009topological}%
  \BibitemOpen
  \bibfield  {author} {\bibinfo {author} {\bibfnamefont {J.}~\bibnamefont {Li}}, \bibinfo {author} {\bibfnamefont {R.-L.}\ \bibnamefont {Chu}}, \bibinfo {author} {\bibfnamefont {J.~K.}\ \bibnamefont {Jain}},\ and\ \bibinfo {author} {\bibfnamefont {S.-Q.}\ \bibnamefont {Shen}},\ }\bibfield  {title} {\bibinfo {title} {Topological anderson insulator},\ }\href {https://doi.org/10.1103/PhysRevLett.102.136806} {\bibfield  {journal} {\bibinfo  {journal} {Physical Review Letters}\ }\textbf {\bibinfo {volume} {102}},\ \bibinfo {pages} {136806} (\bibinfo {year} {2009})}\BibitemShut {NoStop}%
\bibitem [{\citenamefont {Li}\ \emph {et~al.}(2021)\citenamefont {Li}, \citenamefont {Wang}, \citenamefont {Yang},\ and\ \citenamefont {Xu}}]{PhysRevLett.127.263004}%
  \BibitemOpen
  \bibfield  {author} {\bibinfo {author} {\bibfnamefont {K.}~\bibnamefont {Li}}, \bibinfo {author} {\bibfnamefont {J.-H.}\ \bibnamefont {Wang}}, \bibinfo {author} {\bibfnamefont {Y.-B.}\ \bibnamefont {Yang}},\ and\ \bibinfo {author} {\bibfnamefont {Y.}~\bibnamefont {Xu}},\ }\bibfield  {title} {\bibinfo {title} {Symmetry-protected topological phases in a rydberg glass},\ }\href {https://doi.org/10.1103/PhysRevLett.127.263004} {\bibfield  {journal} {\bibinfo  {journal} {Physical Review Letters}\ }\textbf {\bibinfo {volume} {127}},\ \bibinfo {pages} {263004} (\bibinfo {year} {2021})}\BibitemShut {NoStop}%
\bibitem [{\citenamefont {St{\"u}tzer}\ \emph {et~al.}(2018)\citenamefont {St{\"u}tzer}, \citenamefont {Plotnik}, \citenamefont {Lumer}, \citenamefont {Titum}, \citenamefont {Lindner}, \citenamefont {Segev}, \citenamefont {Rechtsman},\ and\ \citenamefont {Szameit}}]{stutzer2018photonic}%
  \BibitemOpen
  \bibfield  {author} {\bibinfo {author} {\bibfnamefont {S.}~\bibnamefont {St{\"u}tzer}}, \bibinfo {author} {\bibfnamefont {Y.}~\bibnamefont {Plotnik}}, \bibinfo {author} {\bibfnamefont {Y.}~\bibnamefont {Lumer}}, \bibinfo {author} {\bibfnamefont {P.}~\bibnamefont {Titum}}, \bibinfo {author} {\bibfnamefont {N.~H.}\ \bibnamefont {Lindner}}, \bibinfo {author} {\bibfnamefont {M.}~\bibnamefont {Segev}}, \bibinfo {author} {\bibfnamefont {M.~C.}\ \bibnamefont {Rechtsman}},\ and\ \bibinfo {author} {\bibfnamefont {A.}~\bibnamefont {Szameit}},\ }\bibfield  {title} {\bibinfo {title} {Photonic topological anderson insulators},\ }\href {https://doi.org/10.1038/s41586-018-0418-2} {\bibfield  {journal} {\bibinfo  {journal} {Nature}\ }\textbf {\bibinfo {volume} {560}},\ \bibinfo {pages} {461} (\bibinfo {year} {2018})}\BibitemShut {NoStop}%
\bibitem [{\citenamefont {Dai}\ \emph {et~al.}(2024)\citenamefont {Dai}, \citenamefont {Ma}, \citenamefont {Mao}, \citenamefont {Ao}, \citenamefont {Jia}, \citenamefont {Zheng}, \citenamefont {Zhai}, \citenamefont {Yang}, \citenamefont {Li}, \citenamefont {Tang}, \citenamefont {Luo}, \citenamefont {Zhang}, \citenamefont {Hu}, \citenamefont {Gong},\ and\ \citenamefont {Wang}}]{daiProgrammableTopologicalPhotonic2024}%
  \BibitemOpen
  \bibfield  {author} {\bibinfo {author} {\bibfnamefont {T.}~\bibnamefont {Dai}}, \bibinfo {author} {\bibfnamefont {A.}~\bibnamefont {Ma}}, \bibinfo {author} {\bibfnamefont {J.}~\bibnamefont {Mao}}, \bibinfo {author} {\bibfnamefont {Y.}~\bibnamefont {Ao}}, \bibinfo {author} {\bibfnamefont {X.}~\bibnamefont {Jia}}, \bibinfo {author} {\bibfnamefont {Y.}~\bibnamefont {Zheng}}, \bibinfo {author} {\bibfnamefont {C.}~\bibnamefont {Zhai}}, \bibinfo {author} {\bibfnamefont {Y.}~\bibnamefont {Yang}}, \bibinfo {author} {\bibfnamefont {Z.}~\bibnamefont {Li}}, \bibinfo {author} {\bibfnamefont {B.}~\bibnamefont {Tang}}, \bibinfo {author} {\bibfnamefont {J.}~\bibnamefont {Luo}}, \bibinfo {author} {\bibfnamefont {B.}~\bibnamefont {Zhang}}, \bibinfo {author} {\bibfnamefont {X.}~\bibnamefont {Hu}}, \bibinfo {author} {\bibfnamefont {Q.}~\bibnamefont {Gong}},\ and\ \bibinfo {author} {\bibfnamefont {J.}~\bibnamefont {Wang}},\ }\bibfield  {title} {\bibinfo {title} {A programmable topological photonic chip},\ }\href
  {https://doi.org/10.1038/s41563-024-01904-1} {\bibfield  {journal} {\bibinfo  {journal} {Nature Materials}\ }\textbf {\bibinfo {volume} {23}},\ \bibinfo {pages} {928} (\bibinfo {year} {2024})}\BibitemShut {NoStop}%
\bibitem [{\citenamefont {Chen}\ \emph {et~al.}(2024)\citenamefont {Chen}, \citenamefont {Gao}, \citenamefont {Cui}, \citenamefont {Mo}, \citenamefont {Chen}, \citenamefont {Zhang}, \citenamefont {Chan},\ and\ \citenamefont {Dong}}]{chenRealizationTimeReversalInvariant2024}%
  \BibitemOpen
  \bibfield  {author} {\bibinfo {author} {\bibfnamefont {X.-D.}\ \bibnamefont {Chen}}, \bibinfo {author} {\bibfnamefont {Z.-X.}\ \bibnamefont {Gao}}, \bibinfo {author} {\bibfnamefont {X.}~\bibnamefont {Cui}}, \bibinfo {author} {\bibfnamefont {H.-C.}\ \bibnamefont {Mo}}, \bibinfo {author} {\bibfnamefont {W.-J.}\ \bibnamefont {Chen}}, \bibinfo {author} {\bibfnamefont {R.-Y.}\ \bibnamefont {Zhang}}, \bibinfo {author} {\bibfnamefont {C.~T.}\ \bibnamefont {Chan}},\ and\ \bibinfo {author} {\bibfnamefont {J.-W.}\ \bibnamefont {Dong}},\ }\bibfield  {title} {\bibinfo {title} {Realization of time-reversal invariant photonic topological anderson insulators},\ }\href {https://doi.org/10.1103/PhysRevLett.133.133802} {\bibfield  {journal} {\bibinfo  {journal} {Physical Review Letters}\ }\textbf {\bibinfo {volume} {133}},\ \bibinfo {pages} {133802} (\bibinfo {year} {2024})}\BibitemShut {NoStop}%
\bibitem [{\citenamefont {Liu}\ \emph {et~al.}(2020)\citenamefont {Liu}, \citenamefont {Yang}, \citenamefont {Ren}, \citenamefont {Xue}, \citenamefont {Lin}, \citenamefont {Hu}, \citenamefont {Sun}, \citenamefont {Peng}, \citenamefont {Zhou}, \citenamefont {Chong} \emph {et~al.}}]{liu2020topological}%
  \BibitemOpen
  \bibfield  {author} {\bibinfo {author} {\bibfnamefont {G.-G.}\ \bibnamefont {Liu}}, \bibinfo {author} {\bibfnamefont {Y.}~\bibnamefont {Yang}}, \bibinfo {author} {\bibfnamefont {X.}~\bibnamefont {Ren}}, \bibinfo {author} {\bibfnamefont {H.}~\bibnamefont {Xue}}, \bibinfo {author} {\bibfnamefont {X.}~\bibnamefont {Lin}}, \bibinfo {author} {\bibfnamefont {Y.-H.}\ \bibnamefont {Hu}}, \bibinfo {author} {\bibfnamefont {H.-x.}\ \bibnamefont {Sun}}, \bibinfo {author} {\bibfnamefont {B.}~\bibnamefont {Peng}}, \bibinfo {author} {\bibfnamefont {P.}~\bibnamefont {Zhou}}, \bibinfo {author} {\bibfnamefont {Y.}~\bibnamefont {Chong}}, \emph {et~al.},\ }\bibfield  {title} {\bibinfo {title} {Topological anderson insulator in disordered photonic crystals},\ }\href {https://doi.org/10.1103/PhysRevLett.125.133603} {\bibfield  {journal} {\bibinfo  {journal} {Physical Review Letters}\ }\textbf {\bibinfo {volume} {125}},\ \bibinfo {pages} {133603} (\bibinfo {year} {2020})}\BibitemShut {NoStop}%
\bibitem [{\citenamefont {Ren}\ \emph {et~al.}(2024)\citenamefont {Ren}, \citenamefont {Yu}, \citenamefont {Wu}, \citenamefont {Qi}, \citenamefont {Wang}, \citenamefont {Yao}, \citenamefont {Ren}, \citenamefont {Guo}, \citenamefont {Jiang}, \citenamefont {Chen} \emph {et~al.}}]{ren2024realization}%
  \BibitemOpen
  \bibfield  {author} {\bibinfo {author} {\bibfnamefont {M.}~\bibnamefont {Ren}}, \bibinfo {author} {\bibfnamefont {Y.}~\bibnamefont {Yu}}, \bibinfo {author} {\bibfnamefont {B.}~\bibnamefont {Wu}}, \bibinfo {author} {\bibfnamefont {X.}~\bibnamefont {Qi}}, \bibinfo {author} {\bibfnamefont {Y.}~\bibnamefont {Wang}}, \bibinfo {author} {\bibfnamefont {X.}~\bibnamefont {Yao}}, \bibinfo {author} {\bibfnamefont {J.}~\bibnamefont {Ren}}, \bibinfo {author} {\bibfnamefont {Z.}~\bibnamefont {Guo}}, \bibinfo {author} {\bibfnamefont {H.}~\bibnamefont {Jiang}}, \bibinfo {author} {\bibfnamefont {H.}~\bibnamefont {Chen}}, \emph {et~al.},\ }\bibfield  {title} {\bibinfo {title} {Realization of gapped and ungapped photonic topological anderson insulators},\ }\href {https://doi.org/10.1103/PhysRevLett.132.066602} {\bibfield  {journal} {\bibinfo  {journal} {Physical Review Letters}\ }\textbf {\bibinfo {volume} {132}},\ \bibinfo {pages} {066602} (\bibinfo {year} {2024})}\BibitemShut {NoStop}%
\bibitem [{\citenamefont {Zangeneh-Nejad}\ and\ \citenamefont {Fleury}(2020)}]{zangeneh2020disorder}%
  \BibitemOpen
  \bibfield  {author} {\bibinfo {author} {\bibfnamefont {F.}~\bibnamefont {Zangeneh-Nejad}}\ and\ \bibinfo {author} {\bibfnamefont {R.}~\bibnamefont {Fleury}},\ }\bibfield  {title} {\bibinfo {title} {Disorder-induced signal filtering with topological metamaterials},\ }\href {https://doi.org/10.1002/adma.202001034} {\bibfield  {journal} {\bibinfo  {journal} {Advanced Materials}\ }\textbf {\bibinfo {volume} {32}},\ \bibinfo {pages} {2001034} (\bibinfo {year} {2020})}\BibitemShut {NoStop}%
\bibitem [{\citenamefont {Liu}\ \emph {et~al.}(2023)\citenamefont {Liu}, \citenamefont {Xie}, \citenamefont {Wang}, \citenamefont {Liu}, \citenamefont {Li}, \citenamefont {Cheng}, \citenamefont {Tian}, \citenamefont {Liu},\ and\ \citenamefont {Chen}}]{liu2023acoustic}%
  \BibitemOpen
  \bibfield  {author} {\bibinfo {author} {\bibfnamefont {H.}~\bibnamefont {Liu}}, \bibinfo {author} {\bibfnamefont {B.}~\bibnamefont {Xie}}, \bibinfo {author} {\bibfnamefont {H.}~\bibnamefont {Wang}}, \bibinfo {author} {\bibfnamefont {W.}~\bibnamefont {Liu}}, \bibinfo {author} {\bibfnamefont {Z.}~\bibnamefont {Li}}, \bibinfo {author} {\bibfnamefont {H.}~\bibnamefont {Cheng}}, \bibinfo {author} {\bibfnamefont {J.}~\bibnamefont {Tian}}, \bibinfo {author} {\bibfnamefont {Z.}~\bibnamefont {Liu}},\ and\ \bibinfo {author} {\bibfnamefont {S.}~\bibnamefont {Chen}},\ }\bibfield  {title} {\bibinfo {title} {Acoustic spin-chern topological anderson insulators},\ }\href {https://doi.org/10.1103/PhysRevB.108.L161410} {\bibfield  {journal} {\bibinfo  {journal} {Physical Review B}\ }\textbf {\bibinfo {volume} {108}},\ \bibinfo {pages} {L161410} (\bibinfo {year} {2023})}\BibitemShut {NoStop}%
\bibitem [{\citenamefont {Meier}\ \emph {et~al.}(2018)\citenamefont {Meier}, \citenamefont {An}, \citenamefont {Dauphin}, \citenamefont {Maffei}, \citenamefont {Massignan}, \citenamefont {Hughes},\ and\ \citenamefont {Gadway}}]{meier2018observation}%
  \BibitemOpen
  \bibfield  {author} {\bibinfo {author} {\bibfnamefont {E.~J.}\ \bibnamefont {Meier}}, \bibinfo {author} {\bibfnamefont {F.~A.}\ \bibnamefont {An}}, \bibinfo {author} {\bibfnamefont {A.}~\bibnamefont {Dauphin}}, \bibinfo {author} {\bibfnamefont {M.}~\bibnamefont {Maffei}}, \bibinfo {author} {\bibfnamefont {P.}~\bibnamefont {Massignan}}, \bibinfo {author} {\bibfnamefont {T.~L.}\ \bibnamefont {Hughes}},\ and\ \bibinfo {author} {\bibfnamefont {B.}~\bibnamefont {Gadway}},\ }\bibfield  {title} {\bibinfo {title} {Observation of the topological anderson insulator in disordered atomic wires},\ }\href {https://doi.org/10.1126/science.aat3406} {\bibfield  {journal} {\bibinfo  {journal} {Science}\ }\textbf {\bibinfo {volume} {362}},\ \bibinfo {pages} {929} (\bibinfo {year} {2018})}\BibitemShut {NoStop}%
\bibitem [{\citenamefont {Li}\ \emph {et~al.}(2024)\citenamefont {Li}, \citenamefont {Xu}, \citenamefont {Wang}, \citenamefont {Tang}, \citenamefont {Zhang}, \citenamefont {Yang}, \citenamefont {Su}, \citenamefont {Wang}, \citenamefont {Mi}, \citenamefont {Sun}, \citenamefont {Liang}, \citenamefont {Chen}, \citenamefont {Li}, \citenamefont {Zhang}, \citenamefont {Linghu}, \citenamefont {Han}, \citenamefont {Liu}, \citenamefont {Feng}, \citenamefont {Liu}, \citenamefont {Xue}, \citenamefont {Zhang}, \citenamefont {Jin}, \citenamefont {Zhu}, \citenamefont {Yu}, \citenamefont {Zhao},\ and\ \citenamefont {Xue}}]{PhysRevResearch.6.L042038}%
  \BibitemOpen
  \bibfield  {author} {\bibinfo {author} {\bibfnamefont {X.}~\bibnamefont {Li}}, \bibinfo {author} {\bibfnamefont {H.}~\bibnamefont {Xu}}, \bibinfo {author} {\bibfnamefont {J.}~\bibnamefont {Wang}}, \bibinfo {author} {\bibfnamefont {L.-Z.}\ \bibnamefont {Tang}}, \bibinfo {author} {\bibfnamefont {D.-W.}\ \bibnamefont {Zhang}}, \bibinfo {author} {\bibfnamefont {C.}~\bibnamefont {Yang}}, \bibinfo {author} {\bibfnamefont {T.}~\bibnamefont {Su}}, \bibinfo {author} {\bibfnamefont {C.}~\bibnamefont {Wang}}, \bibinfo {author} {\bibfnamefont {Z.}~\bibnamefont {Mi}}, \bibinfo {author} {\bibfnamefont {W.}~\bibnamefont {Sun}}, \bibinfo {author} {\bibfnamefont {X.}~\bibnamefont {Liang}}, \bibinfo {author} {\bibfnamefont {M.}~\bibnamefont {Chen}}, \bibinfo {author} {\bibfnamefont {C.}~\bibnamefont {Li}}, \bibinfo {author} {\bibfnamefont {Y.}~\bibnamefont {Zhang}}, \bibinfo {author} {\bibfnamefont {K.}~\bibnamefont {Linghu}}, \bibinfo {author} {\bibfnamefont {J.}~\bibnamefont {Han}}, \bibinfo {author} {\bibfnamefont
  {W.}~\bibnamefont {Liu}}, \bibinfo {author} {\bibfnamefont {Y.}~\bibnamefont {Feng}}, \bibinfo {author} {\bibfnamefont {P.}~\bibnamefont {Liu}}, \bibinfo {author} {\bibfnamefont {G.}~\bibnamefont {Xue}}, \bibinfo {author} {\bibfnamefont {J.}~\bibnamefont {Zhang}}, \bibinfo {author} {\bibfnamefont {Y.}~\bibnamefont {Jin}}, \bibinfo {author} {\bibfnamefont {S.-L.}\ \bibnamefont {Zhu}}, \bibinfo {author} {\bibfnamefont {H.}~\bibnamefont {Yu}}, \bibinfo {author} {\bibfnamefont {S.~P.}\ \bibnamefont {Zhao}},\ and\ \bibinfo {author} {\bibfnamefont {Q.-K.}\ \bibnamefont {Xue}},\ }\bibfield  {title} {\bibinfo {title} {Mapping the topology-localization phase diagram with quasiperiodic disorder using a programmable superconducting simulator},\ }\href {https://doi.org/10.1103/PhysRevResearch.6.L042038} {\bibfield  {journal} {\bibinfo  {journal} {Physical Review Research}\ }\textbf {\bibinfo {volume} {6}},\ \bibinfo {pages} {L042038} (\bibinfo {year} {2024})}\BibitemShut {NoStop}%
\bibitem [{\citenamefont {Barredo}\ \emph {et~al.}(2016)\citenamefont {Barredo}, \citenamefont {{de L{\'e}s{\'e}leuc}}, \citenamefont {Lienhard}, \citenamefont {Lahaye},\ and\ \citenamefont {Browaeys}}]{doi:10.1126/science.aah3778}%
  \BibitemOpen
  \bibfield  {author} {\bibinfo {author} {\bibfnamefont {D.}~\bibnamefont {Barredo}}, \bibinfo {author} {\bibfnamefont {S.}~\bibnamefont {{de L{\'e}s{\'e}leuc}}}, \bibinfo {author} {\bibfnamefont {V.}~\bibnamefont {Lienhard}}, \bibinfo {author} {\bibfnamefont {T.}~\bibnamefont {Lahaye}},\ and\ \bibinfo {author} {\bibfnamefont {A.}~\bibnamefont {Browaeys}},\ }\bibfield  {title} {\bibinfo {title} {An atom-by-atom assembler of defect-free arbitrary two-dimensional atomic arrays},\ }\href {https://doi.org/10.1126/science.aah3778} {\bibfield  {journal} {\bibinfo  {journal} {Science}\ }\textbf {\bibinfo {volume} {354}},\ \bibinfo {pages} {1021} (\bibinfo {year} {2016})}\BibitemShut {NoStop}%
\bibitem [{\citenamefont {Endres}\ \emph {et~al.}(2016)\citenamefont {Endres}, \citenamefont {Bernien}, \citenamefont {Keesling}, \citenamefont {Levine}, \citenamefont {Anschuetz}, \citenamefont {Krajenbrink}, \citenamefont {Senko}, \citenamefont {Vuletic}, \citenamefont {Greiner},\ and\ \citenamefont {Lukin}}]{doi:10.1126/science.aah3752}%
  \BibitemOpen
  \bibfield  {author} {\bibinfo {author} {\bibfnamefont {M.}~\bibnamefont {Endres}}, \bibinfo {author} {\bibfnamefont {H.}~\bibnamefont {Bernien}}, \bibinfo {author} {\bibfnamefont {A.}~\bibnamefont {Keesling}}, \bibinfo {author} {\bibfnamefont {H.}~\bibnamefont {Levine}}, \bibinfo {author} {\bibfnamefont {E.~R.}\ \bibnamefont {Anschuetz}}, \bibinfo {author} {\bibfnamefont {A.}~\bibnamefont {Krajenbrink}}, \bibinfo {author} {\bibfnamefont {C.}~\bibnamefont {Senko}}, \bibinfo {author} {\bibfnamefont {V.}~\bibnamefont {Vuletic}}, \bibinfo {author} {\bibfnamefont {M.}~\bibnamefont {Greiner}},\ and\ \bibinfo {author} {\bibfnamefont {M.~D.}\ \bibnamefont {Lukin}},\ }\bibfield  {title} {\bibinfo {title} {Atom-by-atom assembly of defect-free one-dimensional cold atom arrays},\ }\href {https://doi.org/10.1126/science.aah3752} {\bibfield  {journal} {\bibinfo  {journal} {Science}\ }\textbf {\bibinfo {volume} {354}},\ \bibinfo {pages} {1024} (\bibinfo {year} {2016})}\BibitemShut {NoStop}%
\bibitem [{\citenamefont {Kim}\ \emph {et~al.}(2016)\citenamefont {Kim}, \citenamefont {Lee}, \citenamefont {Lee}, \citenamefont {Jo}, \citenamefont {Song},\ and\ \citenamefont {Ahn}}]{kimSituSingleatomArray2016}%
  \BibitemOpen
  \bibfield  {author} {\bibinfo {author} {\bibfnamefont {H.}~\bibnamefont {Kim}}, \bibinfo {author} {\bibfnamefont {W.}~\bibnamefont {Lee}}, \bibinfo {author} {\bibfnamefont {H.-g.}\ \bibnamefont {Lee}}, \bibinfo {author} {\bibfnamefont {H.}~\bibnamefont {Jo}}, \bibinfo {author} {\bibfnamefont {Y.}~\bibnamefont {Song}},\ and\ \bibinfo {author} {\bibfnamefont {J.}~\bibnamefont {Ahn}},\ }\bibfield  {title} {\bibinfo {title} {In situ single-atom array synthesis using dynamic holographic optical tweezers},\ }\href {https://doi.org/10.1038/ncomms13317} {\bibfield  {journal} {\bibinfo  {journal} {Nature Communications}\ }\textbf {\bibinfo {volume} {7}},\ \bibinfo {pages} {13317} (\bibinfo {year} {2016})}\BibitemShut {NoStop}%
\bibitem [{\citenamefont {Chen}\ \emph {et~al.}(2023)\citenamefont {Chen}, \citenamefont {Bornet}, \citenamefont {Bintz}, \citenamefont {Emperauger}, \citenamefont {Leclerc}, \citenamefont {Liu}, \citenamefont {Scholl}, \citenamefont {Barredo}, \citenamefont {Hauschild}, \citenamefont {Chatterjee}, \citenamefont {Schuler}, \citenamefont {L{\"a}uchli}, \citenamefont {Zaletel}, \citenamefont {Lahaye}, \citenamefont {Yao},\ and\ \citenamefont {Browaeys}}]{chenContinuousSymmetryBreaking2023}%
  \BibitemOpen
  \bibfield  {author} {\bibinfo {author} {\bibfnamefont {C.}~\bibnamefont {Chen}}, \bibinfo {author} {\bibfnamefont {G.}~\bibnamefont {Bornet}}, \bibinfo {author} {\bibfnamefont {M.}~\bibnamefont {Bintz}}, \bibinfo {author} {\bibfnamefont {G.}~\bibnamefont {Emperauger}}, \bibinfo {author} {\bibfnamefont {L.}~\bibnamefont {Leclerc}}, \bibinfo {author} {\bibfnamefont {V.~S.}\ \bibnamefont {Liu}}, \bibinfo {author} {\bibfnamefont {P.}~\bibnamefont {Scholl}}, \bibinfo {author} {\bibfnamefont {D.}~\bibnamefont {Barredo}}, \bibinfo {author} {\bibfnamefont {J.}~\bibnamefont {Hauschild}}, \bibinfo {author} {\bibfnamefont {S.}~\bibnamefont {Chatterjee}}, \bibinfo {author} {\bibfnamefont {M.}~\bibnamefont {Schuler}}, \bibinfo {author} {\bibfnamefont {A.~M.}\ \bibnamefont {L{\"a}uchli}}, \bibinfo {author} {\bibfnamefont {M.~P.}\ \bibnamefont {Zaletel}}, \bibinfo {author} {\bibfnamefont {T.}~\bibnamefont {Lahaye}}, \bibinfo {author} {\bibfnamefont {N.~Y.}\ \bibnamefont {Yao}},\ and\ \bibinfo {author} {\bibfnamefont
  {A.}~\bibnamefont {Browaeys}},\ }\bibfield  {title} {\bibinfo {title} {Continuous symmetry breaking in a two-dimensional {{Rydberg}} array},\ }\href {https://doi.org/10.1038/s41586-023-05859-2} {\bibfield  {journal} {\bibinfo  {journal} {Nature}\ }\textbf {\bibinfo {volume} {616}},\ \bibinfo {pages} {691} (\bibinfo {year} {2023})}\BibitemShut {NoStop}%
\bibitem [{\citenamefont {Browaeys}\ \emph {et~al.}(2016)\citenamefont {Browaeys}, \citenamefont {Barredo},\ and\ \citenamefont {Lahaye}}]{Browaeys_2016_dipole_dipole}%
  \BibitemOpen
  \bibfield  {author} {\bibinfo {author} {\bibfnamefont {A.}~\bibnamefont {Browaeys}}, \bibinfo {author} {\bibfnamefont {D.}~\bibnamefont {Barredo}},\ and\ \bibinfo {author} {\bibfnamefont {T.}~\bibnamefont {Lahaye}},\ }\bibfield  {title} {\bibinfo {title} {Experimental investigations of dipole-dipole interactions between a few rydberg atoms},\ }\href {https://doi.org/10.1088/0953-4075/49/15/152001} {\bibfield  {journal} {\bibinfo  {journal} {Journal of Physics B: Atomic, Molecular and Optical Physics}\ }\textbf {\bibinfo {volume} {49}},\ \bibinfo {pages} {152001} (\bibinfo {year} {2016})}\BibitemShut {NoStop}%
\bibitem [{\citenamefont {Weber}\ \emph {et~al.}(2017)\citenamefont {Weber}, \citenamefont {Tresp}, \citenamefont {Menke}, \citenamefont {Urvoy}, \citenamefont {Firstenberg}, \citenamefont {B{\"u}chler},\ and\ \citenamefont {Hofferberth}}]{Weber_2017}%
  \BibitemOpen
  \bibfield  {author} {\bibinfo {author} {\bibfnamefont {S.}~\bibnamefont {Weber}}, \bibinfo {author} {\bibfnamefont {C.}~\bibnamefont {Tresp}}, \bibinfo {author} {\bibfnamefont {H.}~\bibnamefont {Menke}}, \bibinfo {author} {\bibfnamefont {A.}~\bibnamefont {Urvoy}}, \bibinfo {author} {\bibfnamefont {O.}~\bibnamefont {Firstenberg}}, \bibinfo {author} {\bibfnamefont {H.~P.}\ \bibnamefont {B{\"u}chler}},\ and\ \bibinfo {author} {\bibfnamefont {S.}~\bibnamefont {Hofferberth}},\ }\bibfield  {title} {\bibinfo {title} {Calculation of rydberg interaction potentials},\ }\href {https://doi.org/10.1088/1361-6455/aa743a} {\bibfield  {journal} {\bibinfo  {journal} {Journal of Physics B: Atomic, Molecular and Optical Physics}\ }\textbf {\bibinfo {volume} {50}},\ \bibinfo {pages} {133001} (\bibinfo {year} {2017})}\BibitemShut {NoStop}%
\bibitem [{\citenamefont {Urban}\ \emph {et~al.}(2009)\citenamefont {Urban}, \citenamefont {Johnson}, \citenamefont {Henage}, \citenamefont {Isenhower}, \citenamefont {Yavuz}, \citenamefont {Walker},\ and\ \citenamefont {Saffman}}]{urbanObservationRydbergBlockade2009}%
  \BibitemOpen
  \bibfield  {author} {\bibinfo {author} {\bibfnamefont {E.}~\bibnamefont {Urban}}, \bibinfo {author} {\bibfnamefont {T.~A.}\ \bibnamefont {Johnson}}, \bibinfo {author} {\bibfnamefont {T.}~\bibnamefont {Henage}}, \bibinfo {author} {\bibfnamefont {L.}~\bibnamefont {Isenhower}}, \bibinfo {author} {\bibfnamefont {D.~D.}\ \bibnamefont {Yavuz}}, \bibinfo {author} {\bibfnamefont {T.~G.}\ \bibnamefont {Walker}},\ and\ \bibinfo {author} {\bibfnamefont {M.}~\bibnamefont {Saffman}},\ }\bibfield  {title} {\bibinfo {title} {Observation of {{Rydberg}} blockade between two~atoms},\ }\href {https://doi.org/10.1038/nphys1178} {\bibfield  {journal} {\bibinfo  {journal} {Nature Physics}\ }\textbf {\bibinfo {volume} {5}},\ \bibinfo {pages} {110} (\bibinfo {year} {2009})}\BibitemShut {NoStop}%
\bibitem [{\citenamefont {Semeghini}\ \emph {et~al.}(2021)\citenamefont {Semeghini}, \citenamefont {Levine}, \citenamefont {Keesling}, \citenamefont {Ebadi}, \citenamefont {Wang}, \citenamefont {Bluvstein}, \citenamefont {Verresen}, \citenamefont {Pichler}, \citenamefont {Kalinowski}, \citenamefont {Samajdar}, \citenamefont {Omran}, \citenamefont {Sachdev}, \citenamefont {Vishwanath}, \citenamefont {Greiner}, \citenamefont {Vuleti{\'c}},\ and\ \citenamefont {Lukin}}]{doi:10.1126/science.abi8794}%
  \BibitemOpen
  \bibfield  {author} {\bibinfo {author} {\bibfnamefont {G.}~\bibnamefont {Semeghini}}, \bibinfo {author} {\bibfnamefont {H.}~\bibnamefont {Levine}}, \bibinfo {author} {\bibfnamefont {A.}~\bibnamefont {Keesling}}, \bibinfo {author} {\bibfnamefont {S.}~\bibnamefont {Ebadi}}, \bibinfo {author} {\bibfnamefont {T.~T.}\ \bibnamefont {Wang}}, \bibinfo {author} {\bibfnamefont {D.}~\bibnamefont {Bluvstein}}, \bibinfo {author} {\bibfnamefont {R.}~\bibnamefont {Verresen}}, \bibinfo {author} {\bibfnamefont {H.}~\bibnamefont {Pichler}}, \bibinfo {author} {\bibfnamefont {M.}~\bibnamefont {Kalinowski}}, \bibinfo {author} {\bibfnamefont {R.}~\bibnamefont {Samajdar}}, \bibinfo {author} {\bibfnamefont {A.}~\bibnamefont {Omran}}, \bibinfo {author} {\bibfnamefont {S.}~\bibnamefont {Sachdev}}, \bibinfo {author} {\bibfnamefont {A.}~\bibnamefont {Vishwanath}}, \bibinfo {author} {\bibfnamefont {M.}~\bibnamefont {Greiner}}, \bibinfo {author} {\bibfnamefont {V.}~\bibnamefont {Vuleti{\'c}}},\ and\ \bibinfo {author} {\bibfnamefont
  {M.~D.}\ \bibnamefont {Lukin}},\ }\bibfield  {title} {\bibinfo {title} {Probing topological spin liquids on a programmable quantum simulator},\ }\href {https://doi.org/10.1126/science.abi8794} {\bibfield  {journal} {\bibinfo  {journal} {Science}\ }\textbf {\bibinfo {volume} {374}},\ \bibinfo {pages} {1242} (\bibinfo {year} {2021})}\BibitemShut {NoStop}%
\bibitem [{\citenamefont {Ebadi}\ \emph {et~al.}(2022)\citenamefont {Ebadi}, \citenamefont {Keesling}, \citenamefont {Cain}, \citenamefont {Wang}, \citenamefont {Levine}, \citenamefont {Bluvstein}, \citenamefont {Semeghini}, \citenamefont {Omran}, \citenamefont {Liu}, \citenamefont {Samajdar}, \citenamefont {Luo}, \citenamefont {Nash}, \citenamefont {Gao}, \citenamefont {Barak}, \citenamefont {Farhi}, \citenamefont {Sachdev}, \citenamefont {Gemelke}, \citenamefont {Zhou}, \citenamefont {Choi}, \citenamefont {Pichler}, \citenamefont {Wang}, \citenamefont {Greiner}, \citenamefont {Vuleti{\'c}},\ and\ \citenamefont {Lukin}}]{doi:10.1126/science.abo6587}%
  \BibitemOpen
  \bibfield  {author} {\bibinfo {author} {\bibfnamefont {S.}~\bibnamefont {Ebadi}}, \bibinfo {author} {\bibfnamefont {A.}~\bibnamefont {Keesling}}, \bibinfo {author} {\bibfnamefont {M.}~\bibnamefont {Cain}}, \bibinfo {author} {\bibfnamefont {T.~T.}\ \bibnamefont {Wang}}, \bibinfo {author} {\bibfnamefont {H.}~\bibnamefont {Levine}}, \bibinfo {author} {\bibfnamefont {D.}~\bibnamefont {Bluvstein}}, \bibinfo {author} {\bibfnamefont {G.}~\bibnamefont {Semeghini}}, \bibinfo {author} {\bibfnamefont {A.}~\bibnamefont {Omran}}, \bibinfo {author} {\bibfnamefont {J.-G.}\ \bibnamefont {Liu}}, \bibinfo {author} {\bibfnamefont {R.}~\bibnamefont {Samajdar}}, \bibinfo {author} {\bibfnamefont {X.-Z.}\ \bibnamefont {Luo}}, \bibinfo {author} {\bibfnamefont {B.}~\bibnamefont {Nash}}, \bibinfo {author} {\bibfnamefont {X.}~\bibnamefont {Gao}}, \bibinfo {author} {\bibfnamefont {B.}~\bibnamefont {Barak}}, \bibinfo {author} {\bibfnamefont {E.}~\bibnamefont {Farhi}}, \bibinfo {author} {\bibfnamefont {S.}~\bibnamefont {Sachdev}},
  \bibinfo {author} {\bibfnamefont {N.}~\bibnamefont {Gemelke}}, \bibinfo {author} {\bibfnamefont {L.}~\bibnamefont {Zhou}}, \bibinfo {author} {\bibfnamefont {S.}~\bibnamefont {Choi}}, \bibinfo {author} {\bibfnamefont {H.}~\bibnamefont {Pichler}}, \bibinfo {author} {\bibfnamefont {S.-T.}\ \bibnamefont {Wang}}, \bibinfo {author} {\bibfnamefont {M.}~\bibnamefont {Greiner}}, \bibinfo {author} {\bibfnamefont {V.}~\bibnamefont {Vuleti{\'c}}},\ and\ \bibinfo {author} {\bibfnamefont {M.~D.}\ \bibnamefont {Lukin}},\ }\bibfield  {title} {\bibinfo {title} {Quantum optimization of maximum independent set using {{Rydberg}} atom arrays},\ }\href {https://doi.org/10.1126/science.abo6587} {\bibfield  {journal} {\bibinfo  {journal} {Science}\ }\textbf {\bibinfo {volume} {376}},\ \bibinfo {pages} {1209} (\bibinfo {year} {2022})}\BibitemShut {NoStop}%
\bibitem [{\citenamefont {Chiu}\ \emph {et~al.}(2016)\citenamefont {Chiu}, \citenamefont {Teo}, \citenamefont {Schnyder},\ and\ \citenamefont {Ryu}}]{ChiuRevModPhys}%
  \BibitemOpen
  \bibfield  {author} {\bibinfo {author} {\bibfnamefont {C.-K.}\ \bibnamefont {Chiu}}, \bibinfo {author} {\bibfnamefont {J.~C.~Y.}\ \bibnamefont {Teo}}, \bibinfo {author} {\bibfnamefont {A.~P.}\ \bibnamefont {Schnyder}},\ and\ \bibinfo {author} {\bibfnamefont {S.}~\bibnamefont {Ryu}},\ }\bibfield  {title} {\bibinfo {title} {Classification of topological quantum matter with symmetries},\ }\href {https://doi.org/10.1103/RevModPhys.88.035005} {\bibfield  {journal} {\bibinfo  {journal} {Reviews of Modern Physics}\ }\textbf {\bibinfo {volume} {88}},\ \bibinfo {pages} {035005} (\bibinfo {year} {2016})}\BibitemShut {NoStop}%
\bibitem [{\citenamefont {Resta}(1998)}]{resta1998quantum}%
  \BibitemOpen
  \bibfield  {author} {\bibinfo {author} {\bibfnamefont {R.}~\bibnamefont {Resta}},\ }\bibfield  {title} {\bibinfo {title} {Quantum-mechanical position operator in extended systems},\ }\href {https://doi.org/10.1103/PhysRevLett.80.1800} {\bibfield  {journal} {\bibinfo  {journal} {Physical Review Letters}\ }\textbf {\bibinfo {volume} {80}},\ \bibinfo {pages} {1800} (\bibinfo {year} {1998})}\BibitemShut {NoStop}%
\bibitem [{\citenamefont {Manousakis}(1991)}]{RevModPhys.63.1}%
  \BibitemOpen
  \bibfield  {author} {\bibinfo {author} {\bibfnamefont {E.}~\bibnamefont {Manousakis}},\ }\bibfield  {title} {\bibinfo {title} {The spin-\textonehalf{} heisenberg antiferromagnet on a square lattice and its application to the cuprous oxides},\ }\href {https://doi.org/10.1103/RevModPhys.63.1} {\bibfield  {journal} {\bibinfo  {journal} {Reviews of Modern Physics}\ }\textbf {\bibinfo {volume} {63}},\ \bibinfo {pages} {1} (\bibinfo {year} {1991})}\BibitemShut {NoStop}%
\bibitem [{\citenamefont {Hida}(1992)}]{hida1992crossover}%
  \BibitemOpen
  \bibfield  {author} {\bibinfo {author} {\bibfnamefont {K.}~\bibnamefont {Hida}},\ }\bibfield  {title} {\bibinfo {title} {Crossover between the haldane-gap phase and the dimer phase in the spin-1/2 alternating heisenberg chain},\ }\href {https://doi.org/10.1103/PhysRevB.45.2207} {\bibfield  {journal} {\bibinfo  {journal} {Physical Review B}\ }\textbf {\bibinfo {volume} {45}},\ \bibinfo {pages} {2207} (\bibinfo {year} {1992})}\BibitemShut {NoStop}%
\bibitem [{\citenamefont {Nakamura}\ and\ \citenamefont {Todo}(2002)}]{nakamura2002order}%
  \BibitemOpen
  \bibfield  {author} {\bibinfo {author} {\bibfnamefont {M.}~\bibnamefont {Nakamura}}\ and\ \bibinfo {author} {\bibfnamefont {S.}~\bibnamefont {Todo}},\ }\bibfield  {title} {\bibinfo {title} {Order parameter to characterize valence-bond-solid states in quantum spin chains},\ }\href {https://doi.org/10.1103/PhysRevLett.89.077204} {\bibfield  {journal} {\bibinfo  {journal} {Physical Review Letters}\ }\textbf {\bibinfo {volume} {89}},\ \bibinfo {pages} {077204} (\bibinfo {year} {2002})}\BibitemShut {NoStop}%
\bibitem [{\citenamefont {Tasaki}(2018)}]{tasaki2018topological}%
  \BibitemOpen
  \bibfield  {author} {\bibinfo {author} {\bibfnamefont {H.}~\bibnamefont {Tasaki}},\ }\bibfield  {title} {\bibinfo {title} {Topological phase transition and ${\mathbb{z}}_{2}$ index for $s=1$ quantum spin chains},\ }\href {https://doi.org/10.1103/PhysRevLett.121.140604} {\bibfield  {journal} {\bibinfo  {journal} {Physical Review Letters}\ }\textbf {\bibinfo {volume} {121}},\ \bibinfo {pages} {140604} (\bibinfo {year} {2018})}\BibitemShut {NoStop}%
\bibitem [{\citenamefont {Liang}\ \emph {et~al.}(2024)\citenamefont {Liang}, \citenamefont {Yue}, \citenamefont {Chao}, \citenamefont {Hua}, \citenamefont {Lin}, \citenamefont {Tey},\ and\ \citenamefont {You}}]{liang2024observationanomalousinformationscrambling}%
  \BibitemOpen
  \bibfield  {author} {\bibinfo {author} {\bibfnamefont {X.}~\bibnamefont {Liang}}, \bibinfo {author} {\bibfnamefont {Z.}~\bibnamefont {Yue}}, \bibinfo {author} {\bibfnamefont {Y.-X.}\ \bibnamefont {Chao}}, \bibinfo {author} {\bibfnamefont {Z.-X.}\ \bibnamefont {Hua}}, \bibinfo {author} {\bibfnamefont {Y.}~\bibnamefont {Lin}}, \bibinfo {author} {\bibfnamefont {M.~K.}\ \bibnamefont {Tey}},\ and\ \bibinfo {author} {\bibfnamefont {L.}~\bibnamefont {You}},\ }\href {https://arxiv.org/abs/2410.16174} {\bibinfo {title} {Observation of anomalous information scrambling in a rydberg atom array}} (\bibinfo {year} {2024}),\ \Eprint {https://arxiv.org/abs/2410.16174} {arXiv:2410.16174 [quant-ph]} \BibitemShut {NoStop}%
\bibitem [{\citenamefont {Bahri}\ \emph {et~al.}(2015)\citenamefont {Bahri}, \citenamefont {Vosk}, \citenamefont {Altman},\ and\ \citenamefont {Vishwanath}}]{bahri2015localization}%
  \BibitemOpen
  \bibfield  {author} {\bibinfo {author} {\bibfnamefont {Y.}~\bibnamefont {Bahri}}, \bibinfo {author} {\bibfnamefont {R.}~\bibnamefont {Vosk}}, \bibinfo {author} {\bibfnamefont {E.}~\bibnamefont {Altman}},\ and\ \bibinfo {author} {\bibfnamefont {A.}~\bibnamefont {Vishwanath}},\ }\bibfield  {title} {\bibinfo {title} {Localization and topology protected quantum coherence at the edge of hot matter},\ }\href {https://doi.org/10.1038/ncomms8341} {\bibfield  {journal} {\bibinfo  {journal} {Nature communications}\ }\textbf {\bibinfo {volume} {6}},\ \bibinfo {pages} {7341} (\bibinfo {year} {2015})}\BibitemShut {NoStop}%
\bibitem [{\citenamefont {Zhang}\ \emph {et~al.}(2022{\natexlab{b}})\citenamefont {Zhang}, \citenamefont {Jiang}, \citenamefont {Deng}, \citenamefont {Wang}, \citenamefont {Chen}, \citenamefont {Zhang}, \citenamefont {Ren}, \citenamefont {Dong}, \citenamefont {Xu}, \citenamefont {Gao}, \citenamefont {Jin}, \citenamefont {Zhu}, \citenamefont {Guo}, \citenamefont {Li}, \citenamefont {Song}, \citenamefont {Gorshkov}, \citenamefont {Iadecola}, \citenamefont {Liu}, \citenamefont {Gong}, \citenamefont {Wang}, \citenamefont {Deng},\ and\ \citenamefont {Wang}}]{zhang_digital_2022}%
  \BibitemOpen
  \bibfield  {author} {\bibinfo {author} {\bibfnamefont {X.}~\bibnamefont {Zhang}}, \bibinfo {author} {\bibfnamefont {W.}~\bibnamefont {Jiang}}, \bibinfo {author} {\bibfnamefont {J.}~\bibnamefont {Deng}}, \bibinfo {author} {\bibfnamefont {K.}~\bibnamefont {Wang}}, \bibinfo {author} {\bibfnamefont {J.}~\bibnamefont {Chen}}, \bibinfo {author} {\bibfnamefont {P.}~\bibnamefont {Zhang}}, \bibinfo {author} {\bibfnamefont {W.}~\bibnamefont {Ren}}, \bibinfo {author} {\bibfnamefont {H.}~\bibnamefont {Dong}}, \bibinfo {author} {\bibfnamefont {S.}~\bibnamefont {Xu}}, \bibinfo {author} {\bibfnamefont {Y.}~\bibnamefont {Gao}}, \bibinfo {author} {\bibfnamefont {F.}~\bibnamefont {Jin}}, \bibinfo {author} {\bibfnamefont {X.}~\bibnamefont {Zhu}}, \bibinfo {author} {\bibfnamefont {Q.}~\bibnamefont {Guo}}, \bibinfo {author} {\bibfnamefont {H.}~\bibnamefont {Li}}, \bibinfo {author} {\bibfnamefont {C.}~\bibnamefont {Song}}, \bibinfo {author} {\bibfnamefont {A.~V.}\ \bibnamefont {Gorshkov}}, \bibinfo {author} {\bibfnamefont
  {T.}~\bibnamefont {Iadecola}}, \bibinfo {author} {\bibfnamefont {F.}~\bibnamefont {Liu}}, \bibinfo {author} {\bibfnamefont {Z.-X.}\ \bibnamefont {Gong}}, \bibinfo {author} {\bibfnamefont {Z.}~\bibnamefont {Wang}}, \bibinfo {author} {\bibfnamefont {D.-L.}\ \bibnamefont {Deng}},\ and\ \bibinfo {author} {\bibfnamefont {H.}~\bibnamefont {Wang}},\ }\bibfield  {title} {\bibinfo {title} {Digital quantum simulation of floquet symmetry-protected topological phases},\ }\href {https://doi.org/10.1038/s41586-022-04854-3} {\bibfield  {journal} {\bibinfo  {journal} {Nature}\ }\textbf {\bibinfo {volume} {607}},\ \bibinfo {pages} {468} (\bibinfo {year} {2022}{\natexlab{b}})}\BibitemShut {NoStop}%
\bibitem [{\citenamefont {Mi}\ \emph {et~al.}(2022)\citenamefont {Mi}, \citenamefont {Sonner}, \citenamefont {Niu}, \citenamefont {Lee}, \citenamefont {Foxen}, \citenamefont {Acharya}, \citenamefont {Aleiner}, \citenamefont {Andersen}, \citenamefont {Arute}, \citenamefont {Arya}, \citenamefont {Asfaw}, \citenamefont {Atalaya}, \citenamefont {Bardin}, \citenamefont {Basso}, \citenamefont {Bengtsson}, \citenamefont {Bortoli}, \citenamefont {Bourassa}, \citenamefont {Brill}, \citenamefont {Broughton}, \citenamefont {Buckley}, \citenamefont {Buell}, \citenamefont {Burkett}, \citenamefont {Bushnell}, \citenamefont {Chen}, \citenamefont {Chiaro}, \citenamefont {Collins}, \citenamefont {Conner}, \citenamefont {Courtney}, \citenamefont {Crook}, \citenamefont {Debroy}, \citenamefont {Demura}, \citenamefont {Dunsworth}, \citenamefont {Eppens}, \citenamefont {Erickson}, \citenamefont {Faoro}, \citenamefont {Farhi}, \citenamefont {Fatemi}, \citenamefont {Flores}, \citenamefont {Forati}, \citenamefont {Fowler},
  \citenamefont {Giang}, \citenamefont {Gidney}, \citenamefont {Gilboa}, \citenamefont {Giustina}, \citenamefont {Dau}, \citenamefont {Gross}, \citenamefont {Habegger}, \citenamefont {Harrigan}, \citenamefont {Hoffmann}, \citenamefont {Hong}, \citenamefont {Huang}, \citenamefont {Huff}, \citenamefont {Huggins}, \citenamefont {Ioffe}, \citenamefont {Isakov}, \citenamefont {Iveland}, \citenamefont {Jeffrey}, \citenamefont {Jiang}, \citenamefont {Jones}, \citenamefont {Kafri}, \citenamefont {Kechedzhi}, \citenamefont {Khattar}, \citenamefont {Kim}, \citenamefont {Kitaev}, \citenamefont {Klimov}, \citenamefont {Klots}, \citenamefont {Korotkov}, \citenamefont {Kostritsa}, \citenamefont {Kreikebaum}, \citenamefont {Landhuis}, \citenamefont {Laptev}, \citenamefont {Lau}, \citenamefont {Lee}, \citenamefont {Laws}, \citenamefont {Liu}, \citenamefont {Locharla}, \citenamefont {Martin}, \citenamefont {McClean}, \citenamefont {McEwen}, \citenamefont {Costa}, \citenamefont {Miao}, \citenamefont {Mohseni}, \citenamefont
  {Montazeri}, \citenamefont {Morvan}, \citenamefont {Mount}, \citenamefont {Mruczkiewicz}, \citenamefont {Naaman}, \citenamefont {Neeley}, \citenamefont {Neill}, \citenamefont {Newman}, \citenamefont {O{\textquoteright}Brien}, \citenamefont {Opremcak}, \citenamefont {Petukhov}, \citenamefont {Potter}, \citenamefont {Quintana}, \citenamefont {Rubin}, \citenamefont {Saei}, \citenamefont {Sank}, \citenamefont {Sankaragomathi}, \citenamefont {Satzinger}, \citenamefont {Schuster}, \citenamefont {Shearn}, \citenamefont {Shvarts}, \citenamefont {Strain}, \citenamefont {Su}, \citenamefont {Szalay}, \citenamefont {Vidal}, \citenamefont {Villalonga}, \citenamefont {Vollgraff-Heidweiller}, \citenamefont {White}, \citenamefont {Yao}, \citenamefont {Yeh}, \citenamefont {Yoo}, \citenamefont {Zalcman}, \citenamefont {Zhang}, \citenamefont {Zhu}, \citenamefont {Neven}, \citenamefont {Bacon}, \citenamefont {Hilton}, \citenamefont {Lucero}, \citenamefont {Babbush}, \citenamefont {Boixo}, \citenamefont {Megrant}, \citenamefont
  {Chen}, \citenamefont {Kelly}, \citenamefont {Smelyanskiy}, \citenamefont {Abanin},\ and\ \citenamefont {Roushan}}]{doi:10.1126/science.abq5769}%
  \BibitemOpen
  \bibfield  {author} {\bibinfo {author} {\bibfnamefont {X.}~\bibnamefont {Mi}}, \bibinfo {author} {\bibfnamefont {M.}~\bibnamefont {Sonner}}, \bibinfo {author} {\bibfnamefont {M.~Y.}\ \bibnamefont {Niu}}, \bibinfo {author} {\bibfnamefont {K.~W.}\ \bibnamefont {Lee}}, \bibinfo {author} {\bibfnamefont {B.}~\bibnamefont {Foxen}}, \bibinfo {author} {\bibfnamefont {R.}~\bibnamefont {Acharya}}, \bibinfo {author} {\bibfnamefont {I.}~\bibnamefont {Aleiner}}, \bibinfo {author} {\bibfnamefont {T.~I.}\ \bibnamefont {Andersen}}, \bibinfo {author} {\bibfnamefont {F.}~\bibnamefont {Arute}}, \bibinfo {author} {\bibfnamefont {K.}~\bibnamefont {Arya}}, \bibinfo {author} {\bibfnamefont {A.}~\bibnamefont {Asfaw}}, \bibinfo {author} {\bibfnamefont {J.}~\bibnamefont {Atalaya}}, \bibinfo {author} {\bibfnamefont {J.~C.}\ \bibnamefont {Bardin}}, \bibinfo {author} {\bibfnamefont {J.}~\bibnamefont {Basso}}, \bibinfo {author} {\bibfnamefont {A.}~\bibnamefont {Bengtsson}}, \bibinfo {author} {\bibfnamefont {G.}~\bibnamefont {Bortoli}},
  \bibinfo {author} {\bibfnamefont {A.}~\bibnamefont {Bourassa}}, \bibinfo {author} {\bibfnamefont {L.}~\bibnamefont {Brill}}, \bibinfo {author} {\bibfnamefont {M.}~\bibnamefont {Broughton}}, \bibinfo {author} {\bibfnamefont {B.~B.}\ \bibnamefont {Buckley}}, \bibinfo {author} {\bibfnamefont {D.~A.}\ \bibnamefont {Buell}}, \bibinfo {author} {\bibfnamefont {B.}~\bibnamefont {Burkett}}, \bibinfo {author} {\bibfnamefont {N.}~\bibnamefont {Bushnell}}, \bibinfo {author} {\bibfnamefont {Z.}~\bibnamefont {Chen}}, \bibinfo {author} {\bibfnamefont {B.}~\bibnamefont {Chiaro}}, \bibinfo {author} {\bibfnamefont {R.}~\bibnamefont {Collins}}, \bibinfo {author} {\bibfnamefont {P.}~\bibnamefont {Conner}}, \bibinfo {author} {\bibfnamefont {W.}~\bibnamefont {Courtney}}, \bibinfo {author} {\bibfnamefont {A.~L.}\ \bibnamefont {Crook}}, \bibinfo {author} {\bibfnamefont {D.~M.}\ \bibnamefont {Debroy}}, \bibinfo {author} {\bibfnamefont {S.}~\bibnamefont {Demura}}, \bibinfo {author} {\bibfnamefont {A.}~\bibnamefont {Dunsworth}},
  \bibinfo {author} {\bibfnamefont {D.}~\bibnamefont {Eppens}}, \bibinfo {author} {\bibfnamefont {C.}~\bibnamefont {Erickson}}, \bibinfo {author} {\bibfnamefont {L.}~\bibnamefont {Faoro}}, \bibinfo {author} {\bibfnamefont {E.}~\bibnamefont {Farhi}}, \bibinfo {author} {\bibfnamefont {R.}~\bibnamefont {Fatemi}}, \bibinfo {author} {\bibfnamefont {L.}~\bibnamefont {Flores}}, \bibinfo {author} {\bibfnamefont {E.}~\bibnamefont {Forati}}, \bibinfo {author} {\bibfnamefont {A.~G.}\ \bibnamefont {Fowler}}, \bibinfo {author} {\bibfnamefont {W.}~\bibnamefont {Giang}}, \bibinfo {author} {\bibfnamefont {C.}~\bibnamefont {Gidney}}, \bibinfo {author} {\bibfnamefont {D.}~\bibnamefont {Gilboa}}, \bibinfo {author} {\bibfnamefont {M.}~\bibnamefont {Giustina}}, \bibinfo {author} {\bibfnamefont {A.~G.}\ \bibnamefont {Dau}}, \bibinfo {author} {\bibfnamefont {J.~A.}\ \bibnamefont {Gross}}, \bibinfo {author} {\bibfnamefont {S.}~\bibnamefont {Habegger}}, \bibinfo {author} {\bibfnamefont {M.~P.}\ \bibnamefont {Harrigan}}, \bibinfo
  {author} {\bibfnamefont {M.}~\bibnamefont {Hoffmann}}, \bibinfo {author} {\bibfnamefont {S.}~\bibnamefont {Hong}}, \bibinfo {author} {\bibfnamefont {T.}~\bibnamefont {Huang}}, \bibinfo {author} {\bibfnamefont {A.}~\bibnamefont {Huff}}, \bibinfo {author} {\bibfnamefont {W.~J.}\ \bibnamefont {Huggins}}, \bibinfo {author} {\bibfnamefont {L.~B.}\ \bibnamefont {Ioffe}}, \bibinfo {author} {\bibfnamefont {S.~V.}\ \bibnamefont {Isakov}}, \bibinfo {author} {\bibfnamefont {J.}~\bibnamefont {Iveland}}, \bibinfo {author} {\bibfnamefont {E.}~\bibnamefont {Jeffrey}}, \bibinfo {author} {\bibfnamefont {Z.}~\bibnamefont {Jiang}}, \bibinfo {author} {\bibfnamefont {C.}~\bibnamefont {Jones}}, \bibinfo {author} {\bibfnamefont {D.}~\bibnamefont {Kafri}}, \bibinfo {author} {\bibfnamefont {K.}~\bibnamefont {Kechedzhi}}, \bibinfo {author} {\bibfnamefont {T.}~\bibnamefont {Khattar}}, \bibinfo {author} {\bibfnamefont {S.}~\bibnamefont {Kim}}, \bibinfo {author} {\bibfnamefont {A.~Y.}\ \bibnamefont {Kitaev}}, \bibinfo {author}
  {\bibfnamefont {P.~V.}\ \bibnamefont {Klimov}}, \bibinfo {author} {\bibfnamefont {A.~R.}\ \bibnamefont {Klots}}, \bibinfo {author} {\bibfnamefont {A.~N.}\ \bibnamefont {Korotkov}}, \bibinfo {author} {\bibfnamefont {F.}~\bibnamefont {Kostritsa}}, \bibinfo {author} {\bibfnamefont {J.~M.}\ \bibnamefont {Kreikebaum}}, \bibinfo {author} {\bibfnamefont {D.}~\bibnamefont {Landhuis}}, \bibinfo {author} {\bibfnamefont {P.}~\bibnamefont {Laptev}}, \bibinfo {author} {\bibfnamefont {K.-M.}\ \bibnamefont {Lau}}, \bibinfo {author} {\bibfnamefont {J.}~\bibnamefont {Lee}}, \bibinfo {author} {\bibfnamefont {L.}~\bibnamefont {Laws}}, \bibinfo {author} {\bibfnamefont {W.}~\bibnamefont {Liu}}, \bibinfo {author} {\bibfnamefont {A.}~\bibnamefont {Locharla}}, \bibinfo {author} {\bibfnamefont {O.}~\bibnamefont {Martin}}, \bibinfo {author} {\bibfnamefont {J.~R.}\ \bibnamefont {McClean}}, \bibinfo {author} {\bibfnamefont {M.}~\bibnamefont {McEwen}}, \bibinfo {author} {\bibfnamefont {B.~M.}\ \bibnamefont {Costa}}, \bibinfo {author}
  {\bibfnamefont {K.~C.}\ \bibnamefont {Miao}}, \bibinfo {author} {\bibfnamefont {M.}~\bibnamefont {Mohseni}}, \bibinfo {author} {\bibfnamefont {S.}~\bibnamefont {Montazeri}}, \bibinfo {author} {\bibfnamefont {A.}~\bibnamefont {Morvan}}, \bibinfo {author} {\bibfnamefont {E.}~\bibnamefont {Mount}}, \bibinfo {author} {\bibfnamefont {W.}~\bibnamefont {Mruczkiewicz}}, \bibinfo {author} {\bibfnamefont {O.}~\bibnamefont {Naaman}}, \bibinfo {author} {\bibfnamefont {M.}~\bibnamefont {Neeley}}, \bibinfo {author} {\bibfnamefont {C.}~\bibnamefont {Neill}}, \bibinfo {author} {\bibfnamefont {M.}~\bibnamefont {Newman}}, \bibinfo {author} {\bibfnamefont {T.~E.}\ \bibnamefont {O{\textquoteright}Brien}}, \bibinfo {author} {\bibfnamefont {A.}~\bibnamefont {Opremcak}}, \bibinfo {author} {\bibfnamefont {A.}~\bibnamefont {Petukhov}}, \bibinfo {author} {\bibfnamefont {R.}~\bibnamefont {Potter}}, \bibinfo {author} {\bibfnamefont {C.}~\bibnamefont {Quintana}}, \bibinfo {author} {\bibfnamefont {N.~C.}\ \bibnamefont {Rubin}}, \bibinfo
  {author} {\bibfnamefont {N.}~\bibnamefont {Saei}}, \bibinfo {author} {\bibfnamefont {D.}~\bibnamefont {Sank}}, \bibinfo {author} {\bibfnamefont {K.}~\bibnamefont {Sankaragomathi}}, \bibinfo {author} {\bibfnamefont {K.~J.}\ \bibnamefont {Satzinger}}, \bibinfo {author} {\bibfnamefont {C.}~\bibnamefont {Schuster}}, \bibinfo {author} {\bibfnamefont {M.~J.}\ \bibnamefont {Shearn}}, \bibinfo {author} {\bibfnamefont {V.}~\bibnamefont {Shvarts}}, \bibinfo {author} {\bibfnamefont {D.}~\bibnamefont {Strain}}, \bibinfo {author} {\bibfnamefont {Y.}~\bibnamefont {Su}}, \bibinfo {author} {\bibfnamefont {M.}~\bibnamefont {Szalay}}, \bibinfo {author} {\bibfnamefont {G.}~\bibnamefont {Vidal}}, \bibinfo {author} {\bibfnamefont {B.}~\bibnamefont {Villalonga}}, \bibinfo {author} {\bibfnamefont {C.}~\bibnamefont {Vollgraff-Heidweiller}}, \bibinfo {author} {\bibfnamefont {T.}~\bibnamefont {White}}, \bibinfo {author} {\bibfnamefont {Z.}~\bibnamefont {Yao}}, \bibinfo {author} {\bibfnamefont {P.}~\bibnamefont {Yeh}}, \bibinfo
  {author} {\bibfnamefont {J.}~\bibnamefont {Yoo}}, \bibinfo {author} {\bibfnamefont {A.}~\bibnamefont {Zalcman}}, \bibinfo {author} {\bibfnamefont {Y.}~\bibnamefont {Zhang}}, \bibinfo {author} {\bibfnamefont {N.}~\bibnamefont {Zhu}}, \bibinfo {author} {\bibfnamefont {H.}~\bibnamefont {Neven}}, \bibinfo {author} {\bibfnamefont {D.}~\bibnamefont {Bacon}}, \bibinfo {author} {\bibfnamefont {J.}~\bibnamefont {Hilton}}, \bibinfo {author} {\bibfnamefont {E.}~\bibnamefont {Lucero}}, \bibinfo {author} {\bibfnamefont {R.}~\bibnamefont {Babbush}}, \bibinfo {author} {\bibfnamefont {S.}~\bibnamefont {Boixo}}, \bibinfo {author} {\bibfnamefont {A.}~\bibnamefont {Megrant}}, \bibinfo {author} {\bibfnamefont {Y.}~\bibnamefont {Chen}}, \bibinfo {author} {\bibfnamefont {J.}~\bibnamefont {Kelly}}, \bibinfo {author} {\bibfnamefont {V.}~\bibnamefont {Smelyanskiy}}, \bibinfo {author} {\bibfnamefont {D.~A.}\ \bibnamefont {Abanin}},\ and\ \bibinfo {author} {\bibfnamefont {P.}~\bibnamefont {Roushan}},\ }\bibfield  {title} {\bibinfo
  {title} {Noise-resilient edge modes on a chain of superconducting qubits},\ }\href {https://doi.org/10.1126/science.abq5769} {\bibfield  {journal} {\bibinfo  {journal} {Science}\ }\textbf {\bibinfo {volume} {378}},\ \bibinfo {pages} {785} (\bibinfo {year} {2022})}\BibitemShut {NoStop}%
\bibitem [{\citenamefont {Agarwala}\ and\ \citenamefont {Shenoy}(2017)}]{PhysRevLett.118.236402}%
  \BibitemOpen
  \bibfield  {author} {\bibinfo {author} {\bibfnamefont {A.}~\bibnamefont {Agarwala}}\ and\ \bibinfo {author} {\bibfnamefont {V.~B.}\ \bibnamefont {Shenoy}},\ }\bibfield  {title} {\bibinfo {title} {Topological insulators in amorphous systems},\ }\href {https://doi.org/10.1103/PhysRevLett.118.236402} {\bibfield  {journal} {\bibinfo  {journal} {Physical Review Letters}\ }\textbf {\bibinfo {volume} {118}},\ \bibinfo {pages} {236402} (\bibinfo {year} {2017})}\BibitemShut {NoStop}%
\bibitem [{\citenamefont {Mitchell}\ \emph {et~al.}(2018)\citenamefont {Mitchell}, \citenamefont {Nash}, \citenamefont {Hexner}, \citenamefont {Turner},\ and\ \citenamefont {Irvine}}]{mitchell2018amorphous}%
  \BibitemOpen
  \bibfield  {author} {\bibinfo {author} {\bibfnamefont {N.~P.}\ \bibnamefont {Mitchell}}, \bibinfo {author} {\bibfnamefont {L.~M.}\ \bibnamefont {Nash}}, \bibinfo {author} {\bibfnamefont {D.}~\bibnamefont {Hexner}}, \bibinfo {author} {\bibfnamefont {A.~M.}\ \bibnamefont {Turner}},\ and\ \bibinfo {author} {\bibfnamefont {W.~T.}\ \bibnamefont {Irvine}},\ }\bibfield  {title} {\bibinfo {title} {Amorphous topological insulators constructed from random point sets},\ }\href {https://doi.org/10.1038/s41567-017-0024-5} {\bibfield  {journal} {\bibinfo  {journal} {Nature Physics}\ }\textbf {\bibinfo {volume} {14}},\ \bibinfo {pages} {380} (\bibinfo {year} {2018})}\BibitemShut {NoStop}%
\bibitem [{\citenamefont {Agarwala}\ \emph {et~al.}(2020)\citenamefont {Agarwala}, \citenamefont {Juri\ifmmode \check{c}\else \v{c}\fi{}i\ifmmode~\acute{c}\else \'{c}\fi{}},\ and\ \citenamefont {Roy}}]{AgarwalaPRRAmorphous}%
  \BibitemOpen
  \bibfield  {author} {\bibinfo {author} {\bibfnamefont {A.}~\bibnamefont {Agarwala}}, \bibinfo {author} {\bibfnamefont {V.}~\bibnamefont {Juri\ifmmode \check{c}\else \v{c}\fi{}i\ifmmode~\acute{c}\else \'{c}\fi{}}},\ and\ \bibinfo {author} {\bibfnamefont {B.}~\bibnamefont {Roy}},\ }\bibfield  {title} {\bibinfo {title} {Higher-order topological insulators in amorphous solids},\ }\href {https://doi.org/10.1103/PhysRevResearch.2.012067} {\bibfield  {journal} {\bibinfo  {journal} {Physical Review Research}\ }\textbf {\bibinfo {volume} {2}},\ \bibinfo {pages} {012067} (\bibinfo {year} {2020})}\BibitemShut {NoStop}%
\bibitem [{\citenamefont {Grushin}\ and\ \citenamefont {Repellin}(2023)}]{grushin2023amorphous}%
  \BibitemOpen
  \bibfield  {author} {\bibinfo {author} {\bibfnamefont {A.~G.}\ \bibnamefont {Grushin}}\ and\ \bibinfo {author} {\bibfnamefont {C.}~\bibnamefont {Repellin}},\ }\bibfield  {title} {\bibinfo {title} {Amorphous and polycrystalline routes toward a chiral spin liquid},\ }\href {https://doi.org/10.1103/PhysRevLett.130.186702} {\bibfield  {journal} {\bibinfo  {journal} {Physical Review Letters}\ }\textbf {\bibinfo {volume} {130}},\ \bibinfo {pages} {186702} (\bibinfo {year} {2023})}\BibitemShut {NoStop}%
\bibitem [{\citenamefont {Cassella}\ \emph {et~al.}(2023)\citenamefont {Cassella}, \citenamefont {d\textquoteright Ornellas}, \citenamefont {Hodson}, \citenamefont {Natori},\ and\ \citenamefont {Knolle}}]{cassella2023exact}%
  \BibitemOpen
  \bibfield  {author} {\bibinfo {author} {\bibfnamefont {G.}~\bibnamefont {Cassella}}, \bibinfo {author} {\bibfnamefont {P.}~\bibnamefont {d\textquoteright Ornellas}}, \bibinfo {author} {\bibfnamefont {T.}~\bibnamefont {Hodson}}, \bibinfo {author} {\bibfnamefont {W.~M.}\ \bibnamefont {Natori}},\ and\ \bibinfo {author} {\bibfnamefont {J.}~\bibnamefont {Knolle}},\ }\bibfield  {title} {\bibinfo {title} {An exact chiral amorphous spin liquid},\ }\href {https://doi.org/10.1038/s41467-023-42105-9} {\bibfield  {journal} {\bibinfo  {journal} {Nature Communications}\ }\textbf {\bibinfo {volume} {14}},\ \bibinfo {pages} {6663} (\bibinfo {year} {2023})}\BibitemShut {NoStop}%
\bibitem [{\citenamefont {He}\ \emph {et~al.}(2024)\citenamefont {He}, \citenamefont {Liu}, \citenamefont {Wu},\ and\ \citenamefont {Wang}}]{WangFloquentWang}%
  \BibitemOpen
  \bibfield  {author} {\bibinfo {author} {\bibfnamefont {P.}~\bibnamefont {He}}, \bibinfo {author} {\bibfnamefont {J.-X.}\ \bibnamefont {Liu}}, \bibinfo {author} {\bibfnamefont {H.}~\bibnamefont {Wu}},\ and\ \bibinfo {author} {\bibfnamefont {Z.~D.}\ \bibnamefont {Wang}},\ }\href {https://arxiv.org/abs/2404.18512} {\bibinfo {title} {Floquet amorphous topological orders in a rydberg glass}} (\bibinfo {year} {2024}),\ \Eprint {https://arxiv.org/abs/2404.18512} {arXiv:2404.18512 [cond-mat]} \BibitemShut {NoStop}%
\bibitem [{\citenamefont {Aramthottil}\ \emph {et~al.}(2024)\citenamefont {Aramthottil}, \citenamefont {Sierant}, \citenamefont {Lewenstein},\ and\ \citenamefont {Zakrzewski}}]{aramthottil2024phenomenology}%
  \BibitemOpen
  \bibfield  {author} {\bibinfo {author} {\bibfnamefont {A.~S.}\ \bibnamefont {Aramthottil}}, \bibinfo {author} {\bibfnamefont {P.}~\bibnamefont {Sierant}}, \bibinfo {author} {\bibfnamefont {M.}~\bibnamefont {Lewenstein}},\ and\ \bibinfo {author} {\bibfnamefont {J.}~\bibnamefont {Zakrzewski}},\ }\bibfield  {title} {\bibinfo {title} {Phenomenology of many-body localization in bond-disordered spin chains},\ }\href {https://doi.org/10.1103/PhysRevLett.133.196302} {\bibfield  {journal} {\bibinfo  {journal} {Physical Review Letters}\ }\textbf {\bibinfo {volume} {133}},\ \bibinfo {pages} {196302} (\bibinfo {year} {2024})}\BibitemShut {NoStop}%
\bibitem [{\citenamefont {Su}\ \emph {et~al.}()\citenamefont {Su}, \citenamefont {Sahay}, \citenamefont {Szurek}, \citenamefont {Douglas}, \citenamefont {Markovic}, \citenamefont {Dag}, \citenamefont {Verresen},\ and\ \citenamefont {Greiner}}]{su2025topologicalphasescriticalitymixed}%
  \BibitemOpen
  \bibfield  {author} {\bibinfo {author} {\bibfnamefont {L.}~\bibnamefont {Su}}, \bibinfo {author} {\bibfnamefont {R.}~\bibnamefont {Sahay}}, \bibinfo {author} {\bibfnamefont {M.}~\bibnamefont {Szurek}}, \bibinfo {author} {\bibfnamefont {A.}~\bibnamefont {Douglas}}, \bibinfo {author} {\bibfnamefont {O.}~\bibnamefont {Markovic}}, \bibinfo {author} {\bibfnamefont {C.~B.}\ \bibnamefont {Dag}}, \bibinfo {author} {\bibfnamefont {R.}~\bibnamefont {Verresen}},\ and\ \bibinfo {author} {\bibfnamefont {M.}~\bibnamefont {Greiner}},\ }\href {https://doi.org/10.48550/arXiv.2505.17009} {\bibinfo {title} {Topological phase transitions and mixed state order in a hubbard quantum simulator}},\ \Eprint {https://arxiv.org/abs/2505.17009 [cond-mat]} {2505.17009 [cond-mat]} \BibitemShut {NoStop}%
\bibitem [{\citenamefont {Hauschild}\ and\ \citenamefont {Pollmann}(2018)}]{tenpy}%
  \BibitemOpen
  \bibfield  {author} {\bibinfo {author} {\bibfnamefont {J.}~\bibnamefont {Hauschild}}\ and\ \bibinfo {author} {\bibfnamefont {F.}~\bibnamefont {Pollmann}},\ }\bibfield  {title} {\bibinfo {title} {{Efficient numerical simulations with Tensor Networks: Tensor Network Python (TeNPy)}},\ }\href {https://doi.org/10.21468/SciPostPhysLectNotes.5} {\bibfield  {journal} {\bibinfo  {journal} {SciPost Phys. Lect. Notes}\ ,\ \bibinfo {pages} {5}} (\bibinfo {year} {2018})},\ \bibinfo {note} {code available from \url{https://github.com/tenpy/tenpy}},\ \Eprint {https://arxiv.org/abs/1805.00055} {arXiv:1805.00055} \BibitemShut {NoStop}%
\bibitem [{\citenamefont {Hughes}\ \emph {et~al.}(2011)\citenamefont {Hughes}, \citenamefont {Prodan},\ and\ \citenamefont {Bernevig}}]{TaylorPhysRevB.83.245132}%
  \BibitemOpen
  \bibfield  {author} {\bibinfo {author} {\bibfnamefont {T.~L.}\ \bibnamefont {Hughes}}, \bibinfo {author} {\bibfnamefont {E.}~\bibnamefont {Prodan}},\ and\ \bibinfo {author} {\bibfnamefont {B.~A.}\ \bibnamefont {Bernevig}},\ }\bibfield  {title} {\bibinfo {title} {Inversion-symmetric topological insulators},\ }\href {https://doi.org/10.1103/PhysRevB.83.245132} {\bibfield  {journal} {\bibinfo  {journal} {Physical Review B}\ }\textbf {\bibinfo {volume} {83}},\ \bibinfo {pages} {245132} (\bibinfo {year} {2011})}\BibitemShut {NoStop}%
\bibitem [{\citenamefont {Chiu}\ \emph {et~al.}(2013)\citenamefont {Chiu}, \citenamefont {Yao},\ and\ \citenamefont {Ryu}}]{ChiuPhysRevB2013}%
  \BibitemOpen
  \bibfield  {author} {\bibinfo {author} {\bibfnamefont {C.-K.}\ \bibnamefont {Chiu}}, \bibinfo {author} {\bibfnamefont {H.}~\bibnamefont {Yao}},\ and\ \bibinfo {author} {\bibfnamefont {S.}~\bibnamefont {Ryu}},\ }\bibfield  {title} {\bibinfo {title} {Classification of topological insulators and superconductors in the presence of reflection symmetry},\ }\href {https://doi.org/10.1103/PhysRevB.88.075142} {\bibfield  {journal} {\bibinfo  {journal} {Physical Review B}\ }\textbf {\bibinfo {volume} {88}},\ \bibinfo {pages} {075142} (\bibinfo {year} {2013})}\BibitemShut {NoStop}%
\bibitem [{\citenamefont {Chen}\ \emph {et~al.}(2011{\natexlab{b}})\citenamefont {Chen}, \citenamefont {Gu},\ and\ \citenamefont {Wen}}]{ChenPhysRevB.84.235128}%
  \BibitemOpen
  \bibfield  {author} {\bibinfo {author} {\bibfnamefont {X.}~\bibnamefont {Chen}}, \bibinfo {author} {\bibfnamefont {Z.-C.}\ \bibnamefont {Gu}},\ and\ \bibinfo {author} {\bibfnamefont {X.-G.}\ \bibnamefont {Wen}},\ }\bibfield  {title} {\bibinfo {title} {Complete classification of one-dimensional gapped quantum phases in interacting spin systems},\ }\href {https://doi.org/10.1103/PhysRevB.84.235128} {\bibfield  {journal} {\bibinfo  {journal} {Phys. Rev. B}\ }\textbf {\bibinfo {volume} {84}},\ \bibinfo {pages} {235128} (\bibinfo {year} {2011}{\natexlab{b}})}\BibitemShut {NoStop}%
\bibitem [{\citenamefont {Thorngren}\ and\ \citenamefont {Else}(2018)}]{ElsePhysRevX.8.011040}%
  \BibitemOpen
  \bibfield  {author} {\bibinfo {author} {\bibfnamefont {R.}~\bibnamefont {Thorngren}}\ and\ \bibinfo {author} {\bibfnamefont {D.~V.}\ \bibnamefont {Else}},\ }\bibfield  {title} {\bibinfo {title} {Gauging spatial symmetries and the classification of topological crystalline phases},\ }\href {https://doi.org/10.1103/PhysRevX.8.011040} {\bibfield  {journal} {\bibinfo  {journal} {Physical Review X}\ }\textbf {\bibinfo {volume} {8}},\ \bibinfo {pages} {011040} (\bibinfo {year} {2018})}\BibitemShut {NoStop}%
\end{thebibliography}%

\begin{acknowledgments}
We acknowledge significant help and enlightening discussions with Cheng Chen and Songtao Huang,
and helpful contributions by Yuanjiang Tang, Chao Liang, Xiangliang Li, Xiaoling Wu, Boyang Wang, and Yifan Wang in the early stages of building up the experimental platform.  
We acknowledge ChatGPT for language polishing.
DMRG calculations were performed using the TeNPy tensor network library~\cite{tenpy}. We finally acknowledge the support by Center of High Performance Computing, Tsinghua University.
This work is supported by the Innovation Program for Quantum Science and Technology (2021ZD0302100). LY is also supported by NSFC (Grants
No. 12361131576 and No. 92265205). MKT is supported by NSFC
(Grants. No.12234012 and W2431002).
YFM, KL, and YX are supported by NSFC (Grants. No.12474265 and 11974201) and 
Innovation Program for Quantum Science and Technology (Grant No. 2021ZD0301604).
\end{acknowledgments}

\begin{widetext}
\setcounter{equation}{0} \setcounter{figure}{0} \setcounter{table}{0} %
\renewcommand{\theequation}{S\arabic{equation}} \renewcommand{\thefigure}{S%
\arabic{figure}} \renewcommand{\bibnumfmt}[1]{[S#1]} 
In the supplementary information, we provide the experimental method in Section S1,
discuss experimental imperfections in Section S2, 
and establish the concept of the average inversion SPT phase 
by proving that the topological invariant is quantized by an
average inversion symmetry and analyze the effects of 
van der Waals interactions in Section S3. 

\section{S1. Experimental method}

\subsection{S1.1. Experimental platform}

The realization of the model in Eq.~(1) in the main text is based on a $^{87}$Rb Rydberg atom array. We encode Rydberg state $\left| s \right\rangle$ = $\ket{55S_{1/2}}$ as $\ket{\uparrow}$ and $\left| p \right\rangle$ = $\ket{55P_{1/2}}$ as $\ket{\downarrow}$. The two pseudo-spin states are coupled by a resonant microwave at 22.1 GHz. A 30 Gauss magnetic field, aligned parallel to the plane formed by the two sub-chains, is applied to split different Zeeman states.

\subsection{S1.2. Defect-free atom arrays with arbitrary configurations}

In order to explore the influence of structural disorder on topological properties, we develop a rearrangement algorithm to realize arbitrary array configurations. The static tweezer array is created using an 830 nm laser modulated by a spatial light modulator (SLM). Additionally, we employ an 810 nm laser, controlled by an acousto-optic deflector (AOD) to generate a movable tweezer. 

Our rearrangement strategy is based on the Hungarian algorithm, using a graph-based approach to model the entire problem. First, we prepare a tweezer array consisting of both target sites and reservoir sites. Given that the loading rate is approximately $50\% $, the numbers of these two types of sites are equal. The tweezers track coordinates for the nodes of a graph, while edges connect different nodes, representing the possible paths to transport atoms in our strategy. To prevent potential collisions during rearrangement, paths which are too close to other sites are not employed for rearrangement, with the threshold length set at 3.5 $\mathrm{\mu m}$. 

Once a graph is constructed, the shortest distances or paths between all pairs of nodes are computed using the Dijkstra algorithm. The lengths are stored and used as the cost for the Hungarian algorithm. The AOD is controlled by an AWG, which requires a large amount of data, therefore waveforms for transport during rearrangement along all edges are computed and stored in advance. Based on these waveforms and the associated costs, in each rearrange cycle, we determine the shortest length collection using the Hungarian algorithm and replay the stored waveforms in real time.

\subsection{S1.3. Single site addressing}

The initial product states are prepared with the help of an addressing laser, split from the 1013 nm laser beam and detuned from the transition between the $\ket{\uparrow}$ and the intermediate excited state $\left| e \right\rangle$ = $\ket{6P_{3/2}}$. The addressing laser is locked to the stable 1013 nm laser employed in two-photon excitation to Rydberg states via their beat frequency, allowing for its frequency to be tuned and is set 350 MHz above the resonance.

To achieve two-dimensional single-site addressing, the addressing laser is modulated by a second SLM to generate multiple addressing beams. Each beam is focused to a radius of approximately 1.3 $\mathrm{\mu m}$, with a typical power of 50 mW. We measure the AC Stark shift for each addressing beam by microwave spectroscopy on the $\left| s \right\rangle$ and $\left| p \right\rangle$ transition. The average AC Stark shift across the seven addressed atoms is approximately $\delta_0 \approx 2\pi \times 20$ MHz. Daily alignment is needed to minimize atom loss and reduce AC Stark shift variations caused by mismatch between the addressing laser beams and the tweezers.

\subsection{S1.4. Experimental sequence}

After generating a defect-free atom array, we implement multiple cooling schemes, including polarization gradient cooling (PGC), electromagnetically induced transparency (EIT) cooling, and adiabatic ramp-down of the optical tweezers, to cool the atoms to a temperature of about 3 $\mathrm{\mu}$K. To minimize heating effects caused by optical pumping, we first optically pump the atoms into the $\ket{g}=\ket{5S_{1/2},F=2,m_F=-2}$ state before reducing the tweezer intensity.

Next, the magnetic field is adiabatically increased to approximately 30 Gauss. The optical tweezers are then switched off, and the atoms are excited to the Rydberg $\left| s \right\rangle$ state via two-photon stimulated Raman adiabatic passage (STIRAP) using 420 nm and 1013 nm lasers
respectively coupled resonantly to the transitions
$g \rightarrow e$ and $e \rightarrow s$.
A microwave field at 22.1 GHz is subsequently applied to drive $\left| s \right\rangle$ and $\left| p \right\rangle$ states. Owing to the extremely large polarizability of Rydberg atoms, even a weak microwave field can induce a strong coupling. In our setup, a $-20$ dBm microwave field is already sufficient to achieve a Rabi frequency of 2.5 MHz.

At the end of the above sequence, we measure magnetization $\sigma_i^z$ at each site. Since $[\hat{H},\sigma_i^z] \neq 0$, it is necessary to pause the interaction during readout. To achieve this, we freeze the system using a 10.38 GHz microwave pulse which excites atoms in $\left| p \right\rangle$ to the $n=53$ hydrogenic manifold, in which atoms experience negligible interactions with those remaining in $\left| s \right\rangle$. The pulse duration is 100 ns, which is sufficiently short to minimize unwanted interactions. Subsequently, a de-excitation pulse is applied to transfer atoms in $\left| s \right\rangle$ back to the ground state using a 1013 nm laser resonantly coupling $\left| s \right\rangle$ to the short-lived intermediate $\left| e \right\rangle$ state from which atoms decay back to the ground state.

\subsection{S1.5. Adiabatic preparation of the many-body ground state}
We experimentally prepare the many-body ground states with particle numbers near half-filling. 
Atoms in an initial product state $\left| \psi_0 \right\rangle$, 
e.g., $|1010... 10\rangle$, $|0010... 10\rangle$, or $|1010... 11\rangle$, 
respectively in the subspaces with $N$, $N-1$, and $N+1$ particles, are arrived  in steps: 
First, all atoms are transferred from the ground $g$ state to the $s$ state via STIRAP, followed by a second adiabatic protocol to the $p$ state from a microwave pulse sweeping across the resonance; focused pinning laser beams are subsequently applied to introduce a spatially 
dependent AC Stark shift of $\Delta E/h \approx 20$ MHz (Fig. 3a in the main text) to the $s$ state; finally, a second microwave pulse is 
swept across the resonance to produce the 
product state in which atoms on sites with pinning laser beams remain in the $p$ state while getting de-excited to the $s$ state elsewhere (Fig. 3a in the main text). 

The product state $\left| \psi_0 \right\rangle$ evolves into the many-body ground state following the  
adiabatic ramping down of the pinning laser beams (Fig. 3a in the main text), in a process described by the Hamiltonian
\begin{equation}
\hat{H}_{\mathrm{tot}}=\hat{H}-
\Delta E(t)\sum_{i \in S }\hat{n}_{i},
\end{equation}
where $S$ denotes the set of sites with applied pinning lasers. 
Since this Hamiltonian respects $U(1)$ symmetry, the total particle number is conserved, and the evolving state remains within 
the subspace of fixed total particle number given by the initial state $\left| \psi_0 \right\rangle$. 
At $t=0$, the level shift is significantly larger than  
the energy scale of hopping terms. As a result, 
$\left| \psi_0 \right\rangle$ is a good approximation
to the ground state of $\hat{H}_{\mathrm{tot}}$.
After a $4\text{ } \mathrm{\mu s}$ pulse of exponential ramping down according to $\Delta E(t)=\Delta E e^{-t/\tau}$ for $\tau=0.6\text{ }\mathrm{\mu s}$, we arrive at the ground state experimentally. 


\begin{figure}
	\centering
	\includegraphics[width=\linewidth]{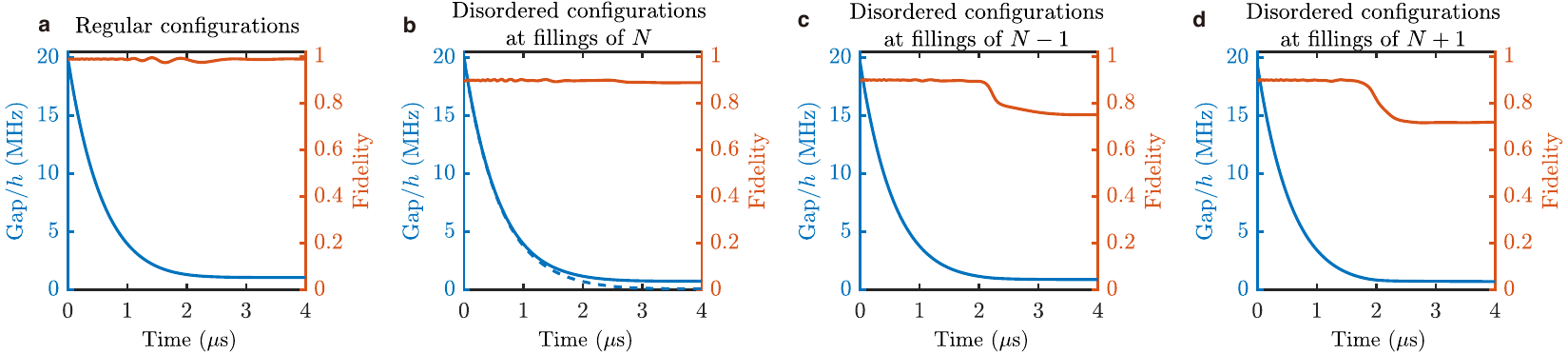}
	\caption{Numerically calculated energy gap and state fidelity between the evolving state and the true ground state for (\textbf{a}) the regular configuration 
		and (\textbf{b-d}) disordered configurations at varying filling numbers and averaged over $15$ random configurations. 
		In \textbf{b}, the solid line represents the energy gap between the third lowest energy state and the lowest energy state,
		while the dashed line corresponds to the gap between the second lowest energy state and the lowest energy state.
		}\label{fig:gap_fidelity}
\end{figure}

The effectiveness of the adiabatic evolution depends on the low energy spectral gap.
Figure~\ref{fig:gap_fidelity} displays the numerically calculated energy gap during the time evolution. 
We see that for the regular configuration, the gap quickly decreases, reaching a minimum value
of $\Delta_{\mathrm{g}}/h =1.06$ MHz. 
For disordered configurations, the minimum energy gaps averaged over random configurations 
are $0.75$ MHz, $0.89$ MHz, and $0.72$ MHz 
for $N$, $N-1$, and $N+1$ particles, respectively, which are slightly smaller than the
minimum energy gap in the regular case. 
The energy gaps are on the order of the energy scale of the hopping term in the Hamiltonian $ \hat{H} $. 
At half-filling, it is defined as  
the gap between the third lowest energy state with respect to the ground state, 
as the ground state is nearly two-fold degenerate.
During the time evolution, although the gap declines, the ramping speed for the pinning laser beams  
decreases correspondingly, thereby ensuring a successful adiabatic evolution. 

Figure~\ref{fig:gap_fidelity} also plots the fidelity between the numerically calculated evolving state and the corresponding true ground state 
during the time evolution without accounting for errors,
where the evolving state is obtained 
by simulating the time evolution of the Hamiltonian using the Krylov method.
We see that
at the end of the adiabatic evolution, the fidelity reaches $0.99$ for the regular configuration.
For disordered cases, their average fidelity values are $0.89$, $0.75$, and $0.72$ for the 
fillings of $N$, $N-1$, and $N+1$ particles, respectively.
To ensure that imperfections of the experiments are well understood, we also perform simulations 
by considering all SPAM errors. 
The results of numerical simulations are found to be in good agreement with the experimental data, 
as demonstrated in Fig.~3 in the main text.

\section{S2. Experimental imperfections}
We compare the experimental results with numerical simulations considering different imperfections in the main text. Here we discuss the main source of errors in our system, including state preparation, detection errors (SPAM), and those due to thermal motion of atoms.

\subsection{S2.1. SPAM errors}
As discussed earlier, atoms are transferred from $\left| g \right\rangle$ to $\left| s \right\rangle$ using STIRAP. However, a small but finite preparation error, measured to be $\eta_{\mathrm{STIRAP}} \approx 5\% - 8\%$, results as atoms may fail to make the transition to the Rydberg state, effectively creating lattice defects. This effect is incorporated into numerical simulations by averaging over different lattice realizations.

For read out, atoms in the $\left| s \right\rangle$ state are de-excited to the ground state and subsequently recaptured in the optical tweezers. Meanwhile, atoms in the $\left| p \right\rangle$ state are repelled by the 830 nm tweezers. As a result, fluorescence detection can be used to determine which state an atom collapses into. However, atoms would be recaptured may get lost due to thermal motion over a finite evolution time, with an error probability of $\eta_{\epsilon} \approx 5 \%$. Conversely, atoms that would be repelled by the tweezers may instead be recaptured due to the finite lifetime of Rydberg states, with a probability of $\eta_{\epsilon'} \approx 5 \%$. These effects are accounted for in numerical simulations through Monte Carlo sampling.

\subsection{S2.2. Errors in preparing the product state}

In addition to the SPAM errors, during the preparation of the many-body ground states, errors also occur in the first step of preparing for the initial product state helped by the addressing laser. As described above, the AC Stark shift is induced by a 1013 nm laser tuned 350 MHz above the resonance between $\left| e \right\rangle$ and $\left| s \right\rangle$. However, the lifetime of the Rydberg $\left| s \right\rangle$ state, $\tau_s$, is reduced due to coupling, albeit off resonantly to the short-lived $\left| e \right\rangle$ state

\begin{equation} \label{eq:tau}
	\tau_s \approx (\frac{\Delta_{\text{addr}}}{\Omega_{\text{addr}}})^2 \tau _{6P},
\end{equation}
while the Rydberg $\left| p \right\rangle$ state remains unaffected as it is not coupled to the short lived  $\left| e \right\rangle$ state by this addressing laser. To mitigate this detrimental effect of the addressing laser, we choose first to transfer all atoms into the Rydberg $\left| p \right\rangle$ state;
during this process, the presence of the dipolar and van der Waals interactions
gives rise to an error $\eta_{\mathrm{MW}} \approx 0.02$. The addressing lasers are then applied to sub-chain $\alpha$, inducing an AC Stark shift of approximately 20 MHz for all $\left| s\right\rangle$ state atoms in sub-chain $\alpha$. Finally we sweep the MW frequency back across the single-atom resonance to transfer atoms in sub-chain $\beta$ to $\left| s \right\rangle$ states, with an associated error $\eta_\beta \approx 0.05$, while atoms in sub-chain $\alpha$ remain in $\left| p \right\rangle$ state with an error $\eta_\alpha \approx 0.04$.

\subsection{S2.3. Thermal motion of atoms}

Since the atoms remain at a finite temperature $T$, their respective velocities ${\bm v}$ vary in each experimental realization, described by a Gaussian distribution with standard deviation $\Delta v= \sqrt{\frac{k_B T}{m}}$ where $m$ is the mass of the atom. Consequently, the interaction strength fluctuates due to its dependence on both the interatomic distance and the relative angle between atoms and the magnetic field.
 
 In a regular lattice, atoms are arranged into a configuration where intercell interactions are highly sensitive to the relative angle between the two-atom separation vector and the applied magnetic field. As a result, these interactions exhibit greater fluctuations. In a structurally disordered lattice, thermal motion introduces an additional dependence on disorder strength. We attribute the different qualities of the correlation functions between regular and disordered lattices to this effect.
 
 To account for thermal motion in the numerical simulations, we assume stochastic initial conditions. At the beginning of the sequence, atoms are assigned random initial positions following the above mentioned Gaussian distribution. Once the optical tweezers are switched off, atoms are given random velocities and allowed to move freely. The final results are obtained by averaging over hundreds of realizations in numerical simulations.

\section{S3. Average SPT phase induced by structural disorder}
In this section, we will establish the concept of the average inversion SPT phase 
by proving that the topological invariant is quantized by an
average inversion symmetry. In addition, we will analyze the effects of 
van der Waals interactions. In the following, we will consider separately single-particle 
and many-particle cases.

\subsection{S3.1. Single-particle case}
In this subsection, we will focus on the Hilbert space with only a single excitation (one atom in the Rydberg 
$p$ state), which is spanned by 
$\gamma=\{ |1 \rangle=\hat{b}_1^{\dagger}|0\rangle, |2 \rangle=\hat{b}_2^{\dagger}|0\rangle,\dots, 
|2N \rangle =\hat{b}_{2N}^{\dagger}|0\rangle\}$. Relative to this basis, the 
single-particle Hamiltonian reads
\begin{equation} \label{eq:SHam}
	[H^{\textrm{S}}]_{ij}=J_{ij}(1-\delta_{ij})-V_i^{\text{vdW}} \delta_{ij},
\end{equation}
which allows us to rewrite conveniently 
$H^{\textrm{S}}=\sum_{i,j} [H^{\textrm{S}}]_{ij} | i\rangle \langle j|$. 
In the following, we will prove that this Hamiltonian's topology can be 
characterized by the polarization (a topological invariant) protected by the average 
inversion symmetry.

\subsubsection{S3.1.1. The polarization as a topological invariant}
Without van der Waals interactions ($V_i^{\text{vdW}}=0$), the single-particle Hamiltonian in 
Eq.~(\ref{eq:SHam}) respects the 
sublattice symmetry, i.e., $\Pi H^{\textrm{S}}\Pi^{-1} =-H^{\textrm{S}}$, 
with $\Pi=\text{diag}\{ (-1)^j\}_{j=1}^{2N}$ the sublattice symmetry operator. Although the $\mathbb{Z}$ classification applies for the topology protected by sublattice symmetry~\cite{ChiuRevModPhys}, 
our model exhibits only two phases:
a topologically trivial phase without edge modes and a nontrivial phase with one edge state at
each edge. As such, we can use the polarization, which is equivalent to the Berry phase 
in a regular lattice, 
defined as~\cite{resta1998quantum}
\begin{equation}
	P_\mathrm{S} = \left[\frac{1}{2\pi} \mathrm{Im} \ln \det (U^\dagger D U)
	-  \frac{1}{2N {d}} \sum_{i =1}^{2N} x_i \right]  \textrm{ mod } 1,
	\label{polar_f}
\end{equation}
to characterize the topology of the Hamiltonian. Here $U=(\ket{u_1},\ket{u_2},\dots, \ket{u_N})$, with $\ket{u_j}$ ($j=1,2,\dots N$) denoting
the lowest $N$ 
eigenstates of the single-particle Hamiltonian $H^\mathrm{S}$ under 
periodic boundary conditions, and $D = \mathrm{diag} \{e^{2\pi i x_j/(N{d})} \}_{j=1}^{2N}$, 
with ${x}_{j}$ the spatial coordinate of the $\lfloor (j+1)/2\rfloor$th unit cell.
It is well known that with the sublattice symmetry, $P_\mathrm{S}$ can only take the 
values of either $0$ or $0.5$~\cite{PhysRevLett.127.263004}, which allows $P_\mathrm{S}$ to function as a 
topological invariant:
For a topologically trivial phase, $P_\mathrm{S}=0$, while for a nontrivial phase, $P_\mathrm{S}=0.5$.

Although the van der Waals term breaks the sublattice symmetry, our model preserves
the inversion symmetry in a regular lattice and an average inversion symmetry in disordered lattices. The average symmetry enables us to generalize the polarization $P_\mathrm{S}$, ensuring its quantization remains valid.

\subsubsection{S3.1.2. Inversion symmetry}
Previous studies have shown that the inversion symmetry can protect the topology
in 1D~\cite{TaylorPhysRevB.83.245132,ChiuPhysRevB2013}. 
We now present a proof showing that, under this symmetry, the polarization $P_\mathrm{S}$ can 
only take the values of either $0$ or $0.5$.
Specifically, consider  
the Hamiltonian $H^{\textrm{S}}$ that respects the inversion symmetry, i.e., 
$[H^{\textrm{S}},U_R]=0$, where $U_R$ is the inversion operator. 
When acting on $| i \rangle$, the operator yields
$U_R | i \rangle = | \mathcal{R} (i)\rangle $,
where $\mathcal{R} (i)$ denotes the site index of the inversion partner of site $i$.
This symmetry ensures that $\ket{u_i}$ in $U$ is also an eigenstate of $U_R$ corresponding to an eigenvalue
$\nu_i$ that takes a value of $1$ or $-1$. 
With this result, we can derive
\begin{equation} \label{eq:uuduu}
	\det{(U^\dagger D U)}=\det{[U^\dagger U_R^\dagger (U_R D U_R^\dagger) U_R U ]}
	=\det{[D_R^\dagger U^\dagger (U_R D U_R^\dagger) U D_R]}
	=\det{[U^\dagger (U_R D U_R^\dagger) U]},
\end{equation}
where we have used the fact $U_R U=U D_R$ in the derivation, with 
$D_R$ an $N\times N$ diagonal matrix of the form
$D_R= \mathrm{diag} \{ \nu_j \}_{j=1}^{N}$.
Let $x_c$ be the coordinate for the inversion center, thus $1/(2N) \sum_{i=1}^{2N} x_i =x_c$.
We can also derive that
\begin{eqnarray} \label{eq:udu}
	U_R D U_R^\dagger &=&\sum_{j=1}^{2N} e^{2\pi i x_j/(N {d})} U_R|j \rangle \langle j | U_R^\dagger 
	=\sum_{j=1}^{2N} e^{2\pi i x_j/(N {d})} |\mathcal{R}(j) \rangle \langle \mathcal{R}(j) | 
	=\sum_{j^\prime=1}^{2N} e^{2\pi i x_{\mathcal{R}(j^\prime)}/(N {d})} | j^\prime \rangle \langle j^\prime |  \nonumber \\
	&=&\sum_{j^\prime=1}^{2N} e^{2\pi i (2x_c - x_{j^\prime}) /(N {d})} | j^\prime \rangle \langle j^\prime | 
	=e^{4\pi i x_c /(N {d})} D^\dagger .
\end{eqnarray}
Substituting into Eq.~(\ref{eq:uuduu}) and subsequently into
Eq.~(\ref{polar_f}), we then obtain
\begin{equation}
	P_\mathrm{S} = \left[\frac{1}{2\pi} \mathrm{Im} \ln \det (U^\dagger D^\dagger U)
	+ x_c/{d} \right] \textrm{ mod } 1.
\end{equation}
Let $\det (U^\dagger D U)=r e^{i 2\pi \theta}$ ($r>0$). Then we obtain
$P_\mathrm{S}=(\theta - x_c/{d})  \textrm{ mod } 1=(-\theta + x_c/{d})  \textrm{ mod } 1$, yielding
$\theta=x_c/{d}+n/2$ for integer $n$. Therefore, 
$P_\mathrm{S}=n/2  \textrm{ mod } 1$ is quantized to $0$ or $0.5$ by the inversion symmetry.

\subsubsection{S3.1.3. Average inversion symmetry}
In the presence of structural disorder, it is clear to see that for a typical sample $C=\{x_j : j=1,2,\dots,2N\}$, 
the Hamiltonian no longer 
preserves the inversion symmetry, i.e.,
$U_R H^{\textrm{S}}(C) U_R^{-1} \neq H^{\textrm{S}}(C)$. As a result,  
$P_\mathrm{S}$, as defined in Eq.~(\ref{polar_f}), is not quantized
at $0$ or $0.5$ any more and can take other values. However, 
if we consider the ensemble of all possible lattice configurations, an average inversion symmetry remains.
Specifically, the Hamiltonian ensemble is defined as  
$\mathcal{E}_H \equiv \left\{ H^{\textrm{S}}(C) : C\in \mathcal{E}_C \right\}$, 
where $\mathcal{E}_C$ is an ensemble containing all possible random lattice configurations.
In our case, any random lattice configuration $C$ and its inversion 
partner $\mathcal{R}C$ emerge in this ensemble with equal probability.
As a result, the Hamiltonian $H(C)$ and its inversion conjugate partner
$U_R H^{\textrm{S}}(C) U_R^{-1}=H^{\textrm{S}}(\mathcal{R}C)$
appear with the same probability. This Hamiltonian ensemble thus preserves an average 
inversion symmetry~\cite{fu2012topology,fulga2014statistical,ma2023average}. 
We thus can identify the topological property of the 
Hamiltonian $H^{\textrm{S}}(C)$ and 
$H^{\textrm{S}}(\mathcal{R}C)$ as a whole, defining the polarization as
\begin{equation} \label{eq:polar-C}
	P_\mathrm{S}(C) = \left \{ \frac{1}{2\pi} \mathrm{Im} 
	\ln \left[ \sum_{S_0 \in \left\{C, \mathcal{R} C \right\} }  e^{-i2\pi  \frac{1}{2N {d}} \sum_{i =1}^{2N} x_i(S_0) } 
	\det \left( U(S_0)^\dagger D(S_0) U(S_0) \right) \right] 	 
	\right\}  \textrm{ mod } 1,
\end{equation}
where $D(S_0)$ and $U(S_0)$ are the corresponding matrices for the lattice configuration $S_0$ with coordinates 
of $x_i(S_0)$. Given that $H^{\textrm{S}}(C) \ket{u_i}=E_i \ket{u_i}$, we have 
$H^{\textrm{S}}(\mathcal{R}C) U_R \ket{u_i} =E_i U_R \ket{u_i}$, indicating that
$U_R \ket{u_i}$ is the corresponding eigenstate of $H^{\textrm{S}}(\mathcal{R}C)$, or $U(\mathcal{R}C)=U_R U(C)$. Thus we find
\begin{eqnarray} \label{eq:uDRCu}
	U_R D(\mathcal{R}C) U_R^\dagger &=& \sum_{j=1}^{2N} e^{2\pi i x_j(\mathcal{R}C)/(N {d})} 
	U_R|j \rangle \langle j | U_R^\dagger \nonumber \\
	&=&\sum_{j=1}^{2N} e^{2\pi i [2x_c-x_{2N+1-j}(C)]/(N {d})} 
	|2N+1-j \rangle \langle 2N+1-j | \nonumber \\
	&=& e^{4\pi x_c i/(N {d}) } D(C)^\dagger,
\end{eqnarray}
where $x_c=[x_j(C)+x_{2N+1-j}(\mathcal{R}C) ]/2$ is the coordinate of the inversion center.
We substitute Eq.~(\ref{eq:uDRCu}) into Eq.~(\ref{eq:polar-C}), and arrive at
\begin{equation}
	P_\mathrm{S}(C) = \left \{ \frac{1}{2\pi} \mathrm{Im} 
	\ln \left[ e^{-i2\pi  \frac{1}{2N {d}} \sum_{i =1}^{2N} x_i(C) } 
	\det \left( U(C)^\dagger D(C) U(C) \right) +\textrm{c.c.}\right] 	 
	\right\}  \textrm{ mod } 1,
\end{equation}
indicating that $P_\mathrm{S}(C)$ can only take the values of either $0$ or $0.5$.

\begin{figure} 
	\centering
	\includegraphics[width=1\linewidth]{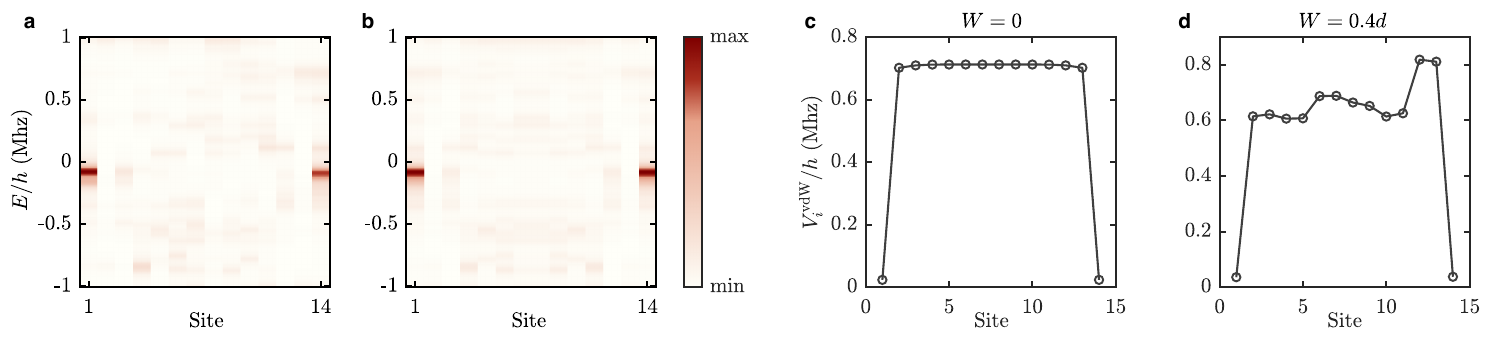} 
	
	\caption{Effects of van der Waals interactions.
		\textbf{a,b,} Numerically calculated local density of states (DOS) for the single-particle Hamiltonian (\ref{eq:SHam})
		in disordered lattice ensemble with $W=0.4 {d}$.
        \textbf{c,d,} Numerically calculated onsite energy $V_i^{\text{vdW} }$ contributed by van der Waals interactions
		under open boundary conditions in a regular lattice and disordered lattices 
		with $W=0.4 {d}$, respectively. 
		In \textbf{a} and \textbf{d}, the results are averaged over $15$ random configurations
		as used in the experiments. 
		In \textbf{b}, the local DOS averaged over the experimentally 
		used $15$ configurations plus their inversion partners. We see that 
        the local DOS is symmetric with respect to the inversion center. 
		$C_6=-48880$ MHz$\cdot \mathrm{\mu m}^6$ is used.} 
	\label{figvdW} 
\end{figure}

\subsubsection{S3.1.4. Statistically degenerate edge states}
Although random displacements break the inversion symmetry such that two edge states have different eigenenergies for each sample, we here will demonstrate that the two edge states remain statistically degenerate over the entire disorder ensemble.

Let $\ket{u_{\mathrm{L}}(C)}$ and $\ket{u_\mathrm{R}(C)}$ be the two edge
states of the single-particle Hamiltonian $H^{\mathrm{S}}(C)$ for a random configuration
$C$ under open boundary conditions. They reside mainly at the left and right edges, respectively,
and their corresponding eigenenergies are $E_\mathrm{L}(C)$ and $E_\mathrm{R}(C)$.
Due to the breaking of the inversion symmetry, $E_\mathrm{L}(C)\neq E_\mathrm{R}(C)$ for each realized configuration. 
For the inversion conjugate Hamiltonian $H^{\mathrm{S}}(\mathcal{R}C)$,
$\ket{u_\mathrm{R}(\mathcal{R}C)}=U_R \ket{u_\mathrm{L}(C)}$, and $\ket{u_\mathrm{L}(\mathcal{R}C)}=U_R\ket{u_\mathrm{R}(C)}$
are its eigenstates corresponding to energy $E_\mathrm{L}(C)$ and $E_\mathrm{R}(C)$, respectively.
They are mainly localized at the right and left edges, respectively. Consider the 
local DOS, $\rho(i=1,E)(C)$ and $\rho(i=2N,E)(C)$ [$\rho(i=1,E)(\mathcal{R}C)$ and $\rho(i=2N,E)(\mathcal{R}C)$], 
at two edges at the energy $E$ for the configuration $C$ (the inversion partner $\mathcal{R}C$), for simplicity, we only include contributions from the edge state: At the left edge, 
\begin{eqnarray}
\rho(i=1,E)(C)&=& |[\ket{u_\mathrm{L}(C)}]_1|^2\delta(E-E_\mathrm{L}(C)), \\
\rho(i=1,E)(\mathcal{R}C) &=& |[\ket{u_\mathrm{L}(\mathcal{R}C)}]_1|^2\delta(E-E_\mathrm{L}(\mathcal{R}C))
= |[\ket{u_\mathrm{R}(C)}]_{2N}|^2\delta(E-E_\mathrm{R}(C)),
\end{eqnarray}
and similarly for the local DOS at the right edge.
For the two configurations, the average of the two local DOS at the left edge is 
$[\rho(i=1,E)(C)+\rho(i=1,E)(\mathcal{R}C)]/2$, which is clearly equal to 
the one at the right edge, $[\rho(i=2N,E)(C)+\rho(i=2N,E)(\mathcal{R}C)]/2$.
Thus, it can be concluded that the two edge states are statistically degenerate.
For better clarity, we plot the numerically computed configuration averaged local DOS 
in Fig.~\ref{figvdW}b, illustrating that it preserves the inversion symmetry, and the
two red peaks
at the left and right edges have the same energy. 

\subsubsection{S3.1.5. Effects of van der Waals interactions}
In the single-particle case, van der Waals interactions contribute an onsite energy of $-V_i^{\text{vdW} }$ 
at site $i$ ($i=1,2,\dots,2N$).
In a regular lattice with periodic boundary conditions, this onsite potential is independent of the site index.
Consequently, it does not alter the eigenstates of the Hamiltonian, leaving the topology unaffected.
However, under open boundary conditions as in our experiment, the potentials at edges differ from in the bulk due 
to the absence of one nearest neighbor for the edge sites, as shown in Fig.~\ref{figvdW}c.
Despite this difference, the total Hamiltonian still respects the inversion symmetry, ensuring that 
two edge states, if they exist, are degenerate.
In the figure, we see that in the bulk $V_i^{\text{vdW} }/h\approx 0.7$ MHz, which is 
comparable to the intracell hopping energy of $0.77$ MHz in the regular case. 
However, at the edges, the potential is only $0.02$ MHz.
This means if there were edge states in the regular lattice, we would have observed a 
peak of occupancy at two edges near zero detuning by microwave spectroscopy. 
However, our experiments (see Fig. 2b in the main text) do not show such a peak, implicating the absence of edge modes. 

In the disordered case, for a typical configuration, $V_i^{\text{vdW} }$ depends on the 
site index even under periodic boundary conditions. However, we have shown in 
the previous subsection that 
the topology of the ensemble of Hamiltonians in disordered lattices can still be characterized 
by the generalized polarization $P_{\textrm{S}}(C)$, and edge states, if they exist, are statistically degenerate. 
Figure~\ref{figvdW}d displays the $V_i^{\text{vdW} }$ averaged over 
the $15$ random configurations used in our experiments. We see that  
similarly to the regular case, $V_i^{\text{vdW} }$ at edge sites is about $0.04$ MHz.
The local DOS in Figs.~\ref{figvdW}a and \ref{figvdW}b show that the peaks at the edges appear at $-0.07$ MHz,
whose absolute value is larger than the averaged onsite energies at the edges, possibly 
due to non-vanishing occupancy probabilities in the bulk sites for edge states.
The existence of edge states is confirmed by our experiments, illustrating a peak of occupancy at two edges 
near zero detuning (see Fig. 2b in the main text).

\subsection{S3.2. Many-particle case}
In this subsection, we will focus on the case at or near half-filling ($N$, $N-1$, or $N+1$ atoms in the Rydberg $p$ state).
In this case it is more convenient to write the Hamiltonian in Eq.~(1) in the main text as a spin model
\begin{equation} \label{eq:HamXXZ-Z}
	\hat{H} = H_{\mathrm{XXZ}}+H_{\mathrm{Z}},
\end{equation}
where 
\begin{eqnarray}
H_{\mathrm{XXZ}} &=&\sum_{i<j}^{2N} \left[ \frac{1}{2}J_{ij}(\sigma_i^{x}\sigma_j^{x}+
\sigma_i^{y}\sigma_j^{y})+\frac{1}{4}V_{ij}^{\text{vdW} } \sigma_i^z \sigma_j^z
\right ],  \\
H_{\mathrm{Z}}&=&\frac{1}{4}\sum_{i} V_i^{\text{vdW} } \sigma_i^z. 
\end{eqnarray}
The Hamiltonian $H_{\mathrm{XXZ}}$ respects 
 two spin-rotation symmetries along $x$ and $y$ respectively represented by 
$\hat{R}_x=\prod_{j=1}^{2N} e^{-i \pi \sigma_j^x/2}$ and 
$\hat{R}_y=\prod_{j=1}^{2N} e^{-i \pi \sigma_j^y/2}$, the time-reversal symmetry represented by 
$\hat{T}=\prod_{j=1}^{2N}\sigma_j^y \kappa$ with $\kappa$ the complex conjugate operator, 
and a symmetry represented by
$\hat{\mathcal{S}}=\prod_{j=1}^{2N} \sigma_j^x \kappa$. The model $\hat{H}$ 
preserves the total
spin along $z$, i.e., $[\hat{H},\sigma^z]=0$ with $\sigma^z=\sum_i \sigma_i^z$,
which corresponds to the $U(1)$ symmetry. 

\subsubsection{S3.2.1. Mapping the Hamiltonian to a fermion interacting model}
We first demonstrate that the hard-core boson model in Eq.~(1) in the main text or the 
spin model in Eq.~(\ref{eq:HamXXZ-Z}) can be mapped to 
a Hamiltonian of interacting fermions~\cite{de2019observation,PhysRevLett.127.263004} through the inverse Jordan-Wigner
transformation, i.e.,
\begin{eqnarray}
	\hat{b}_j^\dagger &=& \mathrm{exp}(-i \pi \sum_{k=1}^{j-1}\hat{n}_k^c) \cdot \hat{c}_j^\dagger,\\
	\hat{b}_j  &=& \mathrm{exp}(i \pi \sum_{k=1}^{j-1}\hat{n}_k^c) \cdot \hat{c}_j.
\end{eqnarray}
Here, $ \hat{c_j}^\dagger(\hat{c}_j) $ denotes a fermionic creation (annihilation) operator
at site $j$, and $ \hat{n}^c_j=\hat{c}_j^\dagger \hat{c}_j$ is the corresponding particle number 
operator. The mapped model reads    
\begin{equation}
    \hat{H}_\mathrm{f} = \sum_{i<j}^{2N} J_{ij}\left[\hat{c}_i^\dagger e^{i \pi \sum_{k=i}^{j-1} \hat{n}^c_k} \hat{c}_j + \mathrm{H.c.}\right] + 
    \sum_{i<j}^{2N} V_{ij}^\mathrm{vdW} (1-\hat{n}^c_i)(1-\hat{n}^c_j).
\end{equation}
In addition to the interactions resulting from the long-range hoppings, the second term clearly introduces interactions between the spinless fermions. Consequently, this system exhibits genuine strong interactions.

\subsubsection{S3.2.2. The $\mathbb{Z}_2$ topological invariant}
To characterize the topology of the spin model Eq.~(12), we define a  
$\mathbb{Z}_2$ topological invariant~\cite{nakamura2002order,tasaki2018topological}
\begin{equation} \label{eq:PM}
	P_\mathrm{M}=\left[ \frac{1}{2\pi} \mathrm{Im} \ln 
	 \bra{\Psi} \hat{\mathcal{P}}_\mathrm{M} \ket{\Psi} \right] 
	\text{ mod } 1,
\end{equation}
in analogy to the case of single particles, where $\ket{\Psi}$ is the ground state of the spin model in the symmetry sector with $\sigma^z=0$ 
under periodic boundary conditions and 
$\hat{\mathcal{P}}_\mathrm{M}= \prod_{j=1}^{2N} e^{- \frac{\pi i}{N{d}} x_j \sigma_{j}^z }$ 
is the twist operator with $x_j$ denoting the position of the $\lfloor (j+1)/2\rfloor$th unit cell. It has been proven
that any one of the above symmetries for $H_{\mathrm{XXZ}}$ ensures that $P_\mathrm{M}$ can only take the values of either $0$
or $0.5$~\cite{PhysRevLett.127.263004}, making it a well-defined topological invariant. For a topologically trivial phase,
$P_\mathrm{M}=0$, while for a nontrivial phase, $P_\mathrm{M}=0.5$. Analogous to the single-particle
case, van der Waals interactions introduce the $H_{\mathrm{Z}}$ term, which breaks all the above protecting symmetries.
In the following, we will prove that either an inversion symmetry or an average inversion symmetry
can protect the quantization of the topological invariant for the ground state in the symmetry 
sector with $\sigma^z=0$. 

\subsubsection{S3.2.3. Inversion symmetry}
For bosonic SPT phases, inversion symmetry plays the role of time-reversal symmetry
based on the crystalline equivalence principle~\cite{ChenPhysRevB.84.235128,ElsePhysRevX.8.011040}, 
leading to a $\mathbb{Z}_2\times \mathbb{Z}_2$ classification in 1D~\cite{ChenPhysRevB.87.155114},
provided the system also respects the $U(1)$ symmetry.
Here we will present a proof demonstrating that inversion symmetry can protect the 
quantization of $P_\mathrm{M}$ in the spin model.
Note that an alternative proof has been provided for a spin-1 model in Ref.~\cite{tasaki2018topological}.

Specifically, suppose that the spin model has an inversion symmetry, i.e., $\mathcal{U}_R \hat{H} \mathcal{U}_R^{-1}=\hat{H}$,
where $\mathcal{U}_R$ realizes the inversion operation or $\mathcal{U}_R \sigma_j^\nu \mathcal{U}_R^{-1}
=\sigma_{2N+1-j}^\nu$ ($\nu=x,y,z$). We note when the $H_{\mathrm{Z}}$ term is sufficiently large, 
$|\Psi \rangle$ as the ground state of this model in the symmetry sector with $\sigma^z=0$ is no longer assured to be the global ground state of the total Hamiltonian $\hat{H}$.
In the regular case as well as for most of the random configurations, we find that this state remains the ground state of the total Hamiltonian.
In fact, ground states are prepared in a fixed symmetry sector in our experiments and we have not encountered any serious trouble with them not being that of the total Hamiltonian.
In addition, we assume that $|\Psi \rangle$ is not degenerate so that $\mathcal{U}_R |\Psi \rangle = \lambda |\Psi \rangle$
with $\lambda=1$ or $-1$.
We now derive that
\begin{eqnarray}
	\bra{\Psi} \hat{\mathcal{P}}_\mathrm{M} \ket{\Psi}
	&=&\bra{\Psi} \mathcal{U}_R \hat{\mathcal{P}}_\mathrm{M} \mathcal{U}_R^{-1} \ket{\Psi} \nonumber \\
	&=&\bra{\Psi}  \prod_{j=1}^{2N} e^{- \frac{\pi i}{N{d}} x_j \sigma_{2N+1-j}^z } \ket{\Psi}  \nonumber \\
	&=&\bra{\Psi}  \prod_{j=1}^{2N} e^{- \frac{\pi i}{N{d}} (2x_c - x_{2N+1-j})  \sigma_{2N+1-j}^z } \ket{\Psi} \nonumber \\
	&=&\bra{\Psi} \hat{\mathcal{P}}_\mathrm{M}^\dagger \ket{\Psi}=(\bra{\Psi} \hat{\mathcal{P}}_\mathrm{M} \ket{\Psi})^*,
\end{eqnarray}
where we have used the fact that $\sigma^z \ket{\Psi} =0$. This indicates that 
$\bra{\Psi} \hat{\mathcal{P}}_\mathrm{M} \ket{\Psi}$ is real so that $P_\mathrm{M}$ 
has to be quantized to $0$ or $0.5$. 

\subsubsection{S3.2.4. Average inversion symmetry}
Similar to the single-particle case, in the presence of structural disorder, for a typical configuration $C$,
the inversion symmetry is broken, i.e.,  
$\mathcal{U}_R \hat{H}(C) \mathcal{U}_R^{-1} \neq \hat{H}(C)$.
As a
result, in the configuration, the topological invariant $P_\mathrm{M}$ is no longer well defined as it is not quantized. 
However,
the Hamiltonian ensemble   
$\mathcal{E}_{\hat{H}} \equiv \left\{ \hat{H}(C) : C\in \mathcal{E}_C \right\}$
possesses an average inversion symmetry. In other words,
the Hamiltonian $\hat{H}(C)$ and its inversion conjugate partner
$\mathcal{U}_R \hat{H}(C) \mathcal{U}_R^{-1}=\hat{H}(\mathcal{R}C)$
appear with the same probability. With this average symmetry, we define a $\mathbb{Z}_2$ topological invariant 
in Eq.~(2) as in the main text. We now prove that it is enforced to be quantized
by the average symmetry for the ground state $\ket{\Psi(C)}$ 
of the configuration $C$ and the ground state $\ket{\Psi(\mathcal{R}C)}$ 
of the configuration $\mathcal{R} C$ in the symmetry sector with $\sigma^z=0$.
Based on the property that $\sigma^z \ket{\Psi(C)} =0$, we show that
\begin{eqnarray}
	\bra{\Psi(\mathcal{R}C)} \hat{\mathcal{P}}_\mathrm{M}(\mathcal{R}C) \ket{\Psi(\mathcal{R}C)}
	&=&
	\bra{\Psi(C)} \mathcal{U}_R \hat{\mathcal{P}}_\mathrm{M}(\mathcal{R}C) \mathcal{U}_R^{-1} \ket{\Psi(C)} \nonumber \\
	&=&\bra{\Psi(C)} \prod_{j=1}^{2N} e^{- \frac{\pi i}{N{d}} x_j(\mathcal{R}C) \sigma_{2N+1-j}^z } \ket{\Psi(C)} \nonumber \\
	&=& \bra{\Psi(C)} \prod_{j=1}^{2N} e^{- \frac{\pi i}{N{d}} [2x_c - x_{2N+1-j}(C) ]  \sigma_{2N+1-j}^z } \ket{\Psi(C)} \nonumber \\
	&=&\bra{\Psi(C)} \hat{\mathcal{P}}_\mathrm{M}(C)^\dagger \ket{\Psi(C)}=(\bra{\Psi(C)} \hat{\mathcal{P}}_\mathrm{M}(C) \ket{\Psi(C)})^*.
\end{eqnarray} 
Thus, $\sum_{S \in \{C, \mathcal{R}C\}}  \bra{\Psi_S} \hat{\mathcal{P}}_\mathrm{M} (S) \ket{\Psi_S}$ is real, and
$P_{\textrm{M}}(C)$ is quantized to $0$ or $0.5$.

In the above proof, we have made use of the fact that $\sigma^z$ is conserved corresponding to 
the $U(1)$ symmetry. 
As such, 
the symmetry class of the system is given by $U(1) \times (\mathbb{Z}_2^T)^{\mathrm{avg}}$, 
where the spatial inversion symmetry is treated as an antiunitary $\mathbb{Z}_2$ internal symmetry through the crystalline equivalence principle~\cite{ChenPhysRevB.84.235128,ElsePhysRevX.8.011040}.
Based on previous classification results~\cite{ma2023average}, the average SPT phase protected by $U(1) \times (\mathbb{Z}_2^T)^{\mathrm{avg}}$ in 1D is given by 
\begin{equation}
\sum_{p=1}^{2} H^{2-p} (\mathbb{Z}_2^T, H^p(U(1), U(1))) = 
H^2 (U(1) \times \mathbb{Z}_2^T, U(1))/ H^2 (\mathbb{Z}_2^T, U(1)) = \mathbb{Z}_2, 
\end{equation}
which is consistent with the $\mathbb{Z}_2$ topological invariant classification.

\subsubsection{S3.2.5. Two-fold statistically degenerate ground states at half-filling} 
We now show that the ground states in the subspace at half-filling with $N$ particles in the Rydberg $p$ state, i.e., in the subspace
with $\sigma^z=0$,  are 
two-fold statistically degenerate under open boundary conditions for a topological phase protected by the average inversion symmetry. 

Let $\ket{\Psi_L(C)}$ and $\ket{\Psi_R(C)}$ be the two ground states of the spin-model $\hat{H}(C)$
under open boundary conditions for a random configuration $C$
with corresponding eigenenergy $E_{gL}(C)$ and $E_{gR}(C)$, respectively.
These two states involve edge excitations at the left and right edges, respectively.
For the inversion partner $\mathcal{R}C$ of the configuration $C$,
due to the fact that 
$\mathcal{U}_R \hat{H}(C) \mathcal{U}_R^{-1} = \hat{H}(\mathcal{R} C)$,
the corresponding two ground states of $\hat{H}(\mathcal{R} C)$ are
$\ket{\Psi_L(\mathcal{R}C)}=\mathcal{U}_R \ket{\Psi_R(C)}$ and 
$\ket{\Psi_R(\mathcal{R}C)}=\mathcal{U}_R \ket{\Psi_L(C)}$.
The states $\ket{\Psi_L(C)}$ and $\ket{\Psi_R(C)}$ can be experimentally prepared 
by initializing into a product state $\ket{1010\dots 10}$ or $\ket{0101\dots 01}$, respectively.
We are interested in the occupancy at the left edge for the state $\ket{\Psi_L(C)}$ 
(i.e., $\bra{\Psi_L(C)} \hat{n}_{1}\ket{\Psi_L(C)}$) and 
the occupancy at the right edge for the state $\ket{\Psi_R(C)}$ (i.e., $\bra{\Psi_R(C)} \hat{n}_{2N}\ket{\Psi_R(C)}$).
For a disorder ensemble $\mathcal{E}_C$, the average occupancy at the left and right edges are given by
\begin{equation}
	\overline{n}_{1}= \frac{1}{N_C}\sum_{C \in \mathcal{E}_C } \bra{\Psi_L(C)} \hat{n}_1\ket{\Psi_L(C)},
\end{equation}
and
\begin{equation}
	\overline{n}_{2N}= \frac{1}{N_C}\sum_{C \in \mathcal{E}_C } \bra{\Psi_R(C)} \hat{n}_{2N}\ket{\Psi_R(C)},
\end{equation}
respectively.
Here $N_C$ denotes the number of elements in the ensemble $\mathcal{E}_C$. 
To show that $\overline{n}_{1}=\overline{n}_{2N}$, we only need to prove that they are almost equal when 
$\mathcal{E}_C=\{ C, \mathcal{R}C\}$,
thanks to the average inversion symmetry. The following derivation constitutes the proof
\begin{eqnarray}
\overline{n}_{1}&=& \frac{1}{2} \left[\bra{\Psi_L(C)} \hat{n}_1\ket{\Psi_L(C)}+
\bra{\Psi_L(\mathcal{R}C)} \hat{n}_1 \ket{\Psi_L(\mathcal{R}C) }\right]  \nonumber \\
&=&\frac{1}{2} \left[\bra{\Psi_L(C)} \hat{n}_1\ket{\Psi_L(C)}+
\bra{\Psi_R(C)} \mathcal{U}_R^{-1} \hat{n}_1 \mathcal{U}_R \ket{\Psi_R(C) }\right] \nonumber \\
&=&\frac{1}{2}\left[\bra{\Psi_L(C)} \hat{n}_1\ket{\Psi_L(C)}+
\bra{\Psi_R(C)} \hat{n}_{2N} \ket{\Psi_R(C) }\right] \nonumber \\
&=&\frac{1}{2}\left[\bra{\Psi_R(\mathcal{R}C)} \hat{n}_{2N}\ket{\Psi_R(\mathcal{R}C)}+
\bra{\Psi_R(C)} \hat{n}_{2N} \ket{\Psi_R(C) }\right] \nonumber \\
&=&\overline{n}_{2N}.
\end{eqnarray}
Thus we see the average occupancy at the two edges for the two ground states $\ket{\Psi_L(C)}$ and $\ket{\Psi_R(C)}$ 
over unlimited number of realizations respects the inversion symmetry. 
Similarly to the local DOS in the single-particle case, we can define a many-body local DOS
at the left and right edges contributed by the ground states $\ket{\Psi_L(C)}$ and $\ket{\Psi_R(C)}$
as
\begin{equation}
	\overline{\rho}(1,E)= \frac{1}{N_C}\sum_{C \in \mathcal{E}_C } \bra{\Psi_L(C)} \hat{n}_1\ket{\Psi_L(C)}\delta(E-E_{gL}(C)),
\end{equation}
and 
\begin{equation}
	\overline{\rho}(2N,E)= \frac{1}{N_C}\sum_{C \in \mathcal{E}_C } \bra{\Psi_R(C)} \hat{n}_{2N}\ket{\Psi_R(C)}\delta(E-E_{gR}(C)),
\end{equation}
respectively. 
Similarly, to show that $\overline{\rho}(1,E)=\overline{\rho}(2N,E)$, we only need to prove that they are equal for
$\mathcal{E}_C=\{ C, \mathcal{R}C\}$. The following constitutes the proof:
\begin{eqnarray}
	\overline{\rho}(1,E)&=& \frac{1}{2} \left[\bra{\Psi_L(C)} \hat{n}_1\ket{\Psi_L(C)}\delta(E-E_{gL}(C))+
	\bra{\Psi_L(\mathcal{R}C)} \hat{n}_1 \ket{\Psi_L(\mathcal{R}C) } \delta(E-E_{gL}(\mathcal{R} C))\right]  \nonumber \\
	&=& \frac{1}{2} \left[\bra{\Psi_R(\mathcal{R}C)} \hat{n}_{2N}\ket{\Psi_R( \mathcal{R}C)}\delta(E-E_{gR}(\mathcal{R}C))+
	\bra{\Psi_R(C)} \hat{n}_{2N} \ket{\Psi_R(C) } \delta(E-E_{gR}(C))\right] \nonumber \\
	&=&\overline{\rho}(2N,E).
\end{eqnarray}
Thus, we conclude the two ground states $\ket{\Psi_L(C)}$ and $\ket{\Psi_R(C)}$ are two-fold statistically degenerate.

\subsubsection{S3.2.6. Effects of van der Waals interactions on ground-state degeneracy}
In the previous parts, we have established the concept of the average inversion SPT phase by showing 
that the $\mathbb{Z}_2$ topological invariant
is enforced to be quantized, and the ground states in the half-filling subspace are two-fold statistically degenerate. 

In the following, we will show
how the energy shifts at the edge sites caused by van der Waals interactions lift the four-fold degeneracy. 
Let $|\Psi(N-1)\rangle$ and $|\Psi(N+1)\rangle$ be the ground states of $H_{\mathrm{XXZ}}$ in the subspace with $N-1$
and $N+1$ particles, respectively, and $|\Psi_{L}(N)\rangle$ and $|\Psi_{R}(N)\rangle$
be the two ground states of $H_{\mathrm{XXZ}}$ in the half-filling subspace.
For $H_{\mathrm{XXZ}}$, when it is topologically nontrivial, the four ground 
states are degenerate. However, the $H_{\mathrm{Z}}$ term modifies the spectra.
For simplicity, we consider its first-order correction for a ground state $|\Psi\rangle$, i.e.,
$\Delta E_{\Psi}=\langle \Psi |H_{\mathrm{Z}}|\Psi\rangle=-\frac{1}{4}\sum_{i=1}^{2N} V_i^{\text{vdW} } 
(2 \langle \Psi| \hat{n}_i |\Psi\rangle-1)$, 
where we have expressed $H_{\mathrm{Z}}=-\frac{1}{4}\sum_{i=1}^{2N} V_i^{\text{vdW}} (2\hat{n}_i-1) $
in terms of $\hat{n}_i=\hat{b}_i^\dagger \hat{b}_i$, the particle number operator at site $i$.
For the four states involved,
in the bulk sites with $2 \le i \le 2N-1$, we approximate   
$\langle \Psi| \hat{n}_i |\Psi\rangle \approx 0.5$ so that 
$\Delta E_{\Psi} \approx -\frac{1}{4} \left[ V_1^{\text{vdW}} (2\hat{n}_1-1)+V_{2N}^{\text{vdW}} (2\hat{n}_{2N}-1)\right ] $. 
For $|\Psi_{L}(N)\rangle$ and $|\Psi_{R}(N)\rangle$, there is a particle occupying either the left or the right 
edge, respectively. We thus have  
$\langle \Psi_{L}(N)| \hat{n}_1 |\Psi_{L}(N)\rangle \approx 1$,
$\langle \Psi_{L}(N)| \hat{n}_{2N} |\Psi_{L}(N)\rangle \approx 0$,
$\langle \Psi_{R}(N)| \hat{n}_1 |\Psi_{R}(N)\rangle \approx 0$, and
$\langle \Psi_{R}(N)| \hat{n}_{2N} |\Psi_{R}(N)\rangle \approx 1$.
As a result, $\Delta E_{\Psi_{L}(N)}\approx -\frac{1}{4}(V_1^{\text{vdW}}-V_{2N}^{\text{vdW}} )$
and $\Delta E_{\Psi_{R}(N)} = -\Delta E_{\Psi_{L}(N)}$.
We see that for a typical realization, $\Delta E_{\Psi_{L}(N)} \neq \Delta E_{\Psi_{R}(N)}$, implying 
that the degeneracy is lifted. However, if we consider the inversion partner, 
then $\Delta E_{\Psi_{L}(N)}$ becomes $\frac{1}{4}(V_1^{\text{vdW}}-V_{2N}^{\text{vdW}})$
and $\Delta E_{\Psi_{R}(N)}$ becomes $-\frac{1}{4}(V_1^{\text{vdW}}-V_{2N}^{\text{vdW}} )$.
The average local DOS at the two edges remains equal at the same energy, which we have proven in the previous subsubsection to 
show that the two states are statistically degenerate.
For $|\Psi(N-1)\rangle$, since no particles occupy the edges,
$\langle \Psi(N-1)| \hat{n}_1 |\Psi(N-1)\rangle \approx 0$ and
$\langle \Psi(N-1)| \hat{n}_{2N} |\Psi(N-1)\rangle \approx 0$.
Thus, $\Delta E_{\Psi(N-1)}=\frac{1}{4}(V_1^{\text{vdW}}+V_{2N}^{\text{vdW}} )$.
For $|\Psi(N+1)\rangle$, there are two particles occupying the edges so that
$\langle \Psi(N+1)| \hat{n}_1 |\Psi(N+1)\rangle \approx 1$ and
$\langle \Psi(N+1)| \hat{n}_{2N} |\Psi(N+1)\rangle \approx 1$.
Thus, $\Delta E_{\Psi(N+1)}=-\frac{1}{4}(V_1^{\text{vdW}}+V_{2N}^{\text{vdW}})$.
We therefore conclude that van der Waals interactions lift the degeneracy between 
the two ground states $\Delta E_{\Psi(N-1)}$ and $\Delta E_{\Psi(N+1)}$, 
with their energy difference of about $\frac{1}{2}|V_1^{\text{vdW}}+V_{2N}^{\text{vdW}}|$.

Figure~\ref{figvdW}d shows that $V_1^{\text{vdW}}/h, V_{2N}^{\text{vdW}}/h \sim 0.04$ MHz,
suggesting an energy difference of about $0.04$ MHz. 
However, our numerical results show that the two ground states exhibit an energy difference 
of about $0.2$ MHz on average.
This can be probably attributed to the distribution of edge states inside the bulk sites, where
energy shifts by van der Waals interactions are much larger than at the edges.
The energy difference between the ground states in the subspace with $N-1$ and 
$N$ particles is about $0.1$ MHz, which is consistent with our experimental observation that
an occupancy peak at two edges arises at the detuning of $-0.1$ MHz (see Fig. 4a in the main text).

\begin{figure}
	\includegraphics[width=0.8\textwidth]{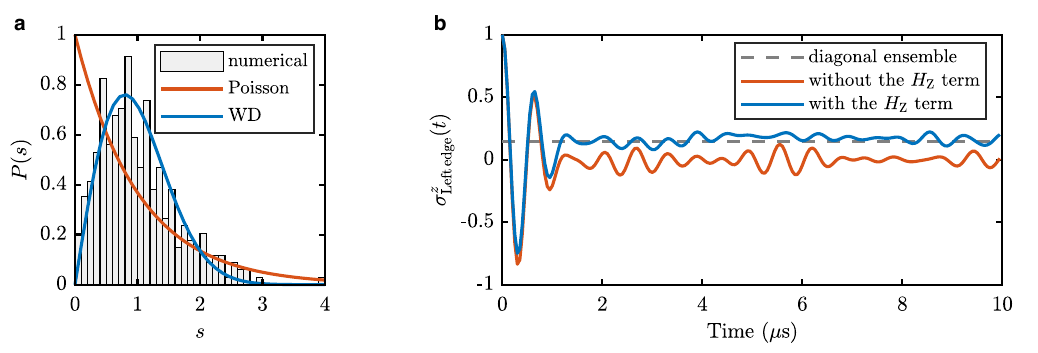}
	\caption{\label{fig:vdwQD} The residual edge spin magnetization for the quench dynamics in the regular configuration.
		\textbf{a,} Level-spacing statistics for the Hamiltonian in a regular lattice at half-filling with inversion symmetry resolved. The expected distributions for an integrable model (Poisson) and the corresponding chaotic or thermalized model (Wigner-Dyson) are plotted. The energy level spacing $s$ is computed using the middle 20\% of all eigenstates and rescaled such to $ \langle s \rangle = 1 $.
		\textbf{b,} Magnetization from numerical simulation of quench dynamics for the left edge spin with and without the $H_{\mathrm{Z}}$ term. The dashed line represents the predicted value in the diagonal ensemble and the right edge spin shows similar long time behavior.}
\end{figure}

\subsubsection{S3.2.7. Effects of van der Waals interactions on quench dynamics}
In Fig.~4 of the main text, we show that there exists a small residual value of magnetization $\langle \sigma_i^z \rangle$ for the edge spins over long time 
in the regular case in a topologically trivial phase. 
Here, we will demonstrate that the small residual value results from the inhomogeneous 
$H_{\mathrm{Z}}$ term in Eq.~(\ref{eq:HamXXZ-Z}) introduced by van der Waals interactions. 
Notably, the $\sigma^z$ terms are nearly zero at the edges while being finite and large in the bulk (see Fig.~\ref{figvdW}c).

Without the $H_{\mathrm{Z}}$ term, we observe that the initially polarized local spins delocalize and spread across the chain due to spin exchange processes, i.e., a particle can hop to other sites in the hard-core boson language.
Over time, the spin distribution becomes homogeneous, leading to an average magnetization of $\langle \sigma_i^z \rangle = 0$ at every site (see Fig.~\ref{fig:vdwQD}b).
In the presence of a homogeneous $H_{\mathrm{Z}}$ term, the physics remains unchanged 
since the dynamics are restricted to the subspace of $\sigma^z=0$. 
This indicates that the slight residual arises from the difference in the $H_{\mathrm{Z}}$ term 
between the edges and the bulk.
Due to this difference in onsite potentials, bosons at the edges prefer to move into the bulk, with the energy decrease compensated for by Ising interactions between the particles.

In the regular lattice without disorder, we find that the energy-level spacing
statistics satisfies the Wigner-Dyson distribution, as shown in Fig.~\ref{fig:vdwQD}a,
indicating that a highly excited state can evolve in time to a thermal equilibrium state.
Based on this fact, for an initial state $ \ket{\Psi(0)}= \sum_{\alpha} c_{\alpha} \ket{E_\alpha} $, 
where $ \ket{E_\alpha} $ is a Hamiltonian eigenstate with energy $E_\alpha$ and $c_{\alpha}$ is the corresponding expansion coefficient, 
the long-time expectation of $\sigma_i^z$ for the initial state is given by the diagonal ensemble average,
\begin{equation}
	\langle \sigma_i^z(t) \rangle_{t\to \infty} \approx \sum_{\alpha} |c_{\alpha}|^2 \langle E_\alpha | \sigma_i^z | E_\alpha \rangle.
\end{equation} 
We find that the expected value of $\langle \sigma^z \rangle$ at the edge is $0.15$, 
closely matching the numerical simulations (see Fig.~\ref{fig:vdwQD}b).


\end{widetext}

\end{document}